\documentclass[a4paper,pdftex]{scrreprt}
\pdfoutput=1

\usepackage{setspace}
\usepackage{graphicx}
\usepackage[monochrome]{color}
\usepackage[numbers]{natbib}
\usepackage[pdftex,colorlinks=false,backref]{hyperref}
\usepackage[squaren]{SIunits}

\usepackage{amsmath,amssymb}
\usepackage{listings}
\newcommand{\plotc}{\includegraphics[clip=true,trim=2cm 1cm 2cm 1cm,width=0.2\textwidth]}
\newcommand{\plota}{\includegraphics[clip=true,trim=2cm 1cm 2cm 1cm,width=0.5\textwidth]}

\usepackage{array}

\graphicspath{{images/}}

\begin{document}

\title{Planck focal plane instruments: advanced modelization and combined analysis}
\author{Andrea Zonca}
\date{}

\begin{titlepage}

\begin{center}
{\Large \sc{Universit\`a degli Studi di Milano}}\\
Dottorato di Ricerca in Fisica, Astrofisica e Fisica Applicata\\
$\,$\\
and\\
$\,$\\
{\Large \sc{Universit\'e Paris Diderot - Paris 7}}\\
Doctorat Champs, Particules, Matieres\\

\vfill

\vspace{1cm}
{\Huge \bf  {Advanced modelling and combined data analysis of Planck focal plane instruments}}\\
\vspace{0.2cm}
Thesis submitted for the degree of Doctor Philophi\ae\\
s.s.d FIS/05
\vspace{2cm}

\vfill

\begin{flushleft}
{\sc Director of the Doctoral School (I)}: Prof.~Gianpaolo Bellini\\
{\sc Director of the Doctoral School (F)}: Prof.~Yves Charon\\
\vspace{1cm}
{\sc Thesis Director (I)}: ~Aniello Mennella\\
{\sc Thesis Director (F)}: ~Jean-Michel Lamarre\\
\end{flushleft}

\vspace{1.cm}

\begin{tabular}{l @{\hspace{7cm}} l}
& {\sc Candidate:}\\
& {\Large Andrea Zonca}\\
& cycle XXI\\
\end{tabular}

\vspace{2.5cm}

\vfill

Academic Year: 2008/2009
\end{center}
\end{titlepage}

\pagenumbering{roman}
\tableofcontents
\listoffigures
\listoftables

\begin{abstract}

This thesis is the result of my work as research fellow at \href{http://www.iasf-milano.inaf.it}{IASF-MI}, Milan section of the Istituto di Astrofisica Spaziale e Fisica Cosmica, part of INAF, Istituto Nazionale di Astrofisica.
This work started in January 2006 in the context of the PhD school program in Astrophysics held at the \href{http://www.fisica.unimi.it}{Physics Department of Universita' degli Studi di Milano} under the supervision of Aniello Mennella.

The main topic of my work is the software modelling of the Low Frequency Instrument (LFI) radiometers. The LFI is one of the two instruments on-board the European Space Agency Planck Mission for high precision measurements of the anisotropies of the Cosmic Microwave Background (CMB).

I was also selected to participate at the International Doctorate in Antiparticles Physics, \href{http://www.fe.infn.it/idapp}{IDAPP}. IDAPP is funded by the Italian Ministry of University and Research (MIUR) and coordinated by Giovanni Fiorentini (Universita' di Ferrara) with the objective of supporting the growing collaboration between the Astrophysics and Particles Physics communities. It is an international program in collaboration with the Paris PhD school, involving Paris VI, VII and XI Universities, leading to a double French-Italian doctoral degree title.

My work was performed with the co-tutoring of Jean-Michel Lamarre, Instrument Scientist of the High Frequency Instrument (HFI), the bolometric instrument on-board Planck. Thanks to this collaboration I had the opportunity to work with the HFI team for four months at the Paris Observatory, so that the focus of my activity was broadened and included the study of cross-correlation between HFI and LFI data.
Planck is the first CMB mission to have on-board the same satellite very different detection technologies, which is a key element for controlling systematic effects and improve measurements quality.

The thesis is organised in four chapters:
    \begin{enumerate}
        \item a short introduction focused on state-of-art CMB phenomenology 
        \item a chapter about the Planck mission mainly focused on the LFI instrument
        \item a chapter about software modelling of the LFI radiometers which includes a detailed description of the LFI instrument. Here I also discuss the experimental data available from the measurements campaigns on radiometer components, the model implementation and its validation against the frequency response measurements
        \item a chapter about the satellite thermal environment, with particular reference to  the stage cooled at 4K, which is of key importance for both instruments. In this chapter I show the result of the analysis of the propagation of temperature fluctuations through the HFI.
        \item a chapter about cross-correlation of HFI and LFI data. In this chapter I describe the implemention of data analysis sessions in the KST data visualization software with the purpose of simplifying and standardsing the cross-correlation analysis
    \end{enumerate}

\end{abstract}

\pagenumbering{arabic}


\chapter{Cosmology}
\label{ch:cosmology}

In this chapter I will give a short overview of the current knowledge on cosmology in order to understand the fundamental role of Cosmic Microwave Background measurements.
I prefer not to follow the historical development, as usual, but focus just on the current state-of-art.

\section{Standard Model of Cosmology}

In the last twenty years, the collection of a huge amount of observational data has greatly contributed to test different theoretical models of the birth and evolution of the Universe and led to the definition of a generally accepted Standard Model of Cosmology.

According to this model the Universe began about 14 billion years ago when it started expanding from a hot and dense state. It was highly homogeneous with fluctuations of the energy density which have subsequently grown by gravitational instability to form the cosmic structures (galaxies, galaxy clusters and superclusters) observed today.

Observations picture a Universe uniform at large scale, this supports the Cosmological Principle which states that the Universe is spatially homogeneous and isotropic at large scale.
Under this assumption it is possible to build a spherically symmetric metric model for the spacetime metric, called Friedmann-Lema\^{\i}tre-Robertson-Walker (FLRW) metric, \cite{liddle}:
    
    \begin{equation}
        ds^2 = - dt^2 + a^2(t) [ \dfrac{dr^2}{1-kr^2} + r^2(d\phi^2 + sin^2\theta d\phi) ]
        \label{eq:friedmann_metric}
    \end{equation}

where: 
    \begin{itemize}
         
\item $r$ is the comoving distance, i.e. the distance in a reference frame expanding together with the Universe,
\item $a(t)$ is the cosmic scale factor, which describes the expansion of the Universe and represents the rate at which two points of fixed coordinates $(r_1,\theta_1,\phi_1)$ and $(r_2,\theta_2,\phi_2)$ increase their mutual distance with time. $a(t)$ is adimensional and usually normalised to be equal to 1 today ($t_0$), therefore a generic distance $r$ evolves following this equation:
\begin{equation}
    r = a(t) * r(t_0) 
\end{equation}

\item $k$ is the curvature parameter which can have only three possible values: $k=0$ for a spatially flat, Euclidean Universe, $k= + 1$ for a closed Universe with positive spherical curvature, and $k = -1$ for an open Universe with negative hyperbolic curvature.

    \end{itemize}

In order to study the dynamics of the homogeneous expanding Universe in the context of General Relativity \cite{wald1984}, it is necessary to solve the Einstein Field Equation \cite{1931Einstein}, whose modern formulation is:
    
\begin{equation}
    G_{\mu \nu} = 8 \pi G T_{\mu\nu}
    \label{eq:einstein}
\end{equation}

few details about this equation:
    \begin{itemize}
    \item it assumes units where $c$, the speed of light, is equal to $1$
        \item  $G_{\mu \nu}$, the Einstein tensor, expresses spacetime curvature
        \item $G$ is the universal gravitational constant
        \item $T_{\mu\nu}$, the stress energy tensor, includes also the contribution of Dark Energy or vacuum energy (\cite{2006ApJ...650....1W}): a hypothetical exotic form of energy that permeates the space and tends to increase the rate of expansion of the Universe; it is the favoured hypothesis for explaining the acceleration rate of the Universe expansion \cite{1998Riess} and \cite{1999Perlmutter} .
        The simplest model of vacuum energy is the cosmological constant (\cite{2001Carroll}), whose stress energy tensor is written in equation~\ref{eq:cco}.
        \begin{equation}
            T_{\mu\nu}^{(vac)} = - \dfrac{\Lambda}{8 \pi G} g_{\mu\nu}
            \label{eq:cco}
        \end{equation}
    \end{itemize}
    
Therefore Einstein field equation, \ref{eq:einstein}, states that the curvature of spacetime is due to the matter/energy content of the Universe, which includes Matter, Radiation and vacuum energy.

FLRW metric (Equation~\ref{eq:friedmann_metric}) has an analytic solution to into Einstein's field equations \ref{eq:einstein} by assuming the energy momentum tensor $T_{\mu\nu}$ isotropic and homogeneous.
The results are called Friedmann equations:
    
    \begin{eqnarray}
        (\dfrac{\dot{a}(t)}{a(t)})^2  & = & \dfrac{8 \pi G}{3} \rho - \dfrac{k c^2}{a^2} \\
        \dfrac{\ddot{a}(t)}{a(t)})  & = & - \dfrac{4 \pi G}{3} (\rho + 3p)
        \label{eq:friedmann_eq}
    \end{eqnarray}

where $\rho$ is the global energy density of the Universe, which includes contribution of Matter, Radiation and Vacuum energy, $k$ is the curvature parameter defined above and $G$ the universal gravitational constant.

Cosmological constant, or vacuum energy, is the simplest model of Dark Energy, a hypothetical exotic form of energy that permeates the space and tends to increase the rate of expansion of the Universe; it is the favoured hypothesis for explaining the recent observations  that the Universe appears to be expanding at an accelerating rate.

Applied to a fluid with a given equation of state, the Friedmann equations give the time evolution and geometry of the universe as a function of the fluid density. 

\subsection{Hubble parameter}

Following FLRW metric, equation~\ref{eq:friedmann_metric}, two points at a distance $d = a(t) r$  will move apart with a velocity $v = \dot{a}r = Hd $. This result was obtained by Hubble (\cite{1929Hubble}) thanks to measurements of the recession velocity of galaxies, and $H$ is Hubble constant \footnote{Measurement from WMAP5, Baryon Acoustic Oscillations and SuperNovae, considering Lambda CDM model}:
    \begin{equation}
        H = 70.1 \pm 1.3 km/s/Mpc 
    \end{equation}

Based on the first equation of Friedmann (\ref{eq:friedmann_eq}) it is possible to define the critical density
\begin{equation}
    \rho_c = \dfrac{3H^2}{8\pi G}
\end{equation}
such that density values $\rho$ above, below or equal to $\rho_c$ refer to closed, open or flat Universe.

$\rho_c$ is also used to normalise the densities in order to obtain relative parameters:
    \begin{itemize}
        \item Matter: $\Omega_m=\dfrac{\rho_m}{\rho_c}$       
         \item Radiation: $\Omega_r=\dfrac{\rho_r}{\rho_c}$

                \item Cosmological constant: $\Omega_\Lambda=\dfrac{\rho_\Lambda}{\rho_{c}}$
        \item Total: $\Omega=\dfrac{\rho}{\rho_c} = \Omega_m + \Omega_r + \Omega_\Lambda$
    \end{itemize}
therefore $\Omega = 1$ means flat, $\Omega > 1$ closed and $\Omega < 1$ open Universe.

Observations \cite{2008arXiv0803.0593N} strongly favour a flat Universe, with a global density near to the critical density.

The equation of state, i.e. the relationship between energy density and pressure, can be written, for  $c=1$, simply $p= w \rho$. Once it is specified, the two Friedmann equations can be solved for the scale factor $a(t)$ and give solutions for a Universe dominated by different type of energies:
    
    \begin{itemize}
        \item Matter domination, $w=\dfrac{1}{3}$, $\rho \propto a(t)^{-4}$, $a(t) \propto t^{2/3}$
        \item Radiation domination, $w=0$, $\rho \propto a(t)^{-3}$, $a(t) \propto t^{(1/2)}$
        \item Cosmological constant domination, $w=-1$, $\rho \text{ constant}$, $a(t) \propto e^{(Ht)}$
    \end{itemize}

Extrapolating the components time dependence back in time, it is clear that the early Universe was Radiation dominated; it followed a Matter domination epoch and, thanks to measurements of the present value of $\Omega_{\Lambda}$ [\cite{2008arXiv0803.0593N}]  , it is possible to state that just in the present epoch the cosmological constant and the Matter are at the same energy density level.

\subsection{Hot Big Bang}

Friedmann's equations and galaxies recession, propagated back in time, suggest an initial hot and dense state, named Big Bang \cite{bigbang}, whose physics is still beyond our understanding, which was rapidly expanding and cooling.

Galaxies recession velocity measurements are a strong evidence of the Big Bang, in this section I will review the other main pieces of evidence of the Hot Big Bang theory:
\begin{itemize}

\item Few minutes after the Big Bang, after the primordial plasma cooled and formed protons and neutrons, light elements started forming by nuclear fusion until the Universe became too cold to allow nuclear reactions. The primordial abundances of light elements (Helium-4, Helium-3, deuterium and lithium-7) can be computed from mathematical models as ratios to the amount of hydrogen, H. Mass ratios predicted with this method agree with the abundances measured in galactic surveys: hydrogen $\sim 76\%$, helium $\sim 24\%$, see \cite{bbn}.

\item The third evidence of a Hot Big Bang is galactic evolution and distribution: population of stars have been evolving so that distant galaxies, observed as they were in the early Universe, appear very different from nearby galaxies, observed in a more recent state. Moreover, Big Bang simulations match well the observed star formation, galaxy and quasar distribution.

\item The evidence that, at least historically, had been the most important is the observation of the Cosmic Microwave Background, a nearly isotropic microwave radiation at few Kelvins which was started propagating when the Universe was very hot. Its photons started propagating 379000 years after the Big Bang at a temperature of few thousands Kelvins.
After $\sim 14 Gyr$ propagating in the expanding Universe these photons have lost most of their energy due to spacetime dilation and are now detectable as microwave black body at $\sim 2.7 K$
\end{itemize}

\subsection{Inflation}
\label{sec:inflation}

The standard Big Bang (BB) model however, cannot explain some features of the present Universe, in particular:

    \begin{itemize}
    \item Flatness: Big Bang expansion cannot constrain the Universe to be so near to flatness at the level of few percent unless very special initial conditions are chosen
    \item Homogeneity and isotropy: CMB photons, since the birth of the Universe, had time just to cover a small part of the Universe ($\sim 10^{14}m$), about $1\degree$ in the sky today, leading to just small part of causally connected regions. This contrasts with the fact that CMB is highly homogeneous at all scales.
    \item Structure formation: the standard BB model provide no explanation of the source of the inhomogeneities which lead to structure formation
    \end{itemize}
    
All these issues have been elegantly explained by inflation \cite{1981Guth}, the theory which assumes the early Universe passed through a phase of exponential expansion driven by a negative-pressure vacuum scalar field, the inflaton.

The flatness issue is solved by rewriting the Friedmann equation (\ref{eq:friedmann_eq}), as:
    \begin{equation}
        (\Omega ^{-1} - 1) \rho a^2 = \dfrac{-3k c^2}{8 \pi G}
    \end{equation}

the right hand side of the equation is constant, therefore during inflation, the $\rho a^2 $ term increases extremely rapidly as the scale factor $a$ grows exponentially, therefore the term $| \Omega^{-1} - 1 |$ must decrease with time.
This means that whatever is the initial value of the Universe energy density $\Omega$, after inflation it is forced to be very close to 1, an almost perfectly flat Universe. Subsequent evolution of the Universe will cause the value to grow, bringing it to the currently observed value of about 1.01.

Homogeneity and isotropy are explained by the fact that the Universe was causally connected and therefore in thermal equilibrium just before inflation.

Finally the inhomogeneities needed for structure formation are due to quantum fluctuations in the primordial plasma which are amplified to the seeds of the current astrophysical structures.

\subsection{Lambda Cold Dark Matter Model}
\label{sec:cdm}

Lambda Cold Dark Matter model, or $\Lambda CDM$, is the Universe cosmological model which currently explains better the observations: it includes the Friedmann metric, the Hot Big Bang model and the inflation as explained in the previous sections.

Its most important features are:
    \begin{itemize}
        \item the Cold Dark Matter (CDM) model, where the dark matter is predicted to be cold (i.e. having a non-relativistic velocity at the epoch of radiation-matter equality), non-baryonic, dissipationless (cannot cool by radiating photons) and collisionless (i.e. interacting only through gravity).
        \item the cosmological constant, $\Lambda$, a dark energy term that allows for the current accelerating expansion of the Universe
    \end{itemize}

The $\Lambda CDM$ model is characterised by six basic parameters, which needs to be measured through observation of the CMB, SuperNovae and Galaxies Sky Surveys.

\subsection{Content of the Universe}

\begin{figure}[h]
    \centering
    \includegraphics[width=.4\textwidth]{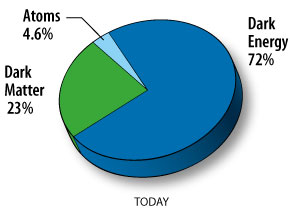}
    \caption{Universe energy composition today}
    \label{fig:wmapcontent}
\end{figure}

    Figure~\ref{fig:wmapcontent} shows the current energy content of the Universe derived from observations: atoms are only a small part, less than 5\%, while Dark Matter and Dark Energy dominate the picture.

\section{Cosmic Microwave Background}
\label{sec:cmb}

CMB is a microwave radiation which pervades all the Universe, it has a thermal blackbody spectrum at the temperature of $\sim 2.7K$, therefore its spectrum peaks at about 160 GHz.

\subsection{Origin}
In the first $~300,000$ years after the Big Bang, thermodynamic equilibrium between matter and radiation was maintained by Thomson scattering between free electrons and photons; after this period the temperature was low enough ($\sim 3000 K$) to allow for the combination of electrons and protons into neutral hydrogen: the reduced density of free electrons made the matter transparent to radiation, which started to propagate freely. 

The time when this event happened is called recombination or decoupling (about 390,000 years after the Big Bang); the photons we receive today have travelled freely from a surface, called last scattering surface, to the Earth. They travelled into a an expanding spacetime and were redshifted from the original $~3000 K$ to  $\sim 2.7 K$, and their temperature will continue to drop as the Universe expands.

\subsection{Anisotropies and power spectrum}

As described in section~\ref{sec:inflation}, inflation amplified quantum fluctuations and generated the  inhomogeneities in the primordial plasma that subsequently seeded structure formation.

\begin{figure}[h]
    \centering
    \includegraphics[width=\textwidth]{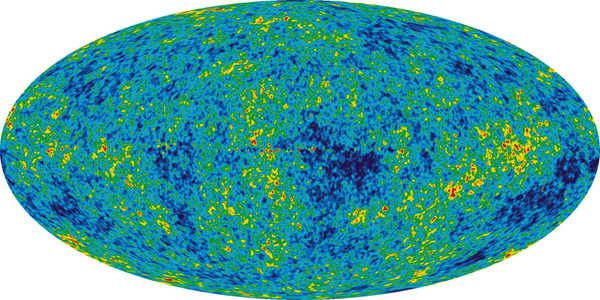}
    \caption{Temperature map of CMB anisotropies measured by WMAP after 5 years of mission}
    \label{fig:wmap5}
\end{figure}

The best measurements of CMB anisotropies until now was given by the satellite WMAP after 5 years of observations, figure~\ref{fig:wmap5}; it shows a very detailed and complex picture of the inhomogeneities of the last scattering surface, their amplitude is:
    \begin{equation}
        \dfrac{\Delta T}{T} \sim 10^{-5}
    \end{equation}
where $T$ is CMB temperature and $\Delta T$ anisotropies temperature amplitude.

The properties of CMB anisotropies are usually represented by expanding it into spherical harmonics:
    \begin{equation}
            \dfrac{\Delta T}{T}(\theta,\phi) = \sum_{\ell,m} a_{\ell m} Y_{\ell m}(\theta,\phi) 
    \end{equation}
where $ Y_{\ell m}(\theta,\phi)$ is the spherical harmonic function of degree $\ell$ and order $m$, $\ell \sim \pi / \theta$ is inversely proportional to the angular scale $\theta$, and $a_{\ell m}$ are the multipole moments which relate the amplitude of the temperature fluctuations to the angular scales.
Due to the cosmological principle, temperature anisotropies do not have a preferred direction in the sky, therefore $a_{\ell m}$ can be averaged over $m$ to obtain:
    \begin{equation}
        C_\ell \equiv \langle |a_{\ell m}|^2 \rangle
    \end{equation}

The CMB angular power spectrum is constituted by the coefficients $C_\ell$ and it is the most important tool for comparing theoretical predictions with CMB measurements.

\begin{figure}[h]
    \centering
    \includegraphics[width=.6\textwidth]{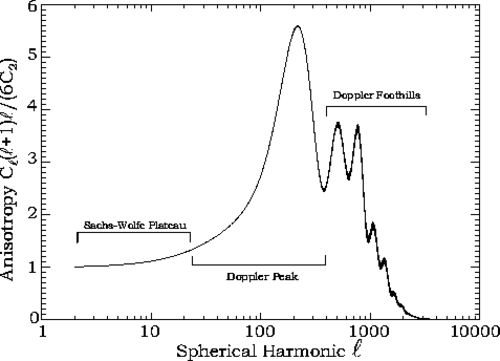}
    \caption{Cosmic Microwave Background power spectrum, main features}
    \label{fig:powerspec}
\end{figure}

\paragraph{Primary anisotropies}

Primary anisotropies originated during or before recombination and were generated by quantum fluctuations amplified by inflation. The inhomogeneities present in the Universe after inflation produce anisotropies through three different effects, see figure~\ref{fig:powerspec}:
    
    \begin{itemize}
        \item \textbf{Acoustic oscillations:} gravity and radiation pressure triggered sound waves that alternatively compressed and rarefied regions of the primordial plasma. After the Universe had cooled enough to allow the formation of neutral atoms, the pattern of density fluctuations caused by the sound waves was frozen into the CMB. 
      The first peak in the CMB angular power spectrum (the so-called
      Doppler peak, at about $1 \degree$) is therefore due to a wave that had a density maximum
      just at the time of last scattering; the secondary peaks at higher multipoles (lower angular scale, so-colled Doppler foothills) are
      high harmonics of the principal oscillations. The effect is directly proportional to the density fluctuations $\Delta\rho$:
          \begin{equation}
              \dfrac{\Delta T}{T} \propto \dfrac{\Delta \rho}{\rho}
            \end{equation}

    \item \textbf{Gravitational perturbations:} Photons coming from high-density regions
      undergo a gravitational redshift, this phenomenon is called Sachs-Wolfe effect (SW). 
      Its effect impacts angular scales larger than the horizon at the last scattering $\theta \gtrsim 2 \degree$  and 
      $ l \lesssim 90 \degree $.
       The global effect of these anisotropies is
                          \begin{equation}
              \dfrac{\Delta T}{T} = - \dfrac{\Delta \Phi}{3}
            \end{equation}
            
    where $\Delta \Phi$ is the gravitational potential.
    
    \item \textbf{Silk damping:} recombination is not an instantaneous process but takes a finite time, during this time the photons had time to diffuse and smooth out anisotropies at multipoles higher than $\sim 1000$, corresponding to angular scales $\theta \lesssim 10 '$
 \end{itemize}

\paragraph{Secondary anisotropies}

Secondary anisotropies are due to interactions that take place between the last scattering surface and the Earth, the most important are:
    \begin{itemize}
        \item gravitational lensing: CMB photons deviation by gravitational attraction of massive objects like galaxies or clusters of galaxies
        \item Sunyaev-Zel\'dovic (SZ) effect: Compton scattering of CMB photons by non-relativistic electron gas present in galaxy clusters
    \end{itemize}

\paragraph{Dipole anisotropy}
\label{calibration}

The Solar System is travelling at about $360 Km/s$ with respect to its Local Group, this is evident and was first measured by the Doppler effect on the CMB which gives a strong signal, at about $3 mK$, superimposed to the primordial and secondary anisotropies.

The dipole anisotropy signal was measured with high precision by COBE DMR (\cite{1996ApJ...470...38L}) at the level of $\sim 0.4 \%$; thanks to its stability, strenght, availabilty through the full sky and the same spectrum as the CMB, the dipole is the natural calibrator for anisotropies experiments.

Thanks to the calibration of the dipole, Planck is expected to have between $1$ and $2 \%$ photometric gain uncertainties for observation lenghts shorter than a day, see \cite{2003A&A...409..375C}

\subsection{CMB Polarisation}

Thomson scattering of the CMB photons during recombination generated linearly polarised photons due to the presence of quadrupole anisotropy. 

Only photons which last scattered in an optically thin region could have possessed a quadrupole anisotropy, and this depends on the duration of the last scattering. For a standard thermal history, the expected level of polarization is $10\%$, therefore the level of E-modes is about one order of magnitude less than temperature signal.

Moreover, polarisation data are complementary to temperature measurements, and could be useful in breaking parameter degeneracies and thus constraining cosmological parameters more accurately.

\subsection{Foregrounds}

Foregrounds is a general term for referring any emission that confuse the primordial CMB signal after the time of last scattering. It contains extragalactic discrete point sources and galactic emissions due to synchrotron, free-free and dust.

\begin{figure}[h]
    \centering
    \includegraphics[width=\textwidth]{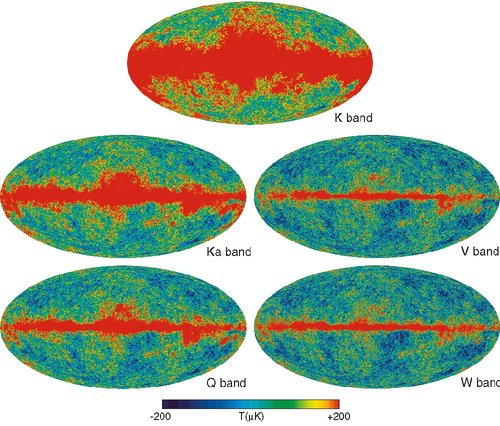}
    \caption{WMAP 5 years temperature maps at all frequencies after dipole removal}
    \label{fig:wmap5maps}
\end{figure}

The best way to understand the impact of foregrounds in CMB measurements is figure~\ref{fig:wmap5maps}, where all 5 temperature maps from WMAP five years are shown. Galactic foreground emission is very strong and widespread, mostly at low frequencies. Sophisticated component separation techniques are used in order to disentangle primordial CMB from foregrounds.

The main help for this task comes from the different spectral indexes of each foreground with respect to the CMB and to other foregrounds; the spectral index $\beta$ is the exponent of the power law dependence for brightness temperature $T_b$ with respect to frequency $\nu$:
    \begin{equation}
        T_b = \nu^{-\beta}
    \end{equation}
multifrequecy measurements are therefore mandatory for separating the different contributions.

\begin{figure}[h]
    \centering
    \includegraphics[width=0.6\columnwidth]{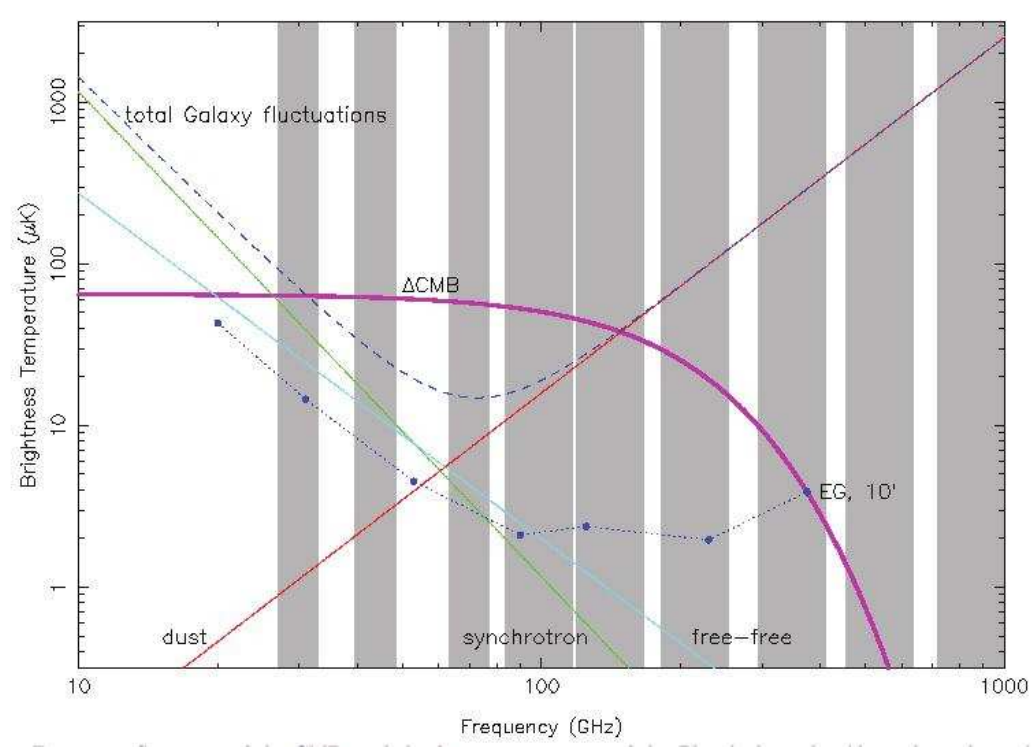}
    \label{fig:foregrounds}
    \caption{Synchrotron, dust, and free-free foregrounds spectra superimposed to {\sc Planck} channels bandwidths and to the expected CMB signal at 1 degree scale; in log-log scale, the exponential dependence of foregrounds emission versus frequency is represented by straight lines}
  \end{figure}

Figure~\ref{fig:foregrounds} displays the application of this concept to the Planck space mission, that will be presented in section \ref{sec:planck}: Planck has a huge frequency spectrum coverage and it will measure parts of the spectrum where the relative contribution of the components is very different, the extreme case is 857 GHz which is going to measure the distribution of dust emission.
 
\clearpage

\subsection{Precision Cosmology with CMB}
\label{sec:precisioncosmology}

A precise measurement of the CMB power spectrum allows to validate the prediction of different cosmological models, as the $\Lambda CDM$ model reviewed in section~\ref{sec:cdm}, and to compute with high precision the related cosmological parameters, six for the $\Lambda CDM$ model.

\begin{figure}[h]
    \centering
    \includegraphics[width=.7\textwidth]{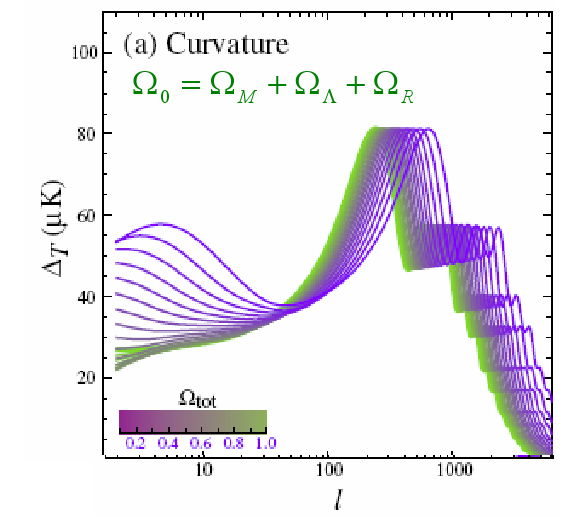}
    \caption{CMB power spectrum simulation for different values of $\Omega$}
    \label{fig:curvature}
\end{figure}

The most simple cosmological parameter measurable with CMB is the total energy density parameter $\Omega$, it is strictly related to the position ($l$) of the first peak; analytically can be computed with the following equation:
    \begin{equation}
        l_{peak} \approx 220 \Omega^{-1/2}
    \end{equation}
The more general approach, which is used to estimate most of the cosmological parameters, is the comparison of the measured power spectrum with predictions given by the cosmological models. For example simulations of CMB power spectra for different value of $\Omega$ are shown in figure~\ref{fig:curvature}: a measurements with low uncertainty is able of constrain more efficiently the value of $\Omega$.

Equivalent procedures can be used to study other cosmological parameters, but more often all the relevant parameters are evaluated simultaneously using Monte Carlo techniques.

The accuracy of the power spectrum estimation can be analytically modelled considering instrument noise and angular resolution, see \cite{imagingfirstlight}:
    \begin{equation}
        \dfrac{\delta C_\ell}{C_\ell} = f_{sky}^{-1/2} \sqrt{\dfrac{2}{2l+1}}[1+\dfrac{A \theta_{pix}\sigma_{pix}^2}{N_{pix} C_\ell W_\ell}]
    \end{equation}
where:
    \begin{itemize}
        \item the term $\dfrac{2}{2l+1}$ is the contribution due to Cosmic Variance, which is the ultimate accuracy limit for CMB power spectrum and is related to the fact that the CMB field is just a single realization of a stochastic process and it is not expected to follow the average of the possible realization, this effect is strong on the largest scales.
        \item  $f_{sky}$ is the fraction of sky covered, which has to consider also eventual cuts of the galaxy plane
        \item  $\theta_{pix}$ is the angular resolution of the optics
        \item  $\sigma_{pix}$ is the noise per pixel
        \item $W_\ell$ is the beam window function
        \item $N_{pix}$ is the number of pixels
        \item $A$ is the surveyed area
    \end{itemize}

\begin{figure}[h]
    \centering
    \includegraphics[width=0.8\textwidth]{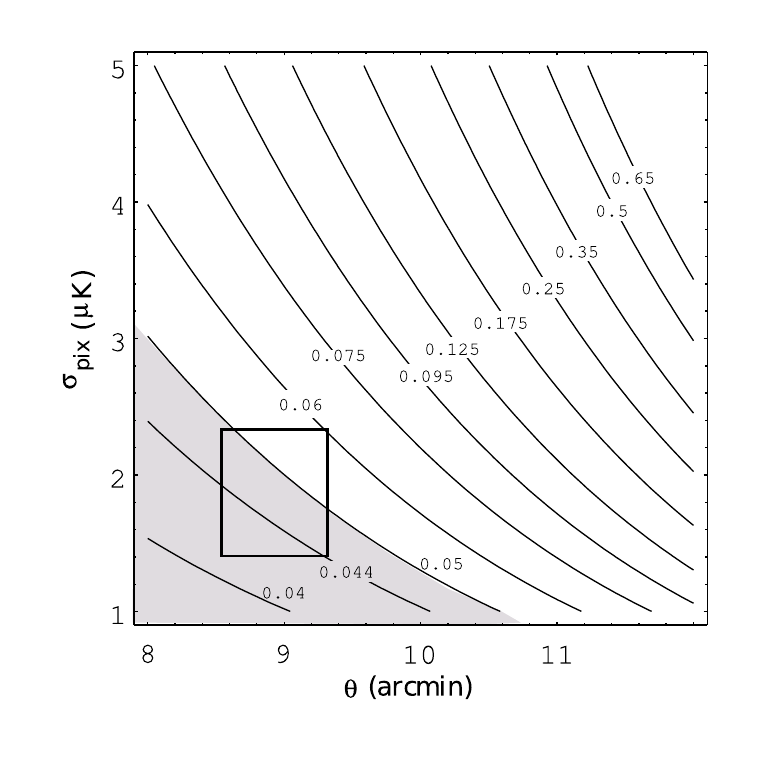}
    \caption{$\dfrac{\delta C_\ell}{C_\ell}$ for $\ell = 1500$ as a function of the angular resolution $\theta$ and of the average noise per pixel $\sigma_{pix}$}
    \label{fig:deltacl}
\end{figure}

Figure~\ref{fig:deltacl} shows the values of $\dfrac{\delta C_\ell}{C_\ell}$ for $\ell = 1500$, the scientific requirement of the Planck mission is a precision of $5\%$, which is represented by the grey area in the figure; the rectangle encloses the expected performance of the 100 GHz Planck HFI channel.
These requirements are strong drivers in instrument design: $\sim 10$ arc-minute angular resolution at 100 GHz requires an antenna of $\sim \lambda/\theta \sim 1.5 m$, $\mu K$ sensitivity requires an array of feed horns cryogenically cooled with at least 1 year of integration time.

This estimation didn't take into account calibration and systematic errors; their impact should be $ \lesssim 1-2 \% \sigma_{pix}$ in order not to dominate the power spectrum; calibration accuracy of the order of $1\%$ is achivable using the CMB dipole signal, systematics due to foregrounds need good frequency coverage, as explained above. 

%

\clearpage

\chapter{Planck Mission}
\label{sec:planck}

%


The European Space Agency Planck satellite is designed to fully extract the cosmological information contained in CMB anisotropy by setting angular resolution, spectral coverage
and sensitivity such that the power spectrum reconstruction will be limited by unavoidable cosmic variance and astrophysical foregrounds. Planck will also provide a precise measurement of
the TE correlation and of the E-mode polarisation power spectrum and possibly offer a first
B-mode detection.

Planck objective is to provide full sky maps with angular resolution between $5'$ and $33'$, spectral coverage between $30$ and $857$ GHz and sensitivity of $\sim 10^{-6}$. Planck performance and design derive, as discussed in section \ref{sec:precisioncosmology}, from the objective of $5\%$ precision on CMB temperature power spectrum at small scales ($1500 < \ell < 2000$)

\begin{figure}[h]
    \centering
    \includegraphics[width=0.6\textwidth]{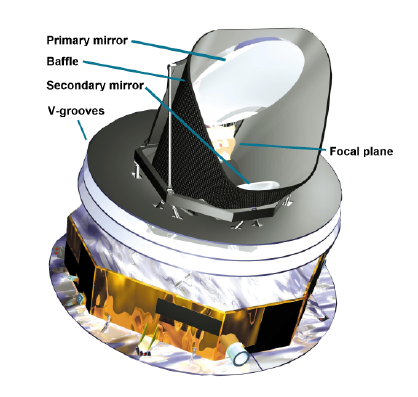}
    \caption{Overview of the Planck Satellite}
    \label{fig:plancksite}
\end{figure}


The satellite, see figure~\ref{fig:plancksite}, will orbit around the Lagrangian point $L_2$ spinning at 1 rpm around the Anti Sun axis; the telescope is pointed at almost $90\degree$ with respect to the spin axis, therefore Planck will map the sky in circles; each circle will be scanned for an hour, 60 times, then the satellite will be repointed by $2.5'$.

The scientific payload is placed in the focal plane of an off-axis aplanatic telescope with a primary of 1.9x1.5m, and it is constituted by two instruments for CMB detection:
\begin{itemize}
    \item HFI, High Frequency Instrument, is an array of 52 bolometers cooled to 0.1 K between 100 and 857 GHz, \cite{2003Lamarre}
    \item LFI, Low Frequency Instrument, is an array of 22 pseudo-correlation differential receivers based on HEMT technology cooled to 20 K and operating between 30 and 70 GHz, \cite{PlanckLFIScientificReq}
\end{itemize}

The Planck cryogenic chain, needed for cooling the scientific instruments, is composed by a passive cooling system and three independent stages of active cooling driven by different coolers, details about Planck thermal design are in section~\ref{sec:plancktherm} 

\section{HFI}
\label{sec:hfi}
The High Frequency Instrument (HFI) is an array of 52 receivers in 6 frequency bands (in the 100-857 GHz range) based on bolometric technology
and cooled to 0.1 K. 
Bolometers are composed by a conductive grid and a piece of semiconductor in thermal contact with a cold bath; when a photon hits the bolometer, it changes its temperature and therefore its resistance. Its resistance is monitored by the readout electronics which records a signal.

Two kinds of bolometers are used in HFI:
 \begin{itemize}

\item Twenty spider-web bolometers in the 143-857 GHz range, absorbing radiation via a spider-web-like structure
 \item Thirty-two polarisation-sensitive bolometers in the 100-353 GHz range,
      absorbing radiation via a pair of linear perpendicular grids. (Each
      grid absorbs only one linear polarisation.) This kind of bolometer, called PSB,
      allows to measure polarised radiation.
      
       \end{itemize}

The temperature of 0.1 K is achieved by three coolers: a hydrogen Sorption cooler, \cite{bhandari2004},  (providing a
stage at 18 K), a Joule-Thomson mechanical refrigerator (precooled by the
18 K cooler and providing 4 K) and an open-loop 3 He/4 He dilution refrigerator, which provides 0.1 K.

The 4K stage is very important because it is a key element in the HFI cryogenic chain and also provides a stable reference load for the LFI pseudo-correlation radiometers, see section~\ref{sec:4kboxtf}.

\begin{figure}[h]
    \centering
    \includegraphics[width=0.7\textwidth]{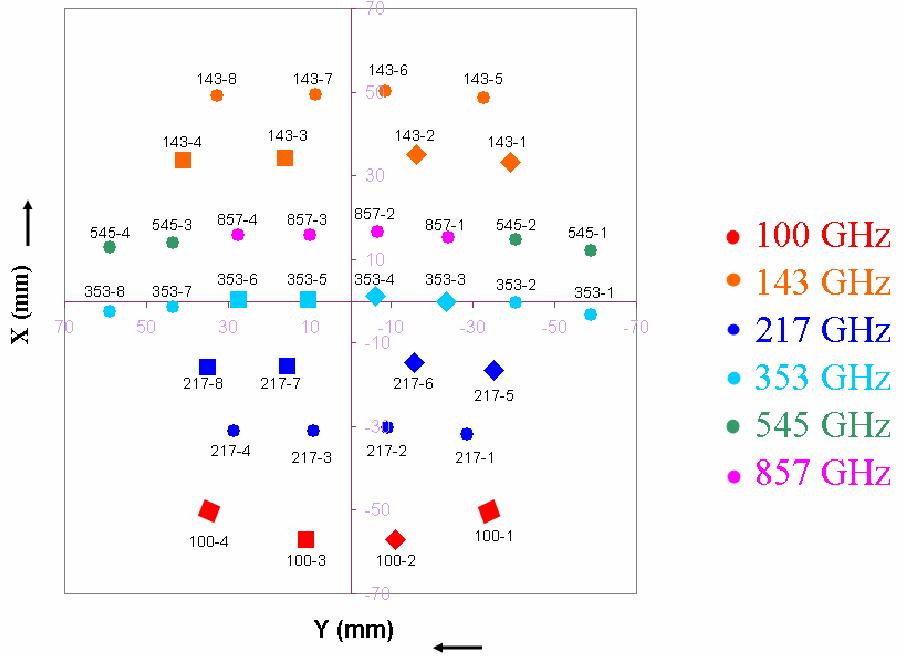}
    \caption{Schematic of the HFI focal plane with horns labels; the two channels of PSB bolometers, square in the schematic, are indicated as $a$ and $b$, e.g. 100-1b}
    \label{fig:hfifocal}
\end{figure}

Figure~\ref{fig:hfifocal} shows the position of HFI horns on the focal plane; HFI naming convention is based on the frequency and an ordinal number for spider web bolometers (see section~\ref{sec:hfi}), polarization sensitive bolometers instead are labelled with the same ordinal number and $a$ or $b$ for distinguishing the ortogonally polarization sensitive channels.

\clearpage

\section{LFI}

LFI is an array of 22 pseudo-correlation radiometers (discussed in the following section \ref{sec:pseudocor}) connected to 11 horns and capable of measuring CMB temperature and polarization in three frequency bands centered at 30, 44 and 70 GHz.

\begin{figure}[h]
    \centering
    \includegraphics[width=0.5\textwidth]{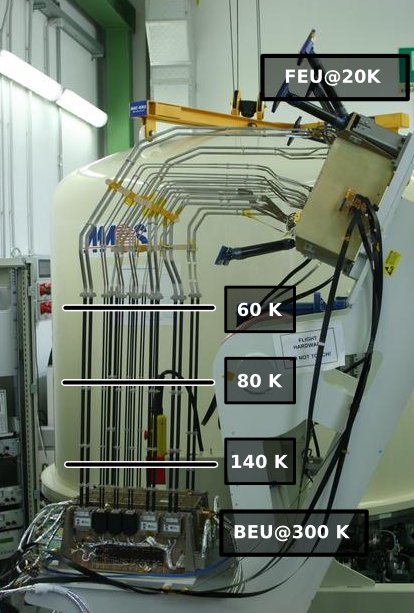}
    \caption{Picture of the LFI instrument during integrated tests in Alcatel Alenia Space Milano}
    \label{fig:lfim}
\end{figure}

LFI, see figure~\ref{fig:lfim} is separated in two sections:
    \begin{itemize}
        \item the Front End Unit, cooled by the Sorption Cooler at $\sim 20 K$, includes the horns and implements the pseudo-correlation strategy as detailed in section~\ref{sec:pseudocor}
        \item the Back End Unit, located into the satellite service module and kept at about 300 K, includes a further amplification stage, the detection system, conversion to digital and compression.
    \end{itemize}
    
The microwave signal is brought from the FEU to the BEU with waveguides about 1.5m long; the three V-grooves are passive radiative cooling system which favour thermal decoupling between the FEU and the BEU by setting the temperature of the interface with the waveguides at 60, 80 and 140 K.

After detection by the Back End Module, sky and reference load signals are integrated and digitized by the Data Acquisition Electronics (DAE). The digital signal is then sent to the Radiometer Electronics Box Assembly (REBA) which downsamples, compresses the data and builds packets that are sent to the Earth thanks to the transmission antenna.

\begin{figure}[h]
    \centering
    \includegraphics[width=\textwidth]{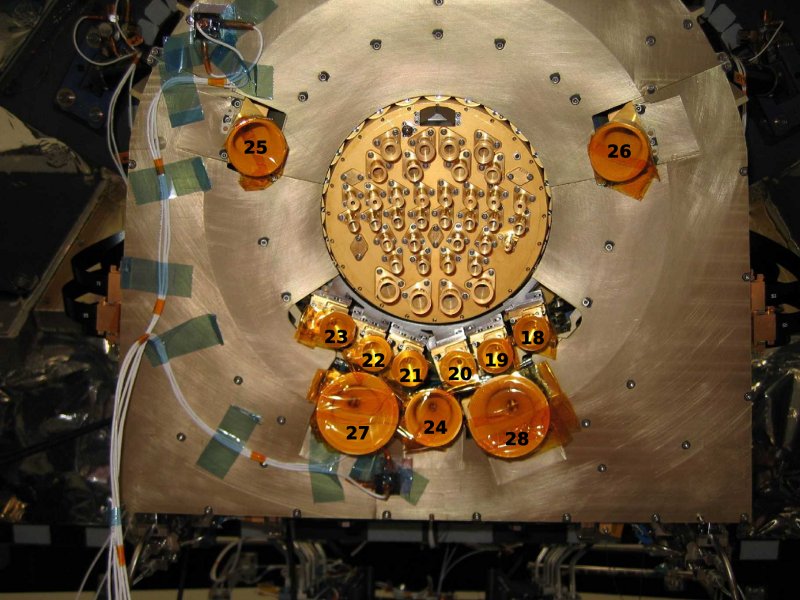}
    \caption{Planck focal plane with LFI horns labels}
    \label{fig:lfifocal}
\end{figure}

Figure~\ref{fig:lfifocal} shows the position on the focal plane of LFI horns.

My work is mainly focused on LFI, therefore general description of the radiometric technology and radiometers frequency response will be presented in the next sections, while the details of each LFI component are presented in the chapter about the radiometer model, see chapter~\ref{ch:lfi_model}.


\section{Radiometric technology}
\label{sec:radiometers}

In this section I will give a brief introduction on the available types of radiometers in order to understand how the pseudo-correlation radiometer was conceived, then I will show  the  radiometer response analytical representation focusing on bandpass response.

\subsection{Total power radiometer} 
\begin{figure}
    \centering
    \includegraphics[width=\textwidth]{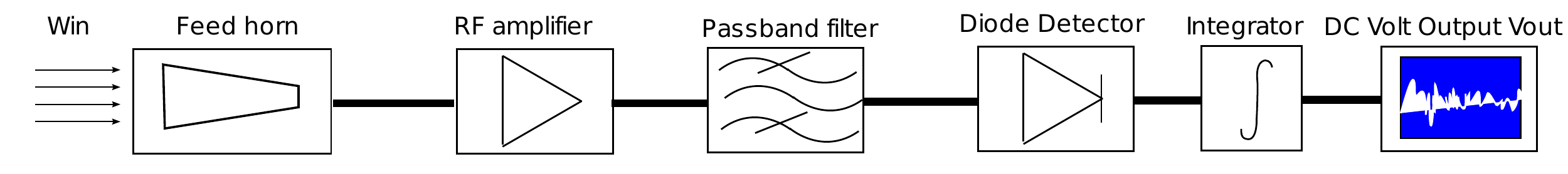}
    \caption{Schematic of a total power radiometer}
    \label{fig:totpow}
\end{figure}

Radiometers are coherent receivers which are able to measure the radiative power of electromagnetic waves. Radiometers operating in the microwave range, in their simplest form, called total power radiometer, are composed by, see figure~\ref{fig:totpow}:
    \begin{itemize}
        \item an antenna, usually a corrugated feed horn, to transfer the electromagnetic waves from vacuum or air to the transmission line reducing losses due to impedance mismatch
        \item one or more amplifying stages, with a total gain in terms of power between $10^6$ and $10^9$, which corresponds to 60-90 dB
        \item a passband filter, needed for select the interesting frequency band for the measurement
        \item a detection stage (usually a square law detector, typically a diode) which transforms the Radio Frequency signal to Direct Current
        \item an integrator, which integrates the DC signal in order to reduce fluctuations
    \end{itemize}
    
The sensitivity of a total power radiometer is given by:
    \begin{equation} 
    \sigma_T \simeq \dfrac{T_{sky} + T_{noise}}{\sqrt{\tau \Delta \nu}}
    \end{equation}
    
where $\tau$ is the integration time, i.e. the observation time, and $\Delta \nu$ is the bandwidth.

\subsection{Dicke switching radiometer} Amplifiers gain fluctuations, usually called $1/f$ because they increase at low frequency, are predominant on signal in this extreme applications; a possible strategy is using a stable thermal source as a reference and quickly switching between the antenna and the reference load.
The Dicke radiometer is basically a total power radiometer with two additional features: an
RF switch and a synchronous demodulator inserted after the square-law detector.

\begin{figure}
    \centering
    \includegraphics[width=\textwidth]{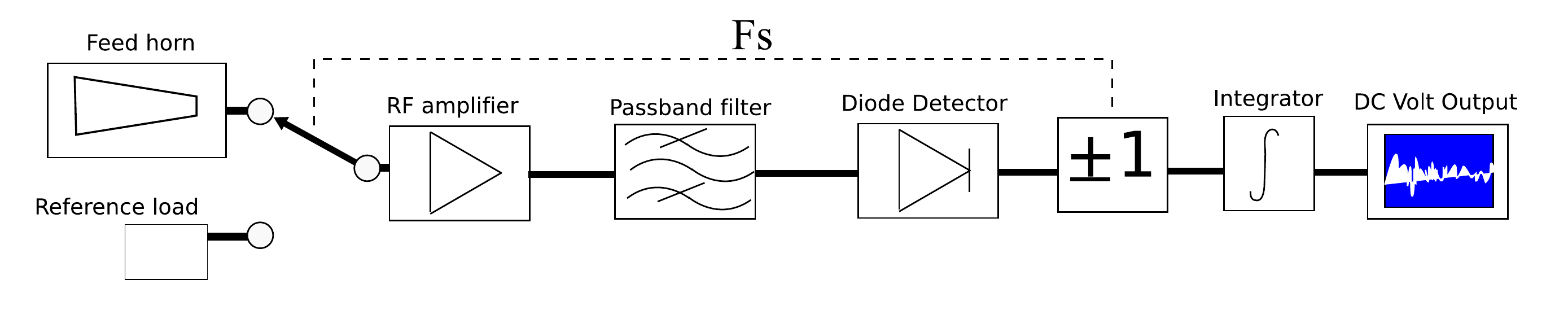}
    \caption{Dicke switching radiometer schematic}
    \label{fig:switching}
\end{figure}

The input of the radiometer (see Figure\ref{fig:switching}) is rapidly switched between the antenna
temperature and the reference temperature. The switch frequency Fs is typically of the order of few GHz.
The output of the square-law detector is multiplied by +1 or –1, depending on the position
of the Dicke switch, before integration.

It is important that the switching frequency is quicker than the typical amplifiers gain fluctuations, in this case:
\begin{equation}
    V_{out} = V_{sky} - V_{ref} = G (T_{sky} + T_{noise}) - G  (T_{ref} + T_{noise}) = G (T_{sky} - T_{ref})
\end{equation}

The paper \cite{2002A&A...391.1185S} shows an analytical tratment of pseudo-correlation radiometers, showing that Dicke switching radiometers reduce strongly the impact of systematic effects. Moreover, if the $T_{sky}$ and $T_{ref}$ are exactly the same, the radiometer is completely not sensitive to gain fluctuations.

\paragraph{Different reference load temperature}\label{sec:erre} If the reference load temperature is different from the sky load temperature is still possible to fight $1/f$ noise with this strategy by multiplying the $T_{ref}$ temperature by a number, $r$, called gain modulation factor:
    
\begin{equation}
    r =\dfrac{\overline{T_{sky}}}{\overline{T_{ref}}}
\end{equation}

where the overline indicates an average computed usually on a time period of one hour.

In this case the diffenced output is:
   \begin{equation}
    V_{diff} = V_{sky} - r V_{ref} = G (T_{sky} + T_{noise}) - r G  (T_{ref} + T_{noise}) = G (T_{sky} - r T_{ref})
\end{equation} 

The main issue of the Dicke switching radiometer is that for half of the observing time the radiometer is receiving the signal of the reference load, and this highly impacts on its sensitivity, see equation \ref{eq:dicke_sens}.

\begin{equation}
    \sigma_T \simeq 2 \dfrac{T_{sky} + T_{noise}}{\sqrt{\tau \Delta \nu}}
    \label{eq:dicke_sens}
\end{equation}

Therefore a Dicke radiometer has a sensitivity two times the sensitivity of a Total power radiometer.

\subsection{Pseudo-correlation radiometer} \label{sec:pseudocor}
\begin{figure}[h]
    \centering
    \includegraphics[width=.8\textwidth]{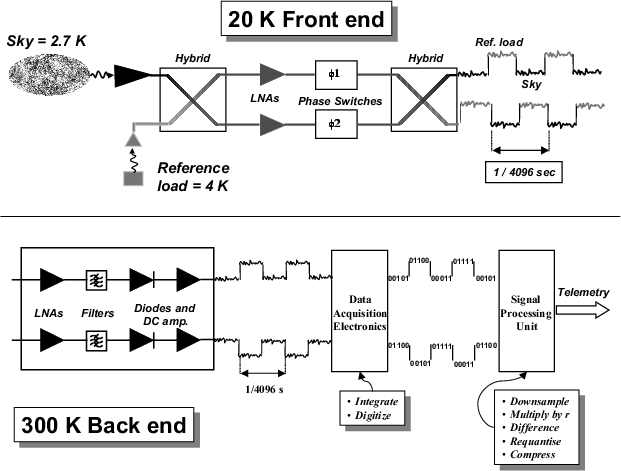}
    \caption{Planck LFI radiometers, in the upper section the cold Front End Module which implements the pseudo-correlation, in the lower section the warm backend}
    \label{fig:lfirad}
\end{figure}

The evolution of the Dicke radiometer is the pseudo-correlation radiometers, in this type of radiometer, sky and reference are simultaneously observered, see for example Planck LFI implementation:  the upper half of figure~\ref{fig:lfirad}, the cryogenic Front End Model, implements the pseudo-correlation, while a second amplification, filtering and detection stage is present in the wark Back End Module. The hybrid outputs the sum and difference which are amplified by two independent amplifiers; the second hybrid splits the amplified signals back into sky and reference.

Both sky and reference streams were amplified by both amplifiers, and are strongly correlated; the correlation is due to the $1/f$ noise of the amplifiers, therefore by taking the difference, similarly to the case of the Dicke radiometer, it can be strongly reduced. 

The advantage of such a configuration is that there is no signal loss and the sensitivity is comparable to the sensitivity of the total power radiometer, in particular:
    \begin{equation}
        \sigma_T \simeq \sqrt{2} \dfrac{T_{sky} + T_{noise}}{\sqrt{\tau \Delta \nu}}
    \end{equation}

The problem of this configuration is that only the amplifying stage involved into the pseudo-correlation is stabilised; the $1/f$ noise of the next stages is not affected. For this reason between the amplifier and the second hybrid a phase switch phase lags at 4 KHz one of the arms of a radiometer. The effect of the switching is that the sky and reference outputs are exchanged continuously; based on the same principle of the Dicke switching radiometer, this technique is used to reduce the $1/f$ fluctuations of the Back End unit.

\subsection{Basic radiometric equations}

In this section I will review the basic analytic modelling of the response of a radiometer and then I will focus on its bandpass response.

If we assume a top-hat frequency response in a band $\Delta\nu$ around the centre frequency, the voltage output from each of the 44 output diodes can be represented mathematically by this equation:

\begin{equation}
   \bar{V}_{out} = \bar{G} \cdot \bar{P}_{in} + \bar{V}_{noise}
   \label{eq:v_out}
\end{equation}
where $\bar{P}_{in}$ represents the input RF power, $\bar{G}$ is a proportionality constant (the so-called \textit{photometric calibration constant}) and $\bar{V}_{noise}$ is a noise contribution from the RF amplifiers and back-end electronics.

    It is common to refer this noise contribution to the input, so that: 

\begin{equation}
   \bar{V}_{out} = \bar{G} \cdot (\bar{P}_{in} + \bar{P}_{noise})
   \label{eq:v_out_2}
\end{equation}
where $\bar{V}_{noise} = \bar{G} \cdot \bar{P}_{noise}$. 

$\bar{P}_{noise}$ can be converted into a temperature, the Noise Temperature $\bar{T}_{noise}$:

\begin{equation}
   \bar{T}_{noise} = \dfrac{\bar{P}_{noise}}{k_B \Delta \nu}
   \label{eq:t_noise}
\end{equation}
where $k_B$ is Boltzmann's constant, $ \Delta \nu $ is the bandwidth, $\bar{T}_{noise}$ is the input load temperature that would cause the $\bar{V}_{noise}$ output signal in an ideal system, i.e. a system with no noise with $\bar{V}_{out} = \bar{G} \cdot \bar{P}_{in}$).

Noise temperature is directly related to radiometer sensitivity by the radiometer equation:
    
\begin{equation}
   \sigma_N = K \dfrac{\bar{T}_{sky}+\bar{T}_{noise}}{\sqrt{\Delta \nu \cdot \tau}}
   \label{eq:radiometer_equation} 
\end{equation}
where:
\begin{itemize}
    \item  $\sigma_N$ is the root mean square of the white noise
    \item $K$ is a factor which is dependent on receiver architecture. For the LFI receivers we have $K=\sqrt{2}$ (\cite{planckbluebook}).
    \item $\Delta \nu$ is the bandwidth
    \item $\tau$ is the integration time
\end{itemize}
 
\subsubsection{Considering in-band behaviour}
\label{sec:inband-behaviour}

If we now consider explicitly the frequency dependence, Eq.~\ref{eq:v_out} reads: 

\begin{equation}
V_{out}(\nu) = G(\nu) \cdot P_{in}(\nu) + V_{noise}(\nu) 
\label{eq:v_out_3}
\end{equation}
where the function $G(\nu)$, i.e. the gain versus frequency, is also referred to as the receiver bandpass.

LFI square law detectors integrate the signal over the band, therefore each channel output signal is:

\begin{equation}
\bar{V}_{out} =\int_{\nu_0 - \frac{\Delta\nu}{2}}^{\nu_0 + \frac{\Delta\nu}{2}} V_{out}(\nu) d\nu
\label{eq:v_out_4} 
\end{equation}

The next equations show all the relations between integrated and frequency dependent quantities:
\begin{eqnarray}
    \bar{V}_{noise} & = & \int_{\nu_0 - \frac{\Delta\nu}{2}}^{\nu_0 + \frac{\Delta\nu}{2}} V_{noise}(\nu) d\nu \\
    \bar{P}_{in} & = & \int_{\nu_0 - \frac{\Delta\nu}{2}}^{\nu_0 + \frac{\Delta\nu}{2}} P_{in}(\nu) d\nu \\
    \bar{G} & = & \dfrac{\int_{\nu_0 - \frac{\Delta\nu}{2}}^{\nu_0 + \frac{\Delta\nu}{2}} G (\nu) \cdot P_{in}(\nu) d\nu }{\bar{P}_{in}}
    \label{eq:all_integrated} 
\end{eqnarray}

$\bar{G}$, the photometric calibration constant, is the conversion factor between a variation in radiometers voltage outputs we measure and the corresponding variation in the input temperature; LFI radiometers, as WMAP ones, provide, by design, only relative measurements of the sky signal.

During Planck Spacecraft operations this calibration factor will be computed using the well known modulation of the Cosmic Microwave Background dipole signal among the whole sky and then used to produce temperature maps of the sky, see \ref{calibration}

As evident from $\bar{G}$ equation (\ref{eq:all_integrated}), calibration is tightly bound to the spectral index, or frequency dependence, of the input source.
Free-free, dust and synchrotron foregrounds, instead, due to their own spectral indexes, have a different coupling with radiometer bandpass and their photometric constant would be different. Therefore, using the photometric constant computed using CMB signal introduces a second order systematic effect on these components.

Knowledge about bandpasses can be used to reduce the effect of this calibration method by using an iterative process to separate different components.

The receiver bandpass, $G(\nu)$ can be determined by performing, for each frequency, two measurements at different input power levels and computing:

\begin{equation}
   G(\nu) = \dfrac{\Delta V_{out}(\nu)}{\Delta P_{in}(\nu)}
   \label{eq:bandpass_calculation}
\end{equation}

\chapter{LFI radiometer model}
\label{ch:lfi_model} 

\section{Introduction}
\subsection{History of the model}
The first radiometer model was developed in 2002 by Paola Battaglia, \cite{thesis_battaglia}, and subsequently improved by Cristian Franceschet, \cite{thesis_franceschet}. 

The model has been developed using the software simulation platform Advanced Design System by Agilent Technologies, \href{http://eesof.tm.agilent.com/products/ads_main.html}{see ADS website}.
The Advanced RF model original aim was to simulate the impact of subcomponents non-idealities, e.g. insertion and return losses, and of systematics, e.g. OrthoMode Transducer asymmetries and thermal fluctuations, on instrument performances.

Advanced RF model started as a complete analytical simulator using components embedded into the Agilent Advanced Design system, configured with expected performance defined in Planck LFI specification documents \cite{PlanckLFIScientificReq}.
Measurements of the true hardware performance were made then available by the manufacturers, therefore some of the analytical models were replaced with data coming from measurements.

During the Planck Low Frequency Instrument Flight Test campaign held in Thales Alenia Space Italia (Vimodrone-Milan) all the measurements available channel by channel were included in the model. This improved model was then used to compare systematically simulations to measurements with two objectives:
    \begin{enumerate}
        \item making a cross validation of the two methods available for estimating bandpasses: RCA frequency response measurements and simulations using bandpasses of single components
        \item obtaining, through simulations, data that were not available from measurements
    \end{enumerate}

The model had several issues related to its implementation into ADS, mainly due to the fact that ADS is a closed environment, it is not possible to run batch simulations on all RCAs and data export must be performed manually. For these reasons it was a very long process to produce a set of simulations of all channels and it was possible to introduce errors. Moreover, being a closed source software, it was not possible to read the source code in order to understand exactly what was the implementation strategy of the embedded models further than the details provided in the documentation.

While continuing the development on the Advanced LFI RF model, I started implementing the model from scratch on the electrical circuit simulator software QUCS \cite{QUcsschema}, website \url{http://qucs.sf.net}. 

The main objective of this work is to take full control of the model, in particular on two aspects:
    \begin{enumerate}
        \item automation - in ADS it is not possible to run many simulations together and export automatically the results, therefore the run operation is quite long and tedious
        \item sourcecode - Cristian Franceschet found a bug in ADS that was not solved, because Agilent wasn't interested in continuing the development of that piece of software. Having source code available allows the users to fix the software themselves if needed 
    \end{enumerate}

I named the new QUCS based model QIMP, which is the acronym for QUCS Integrator of Measured Performances, it became stable and superseded the Advanced LFI RF model in September 2008.

\subsection{Chapter structure}
The first section is a brief introduction to the QUCS software.

The second section is devoted to review the model implementation component by component referring to the hardware setup and to the data available from single component test campaigns.

The third part gathers all the results obtained by simulations.

\section{QUCS}

\begin{figure}
    \centering
    \includegraphics[width=\textwidth]{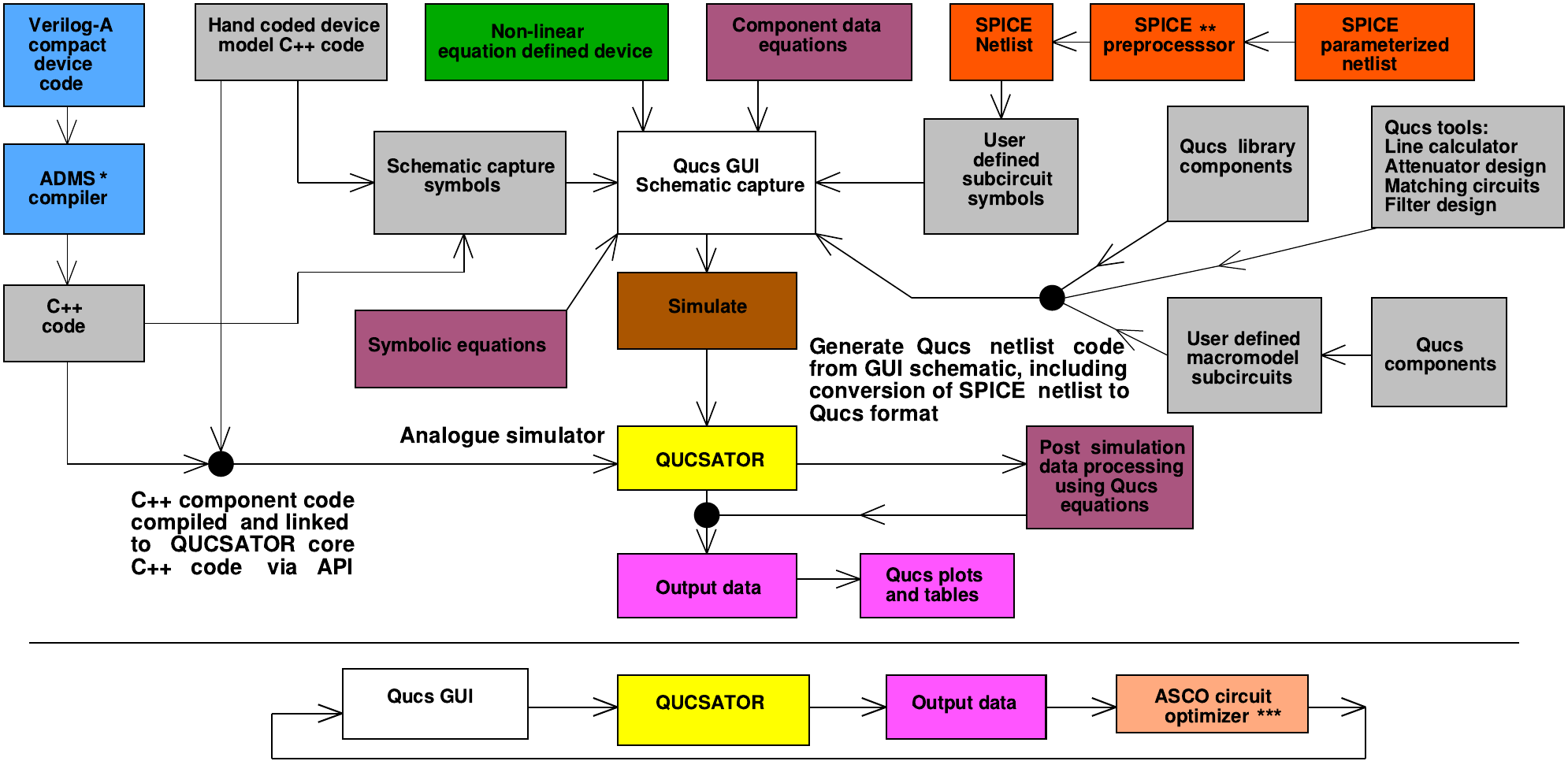}
    \caption{QUCS synopsis: the high level structure, at the bottom of the figure, shows the typical simulation process, in the top more insight into QUCS implementation}
    \label{fig:synopsis_qucs}
\end{figure}

Quite Universal Circuit Simulator (QUCS), \cite{QUcsschema} is an open source electronics circuit simulator software released 
under GPL license.
It gives  the possibility to set up a circuit with a graphical user interface using \href{http://trolltech.com/products}{QT libraries} 
and simulate the signal and noise behaviour of the circuit. It is an alternative to well known \href{http://bwrc.eecs.berkeley.edu/Classes/IcBook/SPICE/}{Berkeley SPICE}   and Agilent ADS. Synopsis of QUCS internals is given in Figure~\ref{fig:synopsis_qucs}.

QUCS is implemented in C++ and use extensively class inference in order to facilitate the implementation
of new components. The main QUCS modules are:
     
     \begin{enumerate}
         \item[\texttt{qucsator}] the command line circuit simulator, which reads a circuit description in a predefined ASCII format (named netlist) and outputs an ASCII format results file
         \item[GUI] Graphical User Interface is completely independent and makes it possible to draw a circuit using a library of devices or file defined components. The GUI can also automatically build the netlist, run \texttt{qucsator} and parse the results file, allowing the user to easily produce tables and plots.
     \end{enumerate}

\clearpage
\section{LFI Radiometer model implementation}

\subsection{The Low Frequency Instrument - Overview}

For a complete introduction on radiometers please refer to section \ref{sec:radiometers}.

\begin{figure}[h!]
    \includegraphics[width=\textwidth]{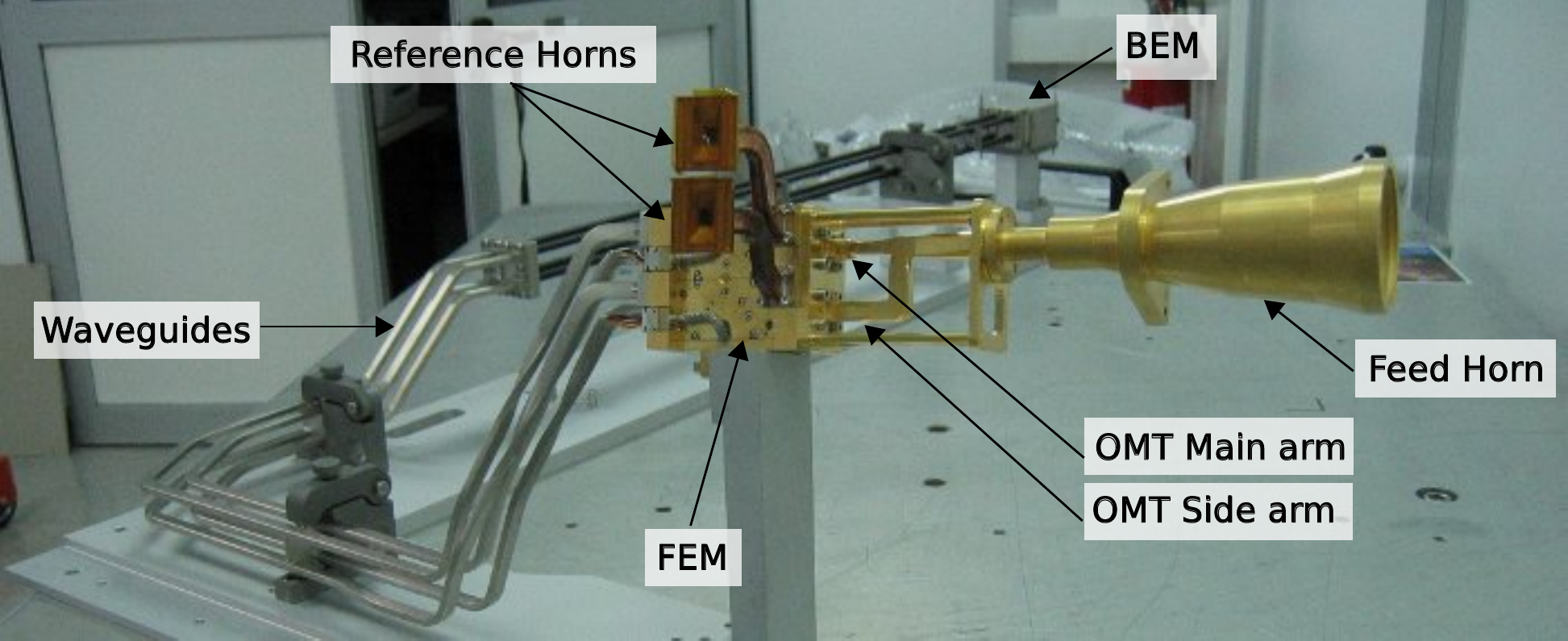}
  \caption{Picture of a 30 GHz Radiometer Chain Assembly (RCA) mounted for testing in Milano}
  \label{fig:rca_photo}
\end{figure}

LFI is a radiometer array mounted in the focal plane of Planck satellite telescope, receiving Cosmic Microwave Background photons at 30, 44 and 70 GHz. 
The complete LFI is an array of 11 Radiometer Chain assemblies (RCA), 6 with a central frequency of 70 GHz, 3 of 44 GHz and 2 of 30 GHz.
Each RCA is composed by (Fig. \ref{fig:rca_photo}):
\begin{itemize}
\item a Feed Horn and an OrthoMode Transducer (polarisation splitter) looking at the Sky signal
\item 2 short waveguides (just for 30 and 44 GHz, 70 GHz reference horn are directly connected to the FEM) connected to two reference loads at about 4 K
\item a cold ($\sim 20 K$) pseudo correlation stage (Front End Module) providing $\sim 35 dB$ amplification
\item 4 waveguides connecting the cold module to the warm backend
\item a second RF amplification, a diode and post-detection DC electronics stage at 300 K (Back End Module)
\end{itemize}

LFI channels naming scheme is based on:
     
     \begin{itemize}
         \item RCA number (18 to 28)
         \item OMT arm (Main or Side)
         \item radiometer number (which is 0 for the radiometer connected to OMT Main arm, 1 for Side) 
         \item detector number (0 or 1)
     \end{itemize}

for example LFI22M-01 refers to RCA 22, OMT Main arm, channel 01 among the 4 RCA 22 channels (00,01,10,11).

\subsection{LFI Radiometer model - Overview}

QIMP (QUCS Integrator of Measured Performances) can simulate the in band behaviour (frequency response) of each RCA using simplified schematics and data of each hardware subcomponent made available by the manufacturers thanks to performance tests.

Although in this thesis the bandpasses provided by the model are often referred to as ``simulated'' bandpasses, we want to stress the fact that they are indeed the product of a combination of \emph{real measurements} performed on the various subsystems to estimate the complete RCA frequency response.

The most important model result is the bandpass which is useful for building CMB polarisation maps and for foregrounds removal; but the model also simulates volt output as a function of sky input temperature, noise temperature, input power at each subcomponent, photometric gain and so on.

The model is built to study the broadband behaviour of the instrument in the frequency domain, therefore it is not able to produce data streams as a function of a input temperature changing with time, but can simulate statically the output with different input temperatures.

\begin{figure*}
    \centering
    \includegraphics[width=\textwidth]{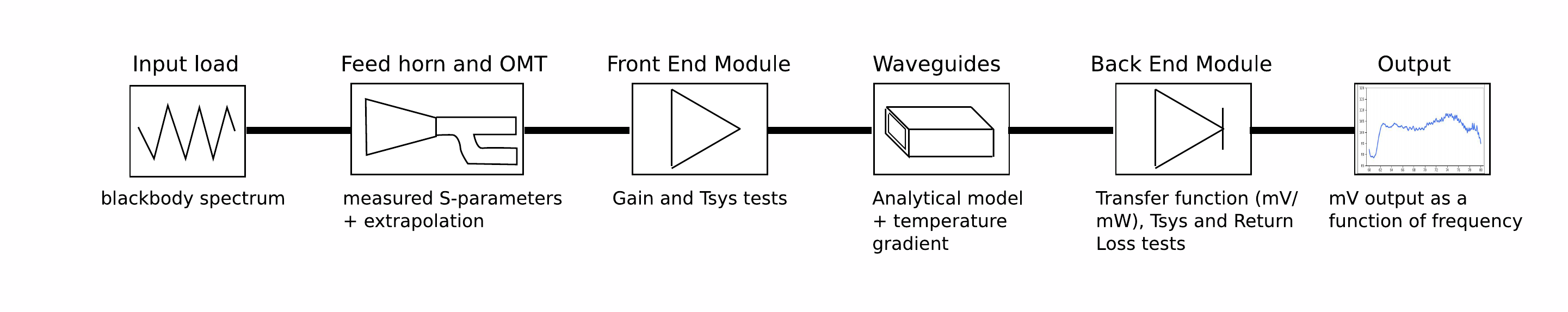}
    \caption{model schematic: FEM is modelled by 4 independent bandpasses, therefore each of the 44 LFI channels is modelled independently as the above model. Each component is represented by its measured behaviour as a function of frequency. Waveguides instead are simulated analytically given their dimensions, length and temperature gradient along their length.}
    \label{fig:model_schematic}
\end{figure*}

The complete model implementation is shown in figure \ref{fig:model_schematic}; the frequency dependence of all parameters is implicitly considered if not otherwise specified.

\subsection{Sky Load}

Sky load is simply modelled as a noisy resistor with tunable temperature; it outputs an uncorrelated signal independent of frequency with a power spectral density of:

\begin{equation}
P_{skyload} = k_B T \quad [Watts/Hz]
\label{eq:sky_load}
\end{equation}

where $k_B$ is Boltzmann's constant and T is the chosen input temperature; the skyload assumes perfect absorption, i.e. no emission from the horn is reflected back.

\subsection{Feed Horn and OrthoMode transducer}

\subsubsection{Description}

The first part of each RCA is a corrugated feed horn which collects radiation coming from the telescope secondary mirror reducing reflection and absorption thanks to an optimised design.
The OrthoMode Transducer receives the sky signal from the horn and splits the orthogonal polarisations, that propagate through two independent radiometers.

Each LFI passive component frequency response has been characterised by tests performed at IFP-CNR (Istituto di Fisica del Plasma - Milano).

\subsubsection{Model}

The Feed Horn and OMT assembly in QIMP has been modelled as a single 2-ports S-parameters component:

\begin{itemize}
    \item The main contribution to reflection (S11) is the OMT, due to the waveguides configuration needed for splitting the incoming wave polarisations, therefore the Feed Horn reflection losses can be neglected and the S11 tests on standalone OMT taken as representative of the behaviour of the FH-OMT assembly.

\item S22 of the assembly, i.e. the reflection losses on FEM side, was measured directly on Feed Horn and OMT together.

\item S21, which is the most important parameter, because influences how much sky signal can reach the radiometer, and includes both reflection and resistive losses; for some devices we have only S21 measurements and not S12; however, looking at devices where both measurements are available, they are very similar, therefore if S12 is not available, it is considered equal to S21.
\end{itemize}

The S11 and S12 parameters characterise respectively the signal reflected back into the Horn by the OMT itself and the signal reflected by all the instrument components after the OMT which travels back towards the Feed Horn. This parameters are included in the model and can be used to perform a study on LFI emission (both in testing configuration, e.g. standing wave between the sky load and the OMT, or in flight, e.g. LFI interference on HFI), but for now they are not used.

\subsubsection{Measurements extrapolation}
\label{sec:omt_extra}
The nominal bandwidth of LFI receivers is 20\% of the central frequency, e.g. for 30 GHz the nominal bandwidth is 6 GHz. Moreover, the standard band of the WR28 waveguides, used for testing is 26.5-40 GHz; for this reason 30 GHz OMTs were only measured between 26.5 and 40 GHz.

However, the fact that the measured bandpass response of the RCAs is still high at 26.5 GHz showed that the radiometer band extends below this limit. For this reason, one of QIMP objectives was to provide the best estimation for the band between:
    \begin{itemize}
        \item 21.3 GHz, which is the low frequency cut of the 30 GHz waveguides
        \item 40 GHz, where the RCA response is already very low due mainly to BEM lowpass filter
    \end{itemize}

OMT design was exactly the same for 30 and 44 GHz devices, it was scaled in order to work efficiently with different wavelengths, therefore it is possible to exploit this similarity and extrapolate the measurements on 30 GHz devices using as a reference the measurements on 44 GHz which were performed on a larger band, from 33 to 50 GHz.

\begin{figure}[h!]
\includegraphics[width=\textwidth]{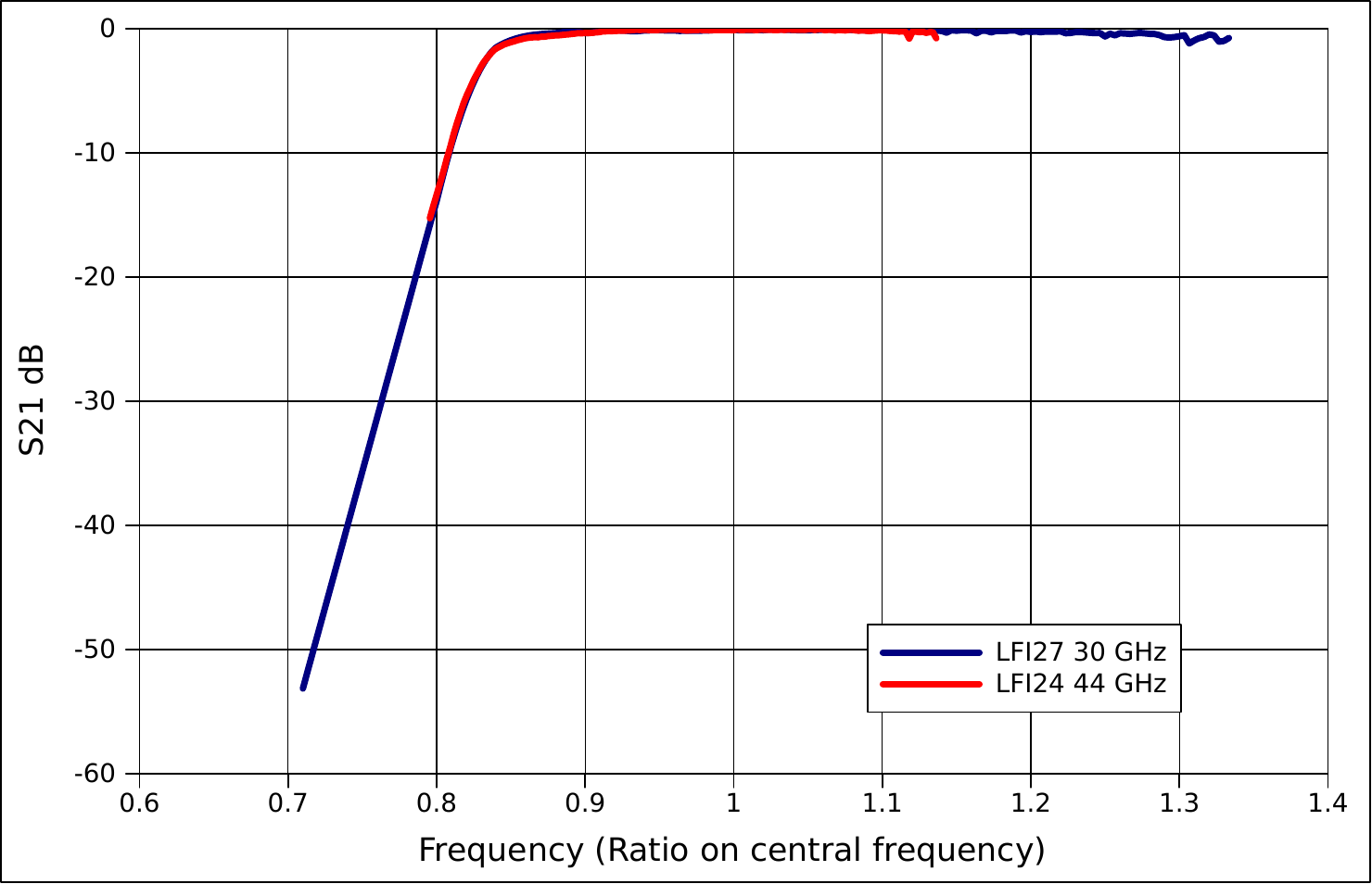}
  \caption{Extrapolation of the 30 GHz measurements performed by comparison with 44 GHz measurements normalised on their own central frequency. The extrapolation was necessary between 21.3 GHz and 26.5 GHz}
  \label{fig:omt_extra}
\end{figure}

The extrapolation was done after normalising the frequency axis on the central frequency, see \ref{fig:omt_extra}, so that all the curves has a central frequency of 1.

\subsection{Front End Module}

\subsubsection{Description}

    The Front End Module is the cryogenic amplification stage where the pseudocorrelation is implemented, see section~\ref{sec:radiometers}.

\begin{figure}
    \centering
    \includegraphics[width=\textwidth]{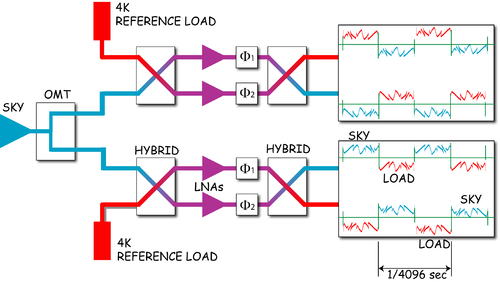}
    \caption{Schematic of LFI radiometers highlighting pseudocorrelation strategy}
    \label{fig:radiom}
\end{figure}

It is constituted by (figure~\ref{fig:radiom}):
    \begin{itemize}
        \item an hybrid which receives the incoming sky and reference signal and outputs the sum and the difference
        \item a cascade of 4 Low Noise Amplifiers based on High Electron Mobility Transistors which amplifies the signal between 30 and 40 dB, i.e. up to a factor of $1e4$ in power and $100$ in voltage amplitude
        \item phase switches: a system through which the wave can travel along two alternative pathways phase-lagged by $180 \degree$. The pathway is selected by two diodes that are alternatively biased at a frequency of 4 KHz; this fast switching is implemented to reduce $\frac{1}{f}$ gain fluctuations of the Back End Module
        \item a second hybrid which splits the signal back into sky and reference components
    \end{itemize}

Originally the Advanced ADS LFI RF model implemented the pseudocorrelation, therefore hybrid, phase switches and LNAs where modelled as distinct components. Ideally this configuration is the most realistic and allows to compute also isolation, i.e. the level of separation of sky and reference channels of the same radiometer.

However, FEM subcomponents are tightly integrated, and we cannot make the assumption that the behaviour of a sub component tested alone is the same when integrated into the FEM; the effect of connection, matching and coupling is strong and FEM designers and builders judged that the tests on the complete FEM performed after integration are more reliable than using sub components measurements.

A test on the complete FEM means measuring gain and noise temperature as a function of frequency of each of the four output ports when a signal is injected into the sky input.

Therefore, in QIMP, the FEM structure is simplified and consists in 4 independent channels, neglecting pseudocorrelation. This simplification doesn't impact on QIMP capability to model bandpass response because frequency response is a property intrinsic to the path of the microwaves into the instrument and it is not influenced by pseudocorrelation.

\subsubsection{FEM bandpass measurements}
\label{sec:fem_tests}
\begin{figure}
    \centering
    \includegraphics[width=0.45\textwidth]{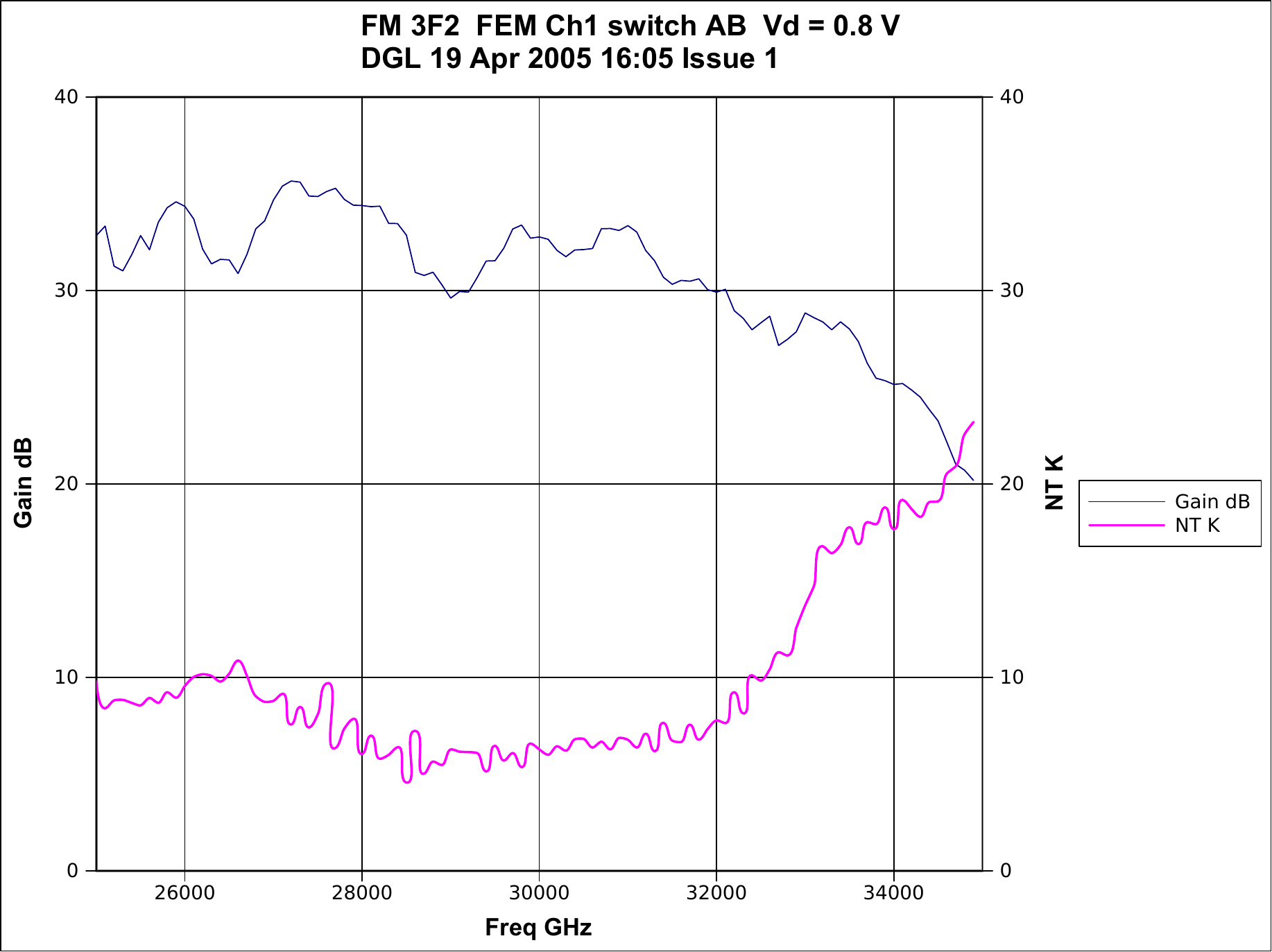}
    \includegraphics[width=0.45\textwidth]{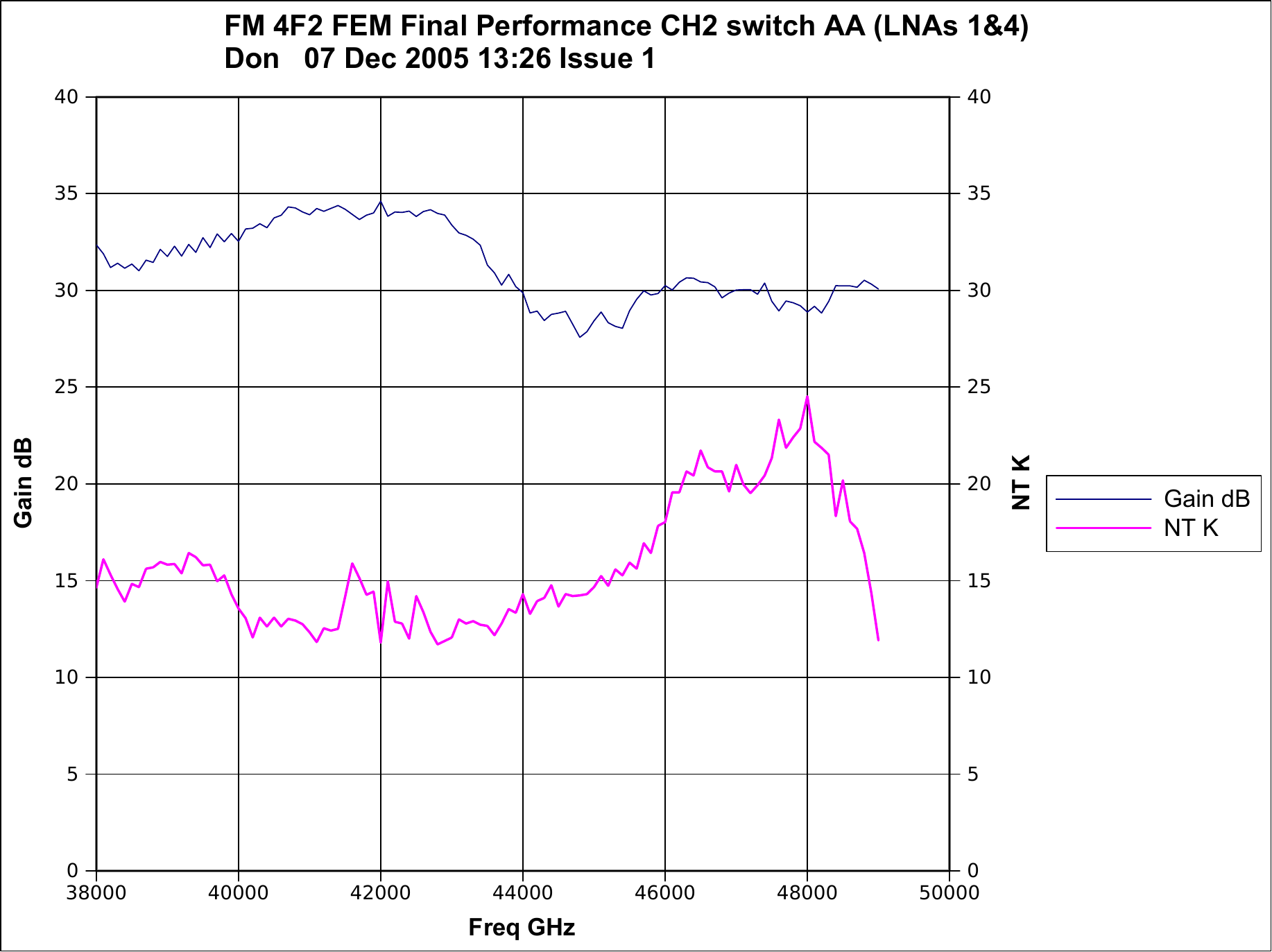} \\
    \includegraphics[width=0.45\textwidth]{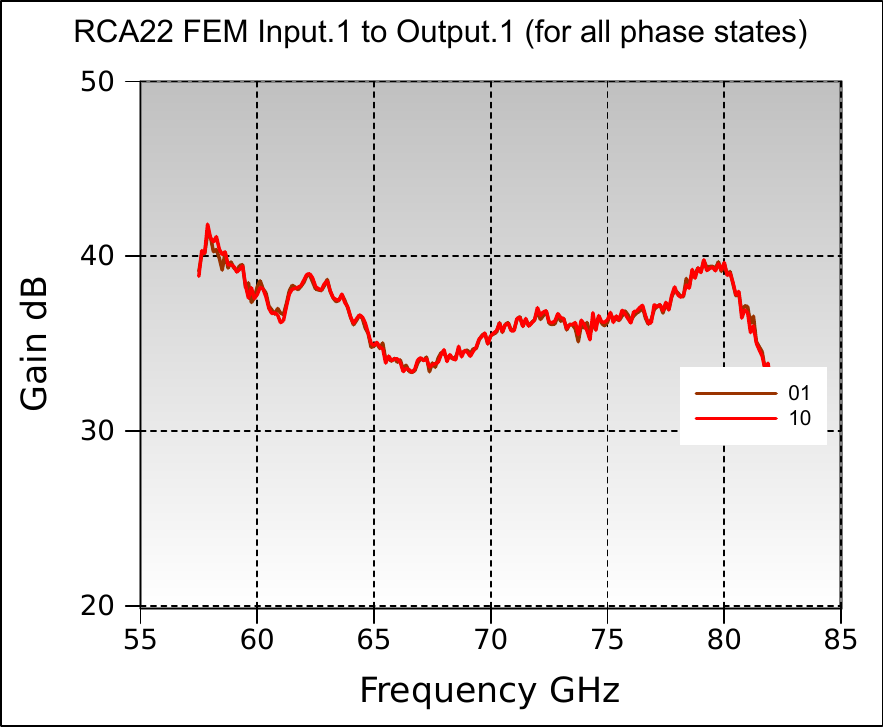} 
    \includegraphics[width=0.45\textwidth]{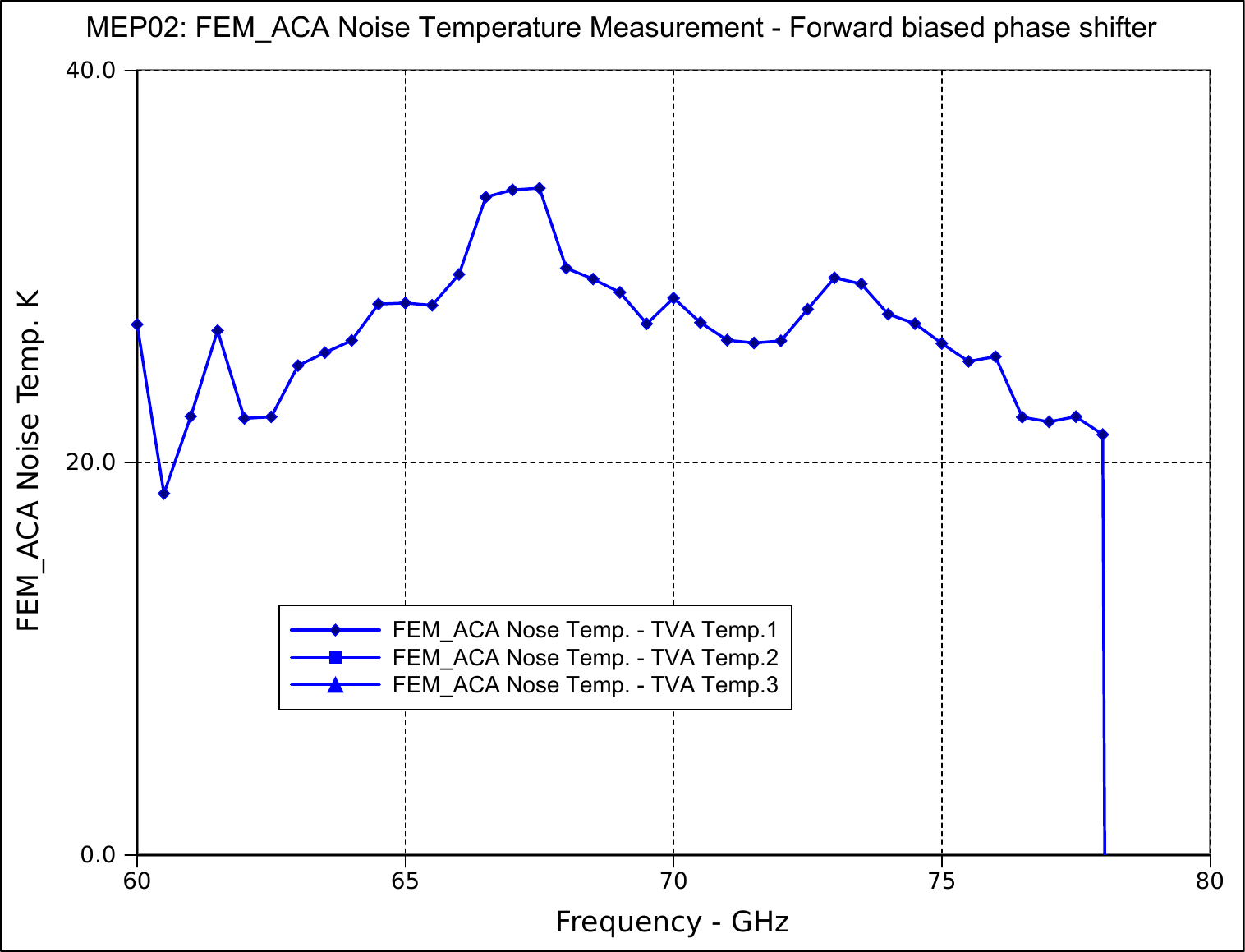} 
    \caption{FEM Gain and Noise Temperature as a function of frequency, an example from LFI27, LFI24 and LFI22, for LFI 22 gain and noise temperature are on different plots because they were measured independently}
    \label{fig:femgnt}
\end{figure}

The most straightforward method for performing frequency response of a single FEM channel to the sky input is:
\begin{enumerate}
\item the FEM is placed in cryogenic environment
\item a known stable input temperature source is attached to the sky load input
\item a mixer and an sweeping frequency source is set up in order to selectively measure noise as a function of frequency
\item stop the 4 KHz switching (otherwise the signal from a single channel switches every 1/4096 s between sky and ref signal)
\item set the phase switches polarisations in a configuration where the current channel receives the sky signal.
Opposite Phase Switches configurations both have the same output correspondence, e.g. arm 1 phase switch direct and arm 2 ph/sw shifted by 180° is equivalent to 1-shifted 2-direct.
In principle the Phase Switches have different attenuation in the abovementioned configurations, but this difference is indeed very small because they were balanced apply the right control currents before performing the tests.

Therefore, Jodrell Bank Observatory, 30 and 44 GHz FEMs manufacturers, performed tests in only one of these configurations, while Elektrobit, manufacturer of 70 GHz RCAs, performed tests in both configurations. However, the difference between the two configurations is as average 0.2 dB, with peaks of 0.9 dB, the gains in both configurations are overplotted on the top of  the RCA22 plot in figure~\ref{fig:femgnt}.

For consistency with the data available for 30 and 44 GHz FEMs, for 70 GHz the mean of the two tests were taken as FEM response.
\item Measure the output signal with a Noise Figure Meter as a function of the signal frequency and amplitude 
\item Perform all the aforementioned steps at 2 different input temperatures; measuring the output at 2 different input power it is possible to compute the gain as $\frac{\Delta V_{out}}{\Delta W_{in}}$ and consequently the $T_{noise}$.
\end{enumerate}

For 70 GHz channels, the cryogenic tests consisted only in gain measurements, therefore the noise temperatures used in the model were taken from the previous test campaigns on the LNA amplification stage, named ACA (Amplifier Chain Assembly), at ambient temperature. Integrated noise temperatures were measured in cryogenic conditions, therefore it is possible to normalise $T_{noise}$ measured at ambient temperature in order to obtain the cryogenic ones when integrated. 

It is important to note that bandpasses, which are the main focus of QIMP, do not depend on either FEM or BEM $T_{noise}$ because they are gain bandpasses, based on the proportionality between output and input, therefore not influenced by static contributions:

\begin{eqnarray}
    V_{out} & = & G_{BEM} * (G_{FEM} * (T_{in} + Tnoise_{FEM})) + Vnoise{BEM} \\
    V_{out} & = & G_{BEM} * G_{FEM} * T_{in} + G_{BEM} * G_{FEM} * Tnoise_{FEM} + Vnoise{BEM}\\
    V_{out} & = & G * T_{in} + Vnoise
\end{eqnarray}

It is clear that only FEM and BEM gains contribute to the $G$ term of the last equation, which is the bandpass response, all noise terms converge to the $Vnoise$ term.

This modelling could be refined when the model will be extensively validated using the results of the LIS tests, which are tests specifically dedicated to understanding the integrated output volt signal behaviour as a function of input temperature.

\begin{figure}
    \centering
    \includegraphics[width=\textwidth]{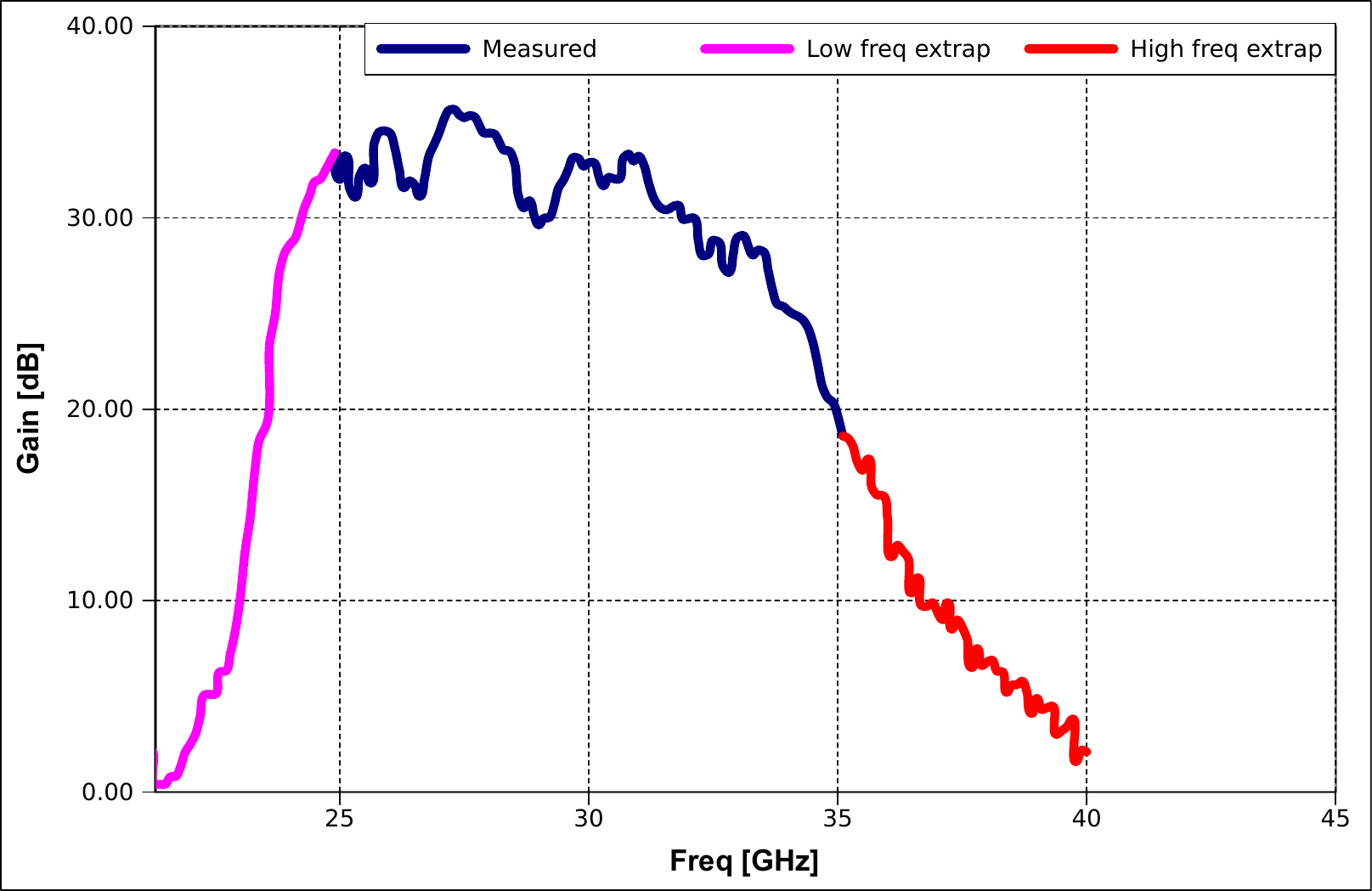}
    \caption{Example extrapolation at FEM level performed on a channel of RCA27}
    \label{fig:fem_extra}
\end{figure}

FEM at 30 GHz were measured just between 25 GHz and 35 GHz, therefore an extrapolation has been necessary in order to have a total simulated bandpasses frequency span between 21.3 and 40 GHz. Extrapolation slope is based on VNA measurements that were performed in warm conditions on a Qualification model FEM; its offset instead was based on the interface with the measurements.Figure~\ref{fig:fem_extra} shows the result of the extrapolation on RCA 27.

\subsubsection{FEM Linear model}
The linear model fundamental FEM equation is:
\begin{equation}
V_{out} = G (T_{in}) + V_{sys}
\end{equation}
The output is the sum of the Gain multiplied by the input power and an offset which doesn't depend on the input temperature and which is due to the thermal noise emission from the electronic components.
Usually instead of referring to the noise contribution as a volt offset, it is divided by the Gain and referred to as the System Temperature ($ T_{sys} $): it is the equivalent input temperature that would produce the same output in an ideal amplifier (i.e. a component without noise). 
\begin{equation}
V_{out} = G (T_{in} + T_{sys})
\end{equation}
It is very convenient because it can be directly compared to the input temperature and understand what is the relative importance of input signal and electronics noise in the output signal. 
During the data analysis Gain and System temperature are computed for each frequency step, typically 0.1 GHz, using the outputs at high and low temperature:
\begin{enumerate}
\item Compute the Gain in dB as [$ 20 log (Vout_{T_{high}}-Vout_{T_{low}})/(Vin_{T_{high}}-Vin_{T_{low}})$]
\item Compute the system temperature $ T_{sys} $ extrapolating linearly the volt output until the intersection with the negative part of the volt axis. 
\end{enumerate}

The next figures are examples of 30,44 and 70 GHz FEM gain and system temperatures.

\subsubsection{Model}

Focusing on a single channel, and ideally freezing the 4 KHz switching when it receives the sky signal, it is possible to study statically the bandpass response.

The purpose of QIMP is to characterise statically the radiometer  properties: in this context, each single channel can be seen as an independent radiometer frozen in one of the phase switches configurations when it is looking to the sky signal.
Actually, there are 2 opposite phase switches configurations where the same BEM channel is connected to the sky, i.e. the phase switches can shift the RF signal phase by $180 \degree$, changing the 'sign' of the incoming radiation, if none or both the phase switches change the sign, the output correspondence to the input signal is the same.
Due to the effect of hybrids, each channel bandpass response includes the contributions from the Low Noise Amplifiers of both FEM legs.

In the QIMP schematic, the FEM has been modelled as an S-parameter component with gain given by the performance tests on each FEM channel and noise figure computed by the system temperature: $ F = 10 log10 (\frac{T_{sys}}{290 K} +1) $.
FEM S11 and S22 were measured only at ambient temperature, in cryogenic conditions they are expected to be about -7 dB, therefore QIMP contains a value of -7 dB flat over the band. FEM S11 is not very important because OMT reflection is low, so the power reflected back negligible. S22 instead would be interesting because it influences the losses due to standing waves into the waveguides. For this reason in the future I plan to investigate if it is possible to predict cryogenic S22 by properly scaling the coefficient  measured at ambient temperature .

Therefore the RCA is studied in an ideal configuration, which could never be obtained in the real hardware: all the 4 RCA channels are looking simultaneously to the sky signal, useful for characterising at once the broadband behaviour of the instrument.
The main disadvantage of this modelling approach is that it is not possible to study isolation between sky and reference signals, which needs the full pseudocorrelation design to be implemented.

Notice that the response to the reference load input is in principle different from the response to the sky, because the receiver is asymmetric upstream the first hybrid. In fact the sky channel is connected to the feed horn - OMT assembly and the reference channel is connected to the 4K reference horn which is connected directly to the 70 GHz FEMs or by means of 2 short waveguides to the 30 and 44 GHz FEMs.

\begin{figure}
    \centering
    \includegraphics[width=\textwidth]{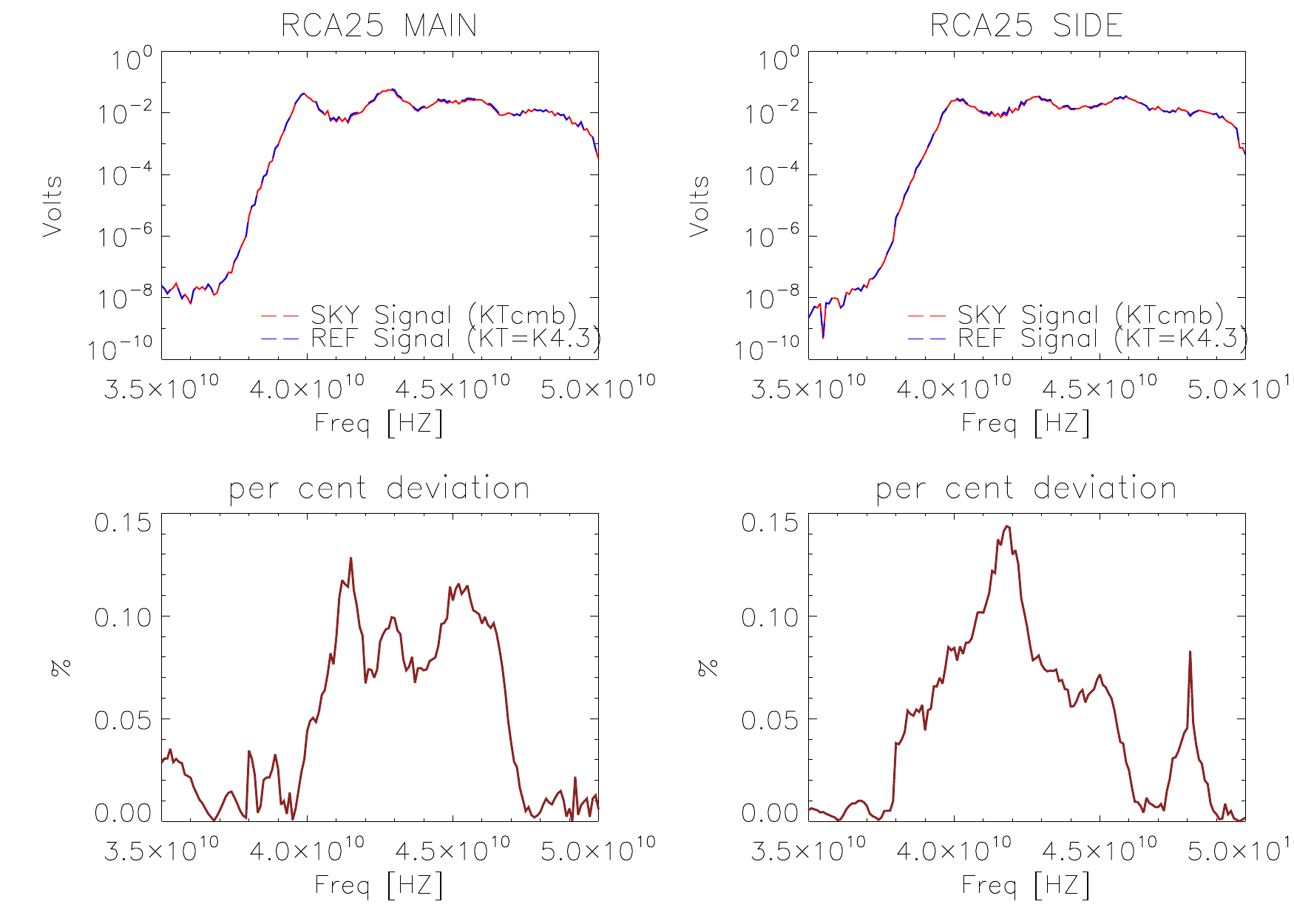}
    \caption{Comparison between sky and reference bandpasses on RCA25}
    \label{fig:ref_bandpass}
\end{figure}

The effect of this asymmetry on RCA25 was analysed by Andrea Catalano using QIMP, the results, in Figure~\ref{fig:ref_bandpass}, show that the difference between sky and reference bandpasses is below $0.15\%$. I extended this analysis to all LFI channels confirming that the impact of the receiver asymmetry on bandpasses is negligible.

\subsection{Waveguides}
\subsubsection{Description}

Waveguides link the 20 K Front End Module to the 300 K Back End Module, and are mechanically and thermally connected to the 3 V-Grooves at 60, 80 and 140 K which thermally decouple the warm stage from the cold.

Waveguides electromagnetic behaviour is strongly influenced by the large temperature gradient.

Each waveguide is made by a twisted copper section connecting the FEM to a straight stainless steel that thermally decouple the 20 K from the 50 K stage. Then a gold plated stainless steel straight section connects this stage to the 300 K BEM.


\subsubsection{Frequency response measurements}
IFP (Istituto di Fisica del Plasma) tested all the waveguides at ambient temperature characterising Return and Insertion losses with a Vector Network Analyser, details about the measurements are available in \cite{prelaunch_wg}. 

Because measurements were performed with impedance-matched connections at the waveguides ends, they are not representative of the standing waves arising when the waveguide is connected to FEM and BEM and therefore are source of important impedance mismatch.

For this reason QIMP implements an analytical waveguide simulator built using many sections of a rectangular waveguide component and ideal transformers. The waveguide model was validated against measurements performing simulations at ambient temperature, after validation the model temperatures were modified in order to simulate the waveguides in nominal conditions, i.e. with the nominal temperatures at FEM, V-grooves and BEM connections and linear interpolation between them. 

\subsubsection{Rectangular waveguide device implementation in QUCS}

In summer 2008, under my request, Bastien Roucaries, one of QUCS developers, implemented a rectangular waveguide component model, which was not present before in QUCS.

The model is based on the analytical waveguide treatment in appendix \ref{sec:wg}.

The rectangular waveguide model was implemented using a S-parameters component. Its implementation has been quite straightforward
thanks to the modular conception of QUCS. The code is constituted by about 600 lines including extensive comments and GUI
part. The core part was done reimplementing six virtual functions of the component base class.

Upon the model developed by Bastien Roucaries I added the embedded computation of the resistivity of gold, stainless steel and aluminium based on device temperature. This feature was implemented using different empirical formulae in different temperature ranges. 

The waveguide component was validated against Agilent ADS results, using a reference waveguide model as follows:
     \begin{itemize}
         \item single rectangular waveguide component
         \item 5.69x2.8 mm size (same as 44 GHz LFI waveguides)
         \item 800 mm length
          \item material is gold
         \item simulation temperature is 20 \degree C
     \end{itemize}

\begin{figure}
    \centering
    \includegraphics[width=\textwidth]{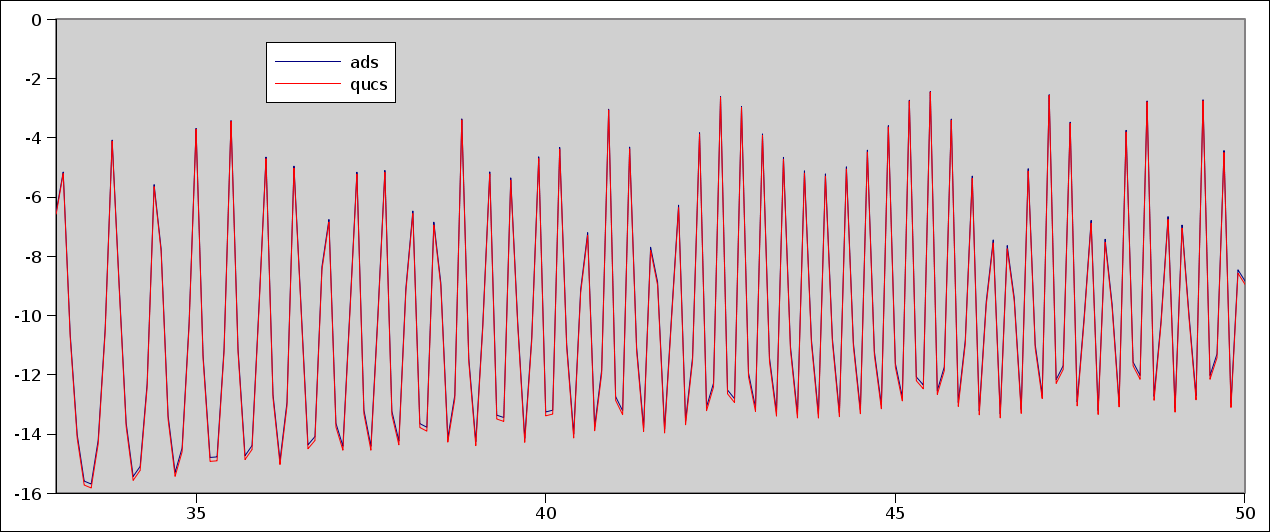}
    \caption{Comparison of S21 [dB] of a gold plated rectangular waveguide as a function of frequency}
    \label{fig:adsqucswg}
\end{figure}

Both S parameters and noise simulations voltage results were compared and the matching is very good, Figure~\ref{fig:adsqucswg} shows S21 as a function of input frequency of ADS and QUCS overplotted.

\subsubsection{Model}

Waveguides were modelled using many sections of the rectangular waveguide component, which consists in an analytical model based on the propagation of the Transverse Electric mode 1,0 (TE10). This was first implemented in ADS and then ported to QUCS.

The transmission line has a reference impedance of $50 \ohm$, while waveguides have a much higher impedance. Interfacing the waveguides directly to the transmission line would lead to strong losses, which are not present in the real hardware, because all components are physically very well matched.

Therefore ideal transformers were used for the purpose of simulating the good matching between waveguides and the rest of the RCA, and only a small mismatch of -35 dB was left, as found during hardware measurements.

\begin{figure}
    \centering
    \includegraphics[width=\textwidth]{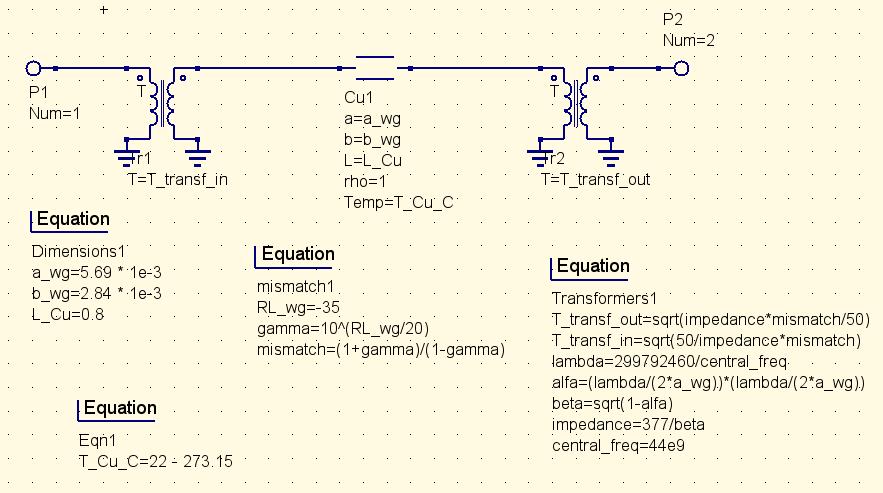}
    \caption{Snapshot of the QUCS implementation of the copper waveguide section}
    \label{fig:wg_cu}
\end{figure}

\begin{figure}
    \centering
    \includegraphics[width=\textwidth]{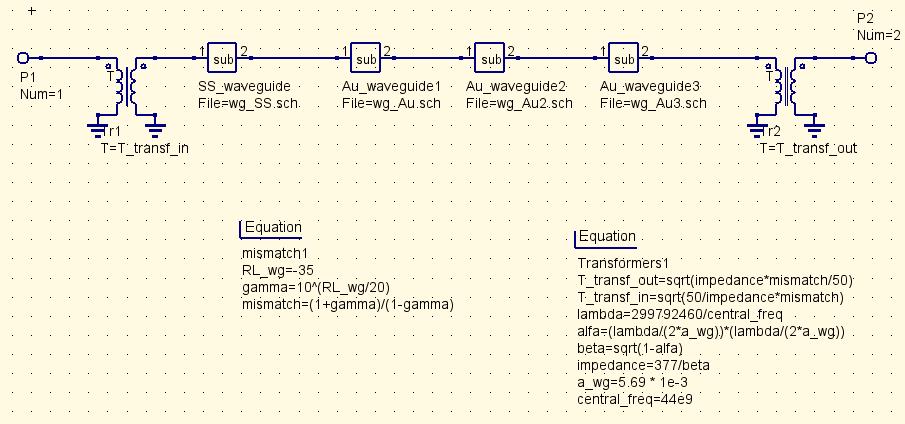}
    \caption{QUCS schematic of the group of stainless steel and gold plated sections enclosed by the ideal transformers}
    \label{fig:wg_ssau}
\end{figure}

\begin{figure}
    \centering
    \includegraphics[width=\textwidth]{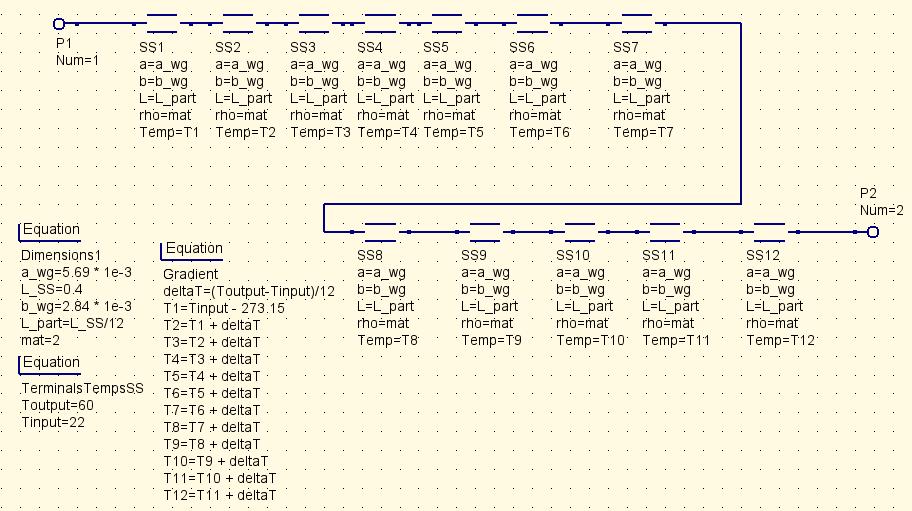}
    \caption{QUCS schematic of the implementation of the stainless steel section, using many rectangular waveguide components at a linearly changing temperature}
    \label{fig:wg_ss}
\end{figure}

    The waveguide model is composed by 2 sections each enclosed by 2 ideal transformers:
    \begin{itemize}
        \item copper section, see figure~\ref{fig:wg_cu} 
        \item stainless steel + 3 sections of gold plated stainless steel (figure~\ref{fig:wg_ssau}) modelled by groups of single waveguide components at different temperatures. In figure~\ref{fig:wg_ss} we show the detail of the stainless steel waveguides modelled joining many waveguide elements each at a different temperature interpolated linearly.
    \end{itemize}

Thanks to this approach it is possible to take into account the influence of temperature gradient on physical temperature and on material resistivity (for more details please refer to Paola Battaglia's work \cite{thesis_battaglia}).

This waveguide model was tested by Cristian Franceschet against the Flight hardware tests using ambient temperature, the results showed a very good matching between measurements and simulations \cite{thesis_franceschet}.

As already mentioned, simulations were performed at ambient temperatures only for validation against the measurements; for producing the complete RCA bandpasses, the nominal temperatures are used, which are $\sim300K$ at the Backend, $\sim140K$, $\sim80K$ and $\sim60K$ at V-grooves interfaces and $\sim20K$ for the entire copper waveguide.

\subsection{Back End Module}
    \subsubsection{Description}
    The Back End Module receives a Microwave signal from the waveguides and outputs a DC volt signal that is then integrated in time and digitised by the Data Acquisition Electronics (DAE).
    Each of the 4 channels of a Back End Module contains:
    \begin{itemize}
    \item a Low Noise Amplifier RF amplification stage
    \item a bandpass filter 
    \item a detector diode which integrates the RF signal to a DC output
    \item a DC amplifier
    \end{itemize}

    \subsubsection{Frequency response measurements}
    The objective of the bandpass characterisation of the BEM was to correlate the DC output (V) to the power RF input (W) as a function of the input signal frequency.
    The DC output in Volts comes from a diode and therefore it is proportional, considering a BEM linear behaviour, to the input power plus a noise term.
    
    The tests were performed by introducing a monochromatic input wave of known power (for example for 30 and 44 GHz was -60 dBm, which is the typical power entering the BEM when observing Tsky $\sim 3K$) sweeping over the interesting bandwidth.

\begin{figure}
    \centering
    \includegraphics[width=\textwidth]{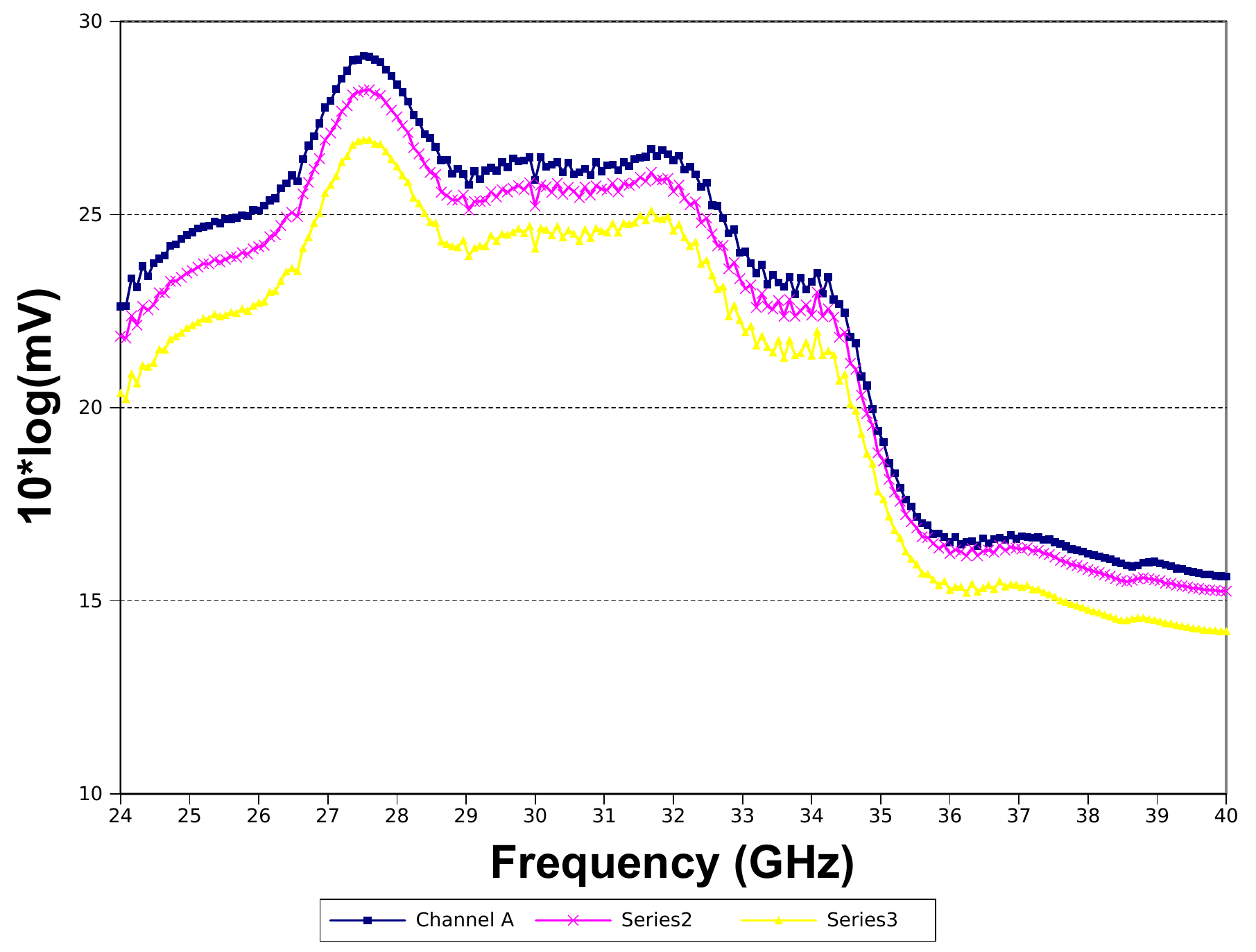}
    \caption{BEM input power to output tension transfer function in dB at high ($48\degree C$), nominal($26\degree C$) and low$5\degree C$  temperature}
    \label{fig:bem_rf2dc}
\end{figure}

BEM response can be modelled as:
    
    \begin{equation}
        V_{out} = G_{BEM} * W_{in} + V_{noise}
    \end{equation}
    
Out of band the Gain is about zero, therefore the output is a good estimation of the BEM noise voltage; therefore the BEM gain can be measured by setting the out of band output voltage as zero level and computing the transfer function between output voltage and input power, see figure~\ref{fig:bem_rf2dc} for LFI27. 

The logarithmic V over W gain without the offset and the BEM noise temperature (which works as an offset independent from the frequency) characterise the BEM frequency response.

\subsubsection{Model}

The RF signal coming from the waveguides is summed with a noise generator based on the measured BEM noise temperature and then amplified frequency by frequency by the V/W transfer function.
The output is a bandpass response of the complete RCA which can be then integrated over the frequency in order to have the output in Volt that can be compared to the hardware output volt.

\paragraph{BEM non-linear model}
During the Flight hardware test campaign, 30 and 44 GHz RCAs showed signal compression at BEM level, i.e. the BEM volt output doesn't increase linearly with the input temperature. This effect has a big impact on noise temperature computation, because noise temperature is computed by linear fitting volt outputs at different temperatures, computing the gain and the offset.

The intercept with the negative part of the temperature axis is equal to the noise temperature; if the radiometer response is not linear, but we are assuming that it is linear, its slope is underestimated and its noise temperature is overestimated.

Therefore, Fabrizio Villa and Luca Terenzi (\cite{gainmodel}) applied a non linear analytic model to LFI radiometers that was used to fit the test data in order to evaluate the overall RCA Gain, System temperature and the BEM compression factor.

The model links the input temperature, $T_A$, to the output voltage $V_{out}$ through the following relationship:
\begin{equation}
    V_{out} = G_{lin} \dfrac{1}{1+b\cdot G_{lin} \cdot (T_A + T_N))} \cdot (T_A + T_N)
\end{equation}

where $G_{lin}$ is the uncompressed gain, $b$ is the compression factor, $T_A$ is antenna temperature and $T_N$ noise temperature.
The compression factor is a single scalar value per channel which represents the BEM Gain dependence on the input power.
The non linear behaviour arises from a compression effect both in the RF amplification stage and in the diode.

Considering that the BEM response is non linear, it is necessary to admit that the measured data we have from the performance campaign are already affected by non linearity and we need first of all to compute an equivalent linear transfer function, $G_{lin}$ from these data.

Using equation~\ref{eq:testtolin} it is possible to compute the linear gain $G_{lin}$ knowing the test input power, the measured compression factor $b$.

\begin{equation}
    G_{lin}=\frac{G_{test}}{1-bP_{in}G_{test}}
    \label{eq:testtolin}
\end{equation}

The BEM non-linear model is not yet been implemented in QUCS, but it could be easily performed. However, due to the fact we do not have extensive measurements of BEM frequency response with different input power level, it is still under discussion whether the effect of the compression on the bandpass is a flattening, a negative offset or a combination of both.
Therefore it is necessary to implement both models into QIMP and deeply study their impact on bandpass response.

\paragraph{QUCS implementation} 

Due to a bug in the ADS software related to a cascade of multiple noisy components, the Advanced LFI RF model implementation of the BEM was very complex and more difficult to manage and improve.

In QIMP instead the implementation is by a single S-Parameters device; its return loss as a function of frequency and its integrated noise temperature must be derived from BEM measurements.

Because measurements provide output voltages versus input power, so that:

\begin{equation}
        G_{BEM} = \dfrac{V_{out}}{W_{in}}
\end{equation} 

we need to manipulate the previous equation in order to convert the power measurements to amplitudes, i.e.:
    
    \begin{eqnarray}
            G_{BEM} = \dfrac{V_{out}}{V_{in}^2/50} \\
        \sqrt{G_{BEM}} = \dfrac{\sqrt{V_{out}}}{\sqrt{V_{in}^2/50}} \\
        \sqrt{V_{out}} = \sqrt{G_{BEM}} \cdot \sqrt{50} V_{in}
    \end{eqnarray}
    
where the factor 50 is due to the standard $50 \Omega$ transmission line impedance.
    
Therefore we have a simple linear relation between the input voltages and the square root of the output voltage.

It is therefore just necessary to create an equation object and apply the square to have the output voltage $V_{out}$.

\subsection{QIMP simulations}
\begin{figure}
    \centering
    \includegraphics[width=\textwidth]{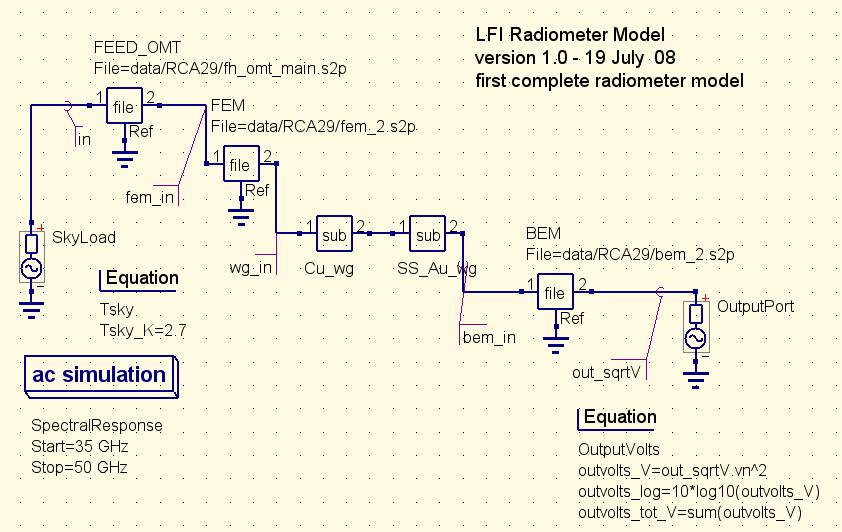}
    \caption{Complete schematic of the LFI Radiometer model implemented in QUCS}
    \label{fig:rad_mod}
\end{figure}

The QUCS simulation is run on the complete schematic defined following the previous sections, see figure~\ref{fig:rad_mod}.

The result is a voltage output in Volts as a function of frequency; the simulations approach is very similar to swept source measurements where a monochromatic input power is swept through the bandpass and the BEM voltage output is recorded; the main difference is that in the test there is also a small contribution to the input due to the background temperature of the sky load all over the receiver band. However, this contribution is negligible because the injected microwave signal is about 3 orders of magnitude higher than the background temperature.

All the RCA bands are swept with a 0.1 GHz step: for each step the power input to the Feed Horn is a rectangular step (in the frequency domain) centred in the sweeping frequency and 0.1 GHz wide, with the power level determined by the input black-body thermal emission spectrum.

Taking into account the frequency response of each component and multiple reflections, including the waveguides model, the LFI Radiometer model computes how the input frequency spectrum is modified by the effect of each component.

\begin{figure}
    \centering
    \includegraphics[width=\textwidth]{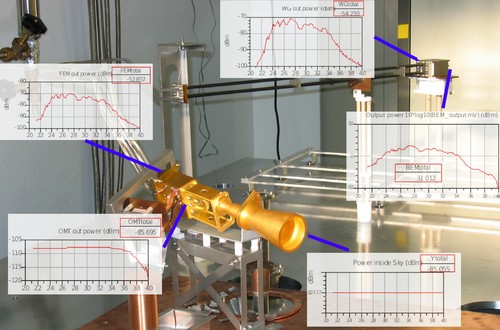}
    \caption{Bandpasses shown along the RCA}
    \label{fig:rca_with_bandpass}
\end{figure}

An example is provided in figure~\ref{fig:rca_with_bandpass} that shows plots of the bandpass along a 30 GHz RCA: it is important to note that neither the bandshape nor the total power in any point along the RCA before the BEMs have been measured experimentally during the test campaigns.

\begin{figure}
    \centering
    \includegraphics[width=.45\textwidth]{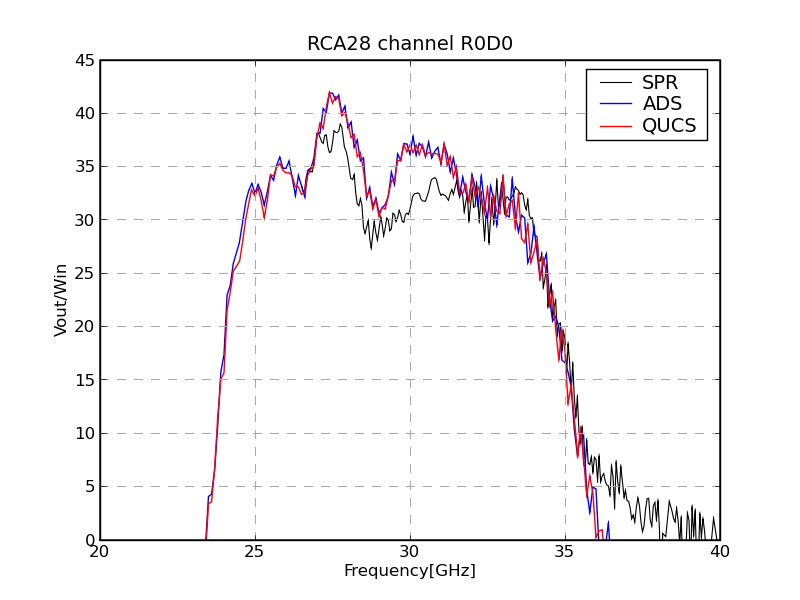}
    \includegraphics[width=.45\textwidth]{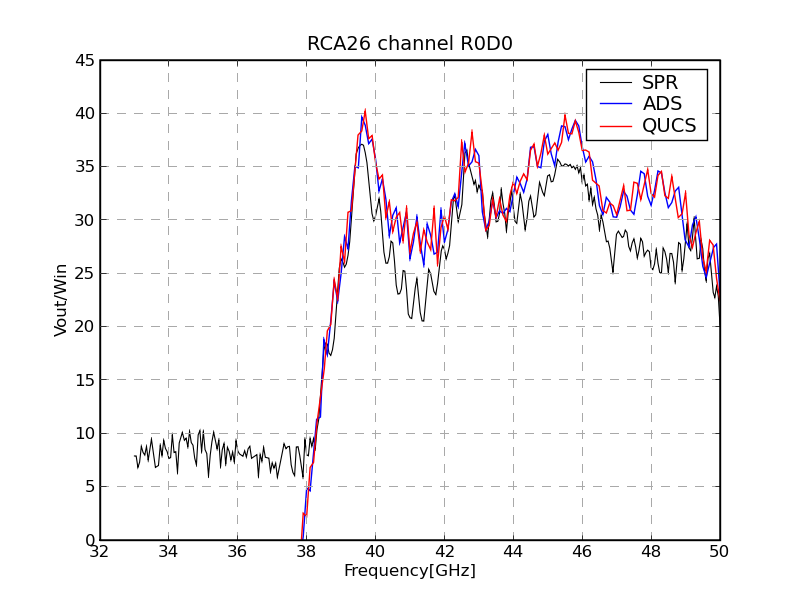}
    \caption{Comparison of ADS, QUCS and measurements bandpasses, LFI28M-00 (left) and LFI26M-00 (right)}
    \label{fig:adsqucs_3044}
\end{figure}

\begin{figure}
    \centering
    \includegraphics[width=.45\textwidth]{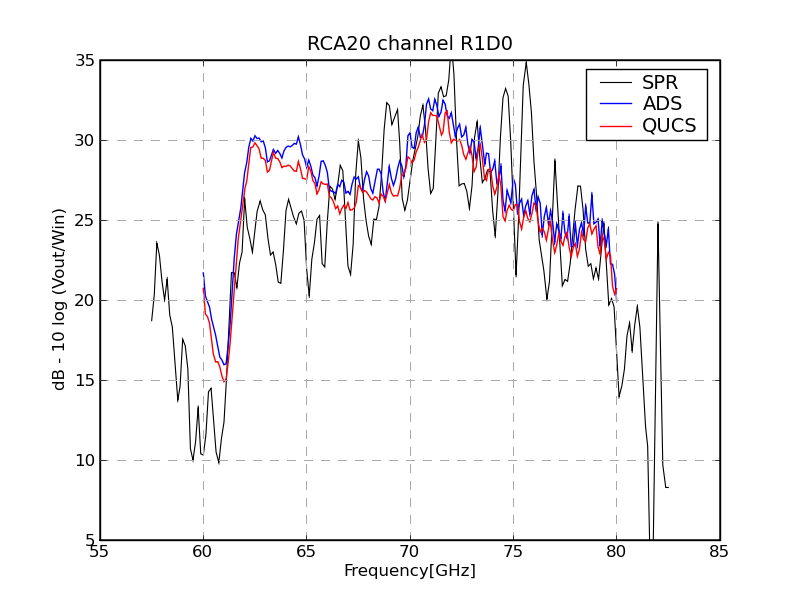}
    \caption{Comparison of ADS, QUCS and measurements bandpasses, LFI20S-10}
    \label{fig:adsqucs_70}
\end{figure}

The first objective of the model port to QUCS was to reproduce exactly the same results concerning output bandpasses; after the waveguide model was implemented into QIMP, the result was reached successfully, in the figures \ref{fig:adsqucs_3044} and \ref{fig:adsqucs_70} I show an example of the comparison between ADS model, QUCS model and measurements for each frequency 


\subsection{Automation}

QIMP includes also a set of Python programs needed for automating most of the procedure of running simulations. I will just review their main features.

They are based on the software package SciPy, see \url{http://www.scipy.org}, a general purpose scientific and numerical software for Python, in particular on:
    \begin{itemize}
        \item NumPy: multidimensional array package
        \item Matplotlib: 2D plotting library 
        \item Ipython: interactive Python console
    \end{itemize}

The programs are implemented following Object Oriented Programming and are stricly independent, each stage runs separately by reading inputs and writing results on standard ASCII files, the stages are:

\begin{itemize}
    \item[ADS2QUCS:] Format conversion from ADS to Touchstone, merging of new input data, quick modification of input data
    \item[lfisim:] Batch simulation run using \texttt{qucsator}, QUCS GUI is never launched, and export of bandshapes from QUCS to ASCII format; a simulation of all 44 channels takes about 5 minutes.
    \item[ba\_lib:] Bandshape analysis tool, for batch processing and interactive analysis: 
      \begin{itemize}
        \item gain bandshape from simulations at 2 different temperatures
        \item normalization
        \item plotting
        \item comparison with Swept source measurements and ADS simulation
        \item export to text format
      \end{itemize}
\end{itemize}

Moreover, thanks to Ipython, it is easy to switch from batch processing to an interactive data analysis session which is very useful when investigating a specific issue.

\clearpage

\section{Frequency response measurements}
\label{sec:swept}

The most important validation of the QIMP model relies on the comparison between the modelled and measured bandpasses.

This section contains a description of the measurements setup and results which are important for understanding the results of the comparison.

\subsection{30 and 44 GHz Radiometers}
\label{sec:swept3044}

\subsubsection{Experimental setup}
\label{sec:swept3044setup}

30 and 44 GHz RCAs were assembled and tested in Thales Alenia Space in Vimodrone during Spring 2006; each RCA was tested separately at ambient and cryogenic temperature. By requirements, the aim of the test was to measure relative frequency response of the instrument, i.e. what is the response relative to its own maximum.

\begin{figure}[h]
    \centering
    \includegraphics[width=\columnwidth]{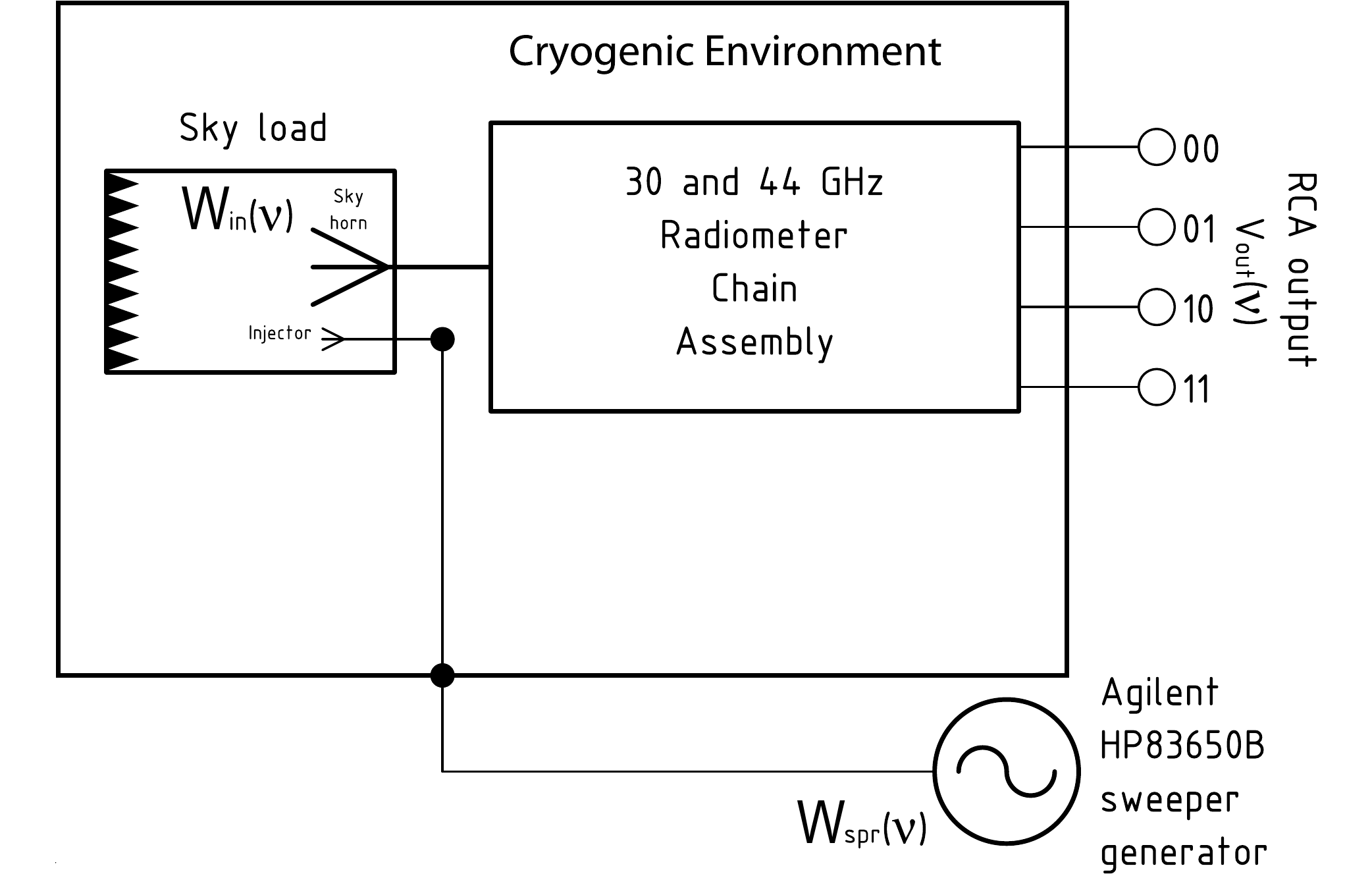}
    \caption{Schematic of the experimental setup for 30/44 GHz Swept Source Tests: a monochromatic input microwave is injected into the sky load, its frequency changes with time through the radiometer band. From the recorded output voltage the bandpass has been determined using Eq.~\ref{eq:bandpass_calculation}.}
    \label{fig:sprsetup}
\end{figure}

    The bandpass was measured by injecting a calibrated monochromatic source into the sky horn sweeping through the operational band and recording the DC output as a function of the input frequency, see figure~\ref{fig:sprsetup}.

The Agilent synthesised microwave generator HP83650B was located at ambient temperature outside of the cryogenic chamber; its output on RF coaxial cable was transferred to rectangular waveguide (WR28 for 30 GHz and WR22 for 44 GHz) by an Agilent transition. The transition was connected with a short waveguide section to the cryogenic chamber by a RF vacuum feed-through with a kapton window. 
Inside the chamber, another straight stainless steel waveguide section acts as a thermal break, to reduce thermal inflow from the outside environment; the last section of the transmission line is a 1-meter long flexible waveguide which allows the connection to the sky load.

The sky load is a cylindrical black cavity 20cm high and with a diameter of 20cm; its back wall, positioned in front of the horn, is a bed of Eccosorb pyramids 5 mm large and 30 mm high. Sky load is composed by modular absorbing panels screwed to an aluminium plate, they are free to float on it in order to compensate for different thermal contractions.

\begin{figure}[h]
    \centering
    \includegraphics[width=\columnwidth]{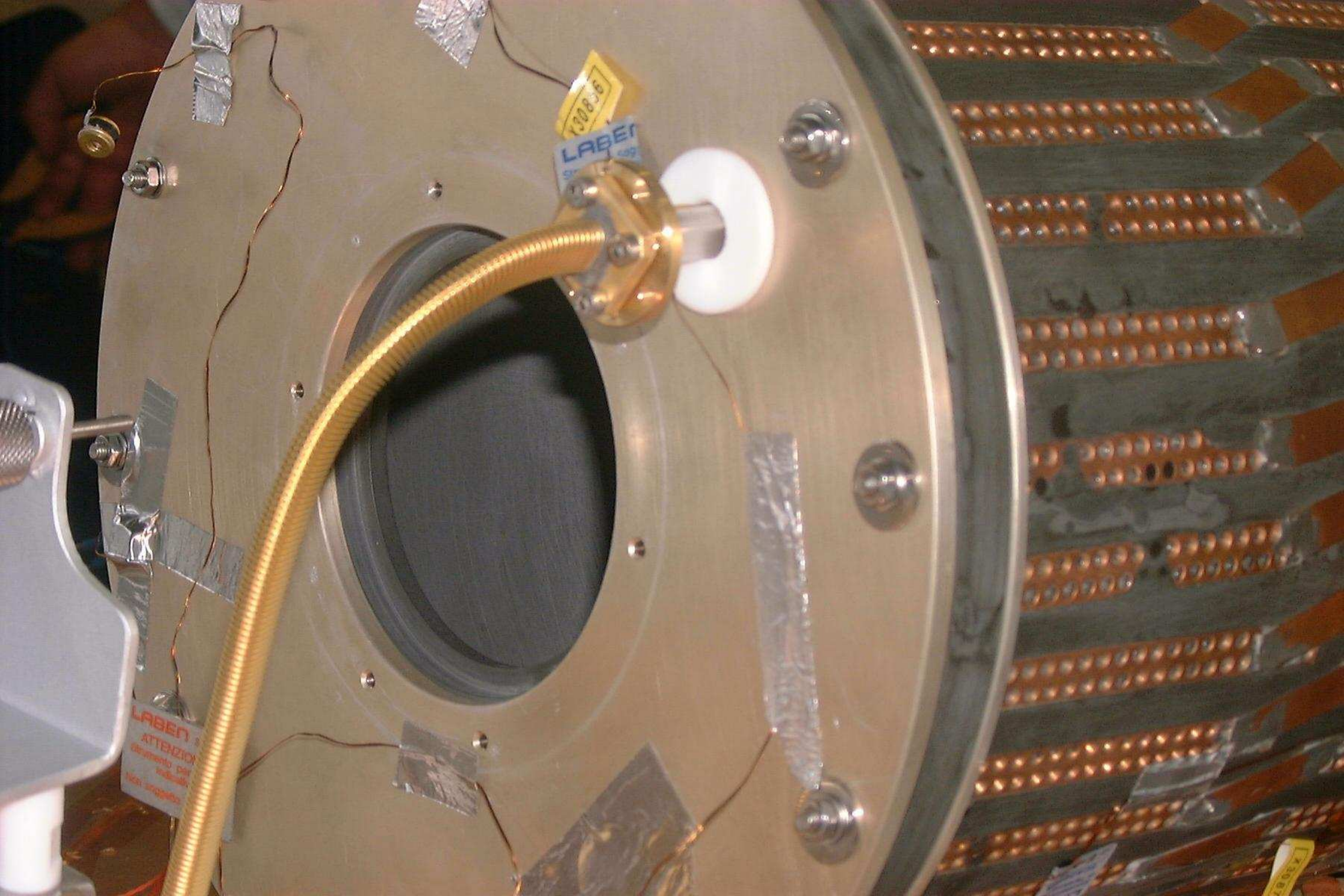}
    \caption{Picture of the injector connected to the sky load for a 30 GHz RCA; the large hole in the centre of the sky load cylindrical base is dedicated to the feed horn}
    \label{fig:injector}
\end{figure}

The injector, see figure~\ref{fig:injector}, is a open ended waveguide located on the same side of the horn with the axis parallel to horn axis; therefore the signal was reflected into the horn by the sky load back wall. The sky load return loss is about 60 dB, so that with the input power set to -35 dBm, the expected power delivered to the Feed Horn is about -95 dBm.
Not knowing exactly the input power permits only to compute a relative bandpass, which is coherent with test requirements, but forces the comparison with modelling to be based just on normalised bandpasses.

\subsubsection{Results}
\label{sec:swept3044res}

\begin{figure}[h]
    \centering
    \resizebox{\hsize}{!}{\includegraphics{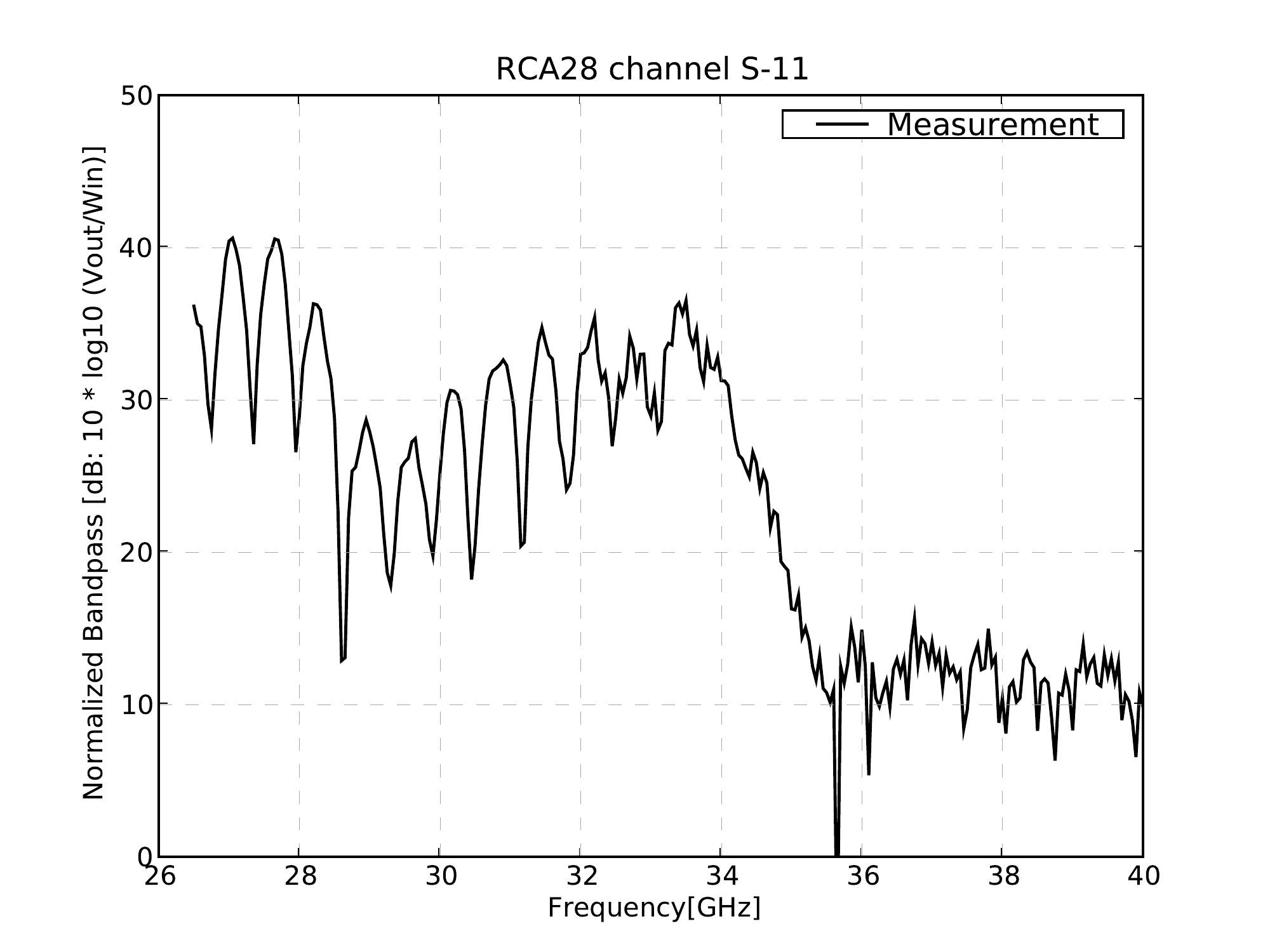}}
    \resizebox{\hsize}{!}{\includegraphics{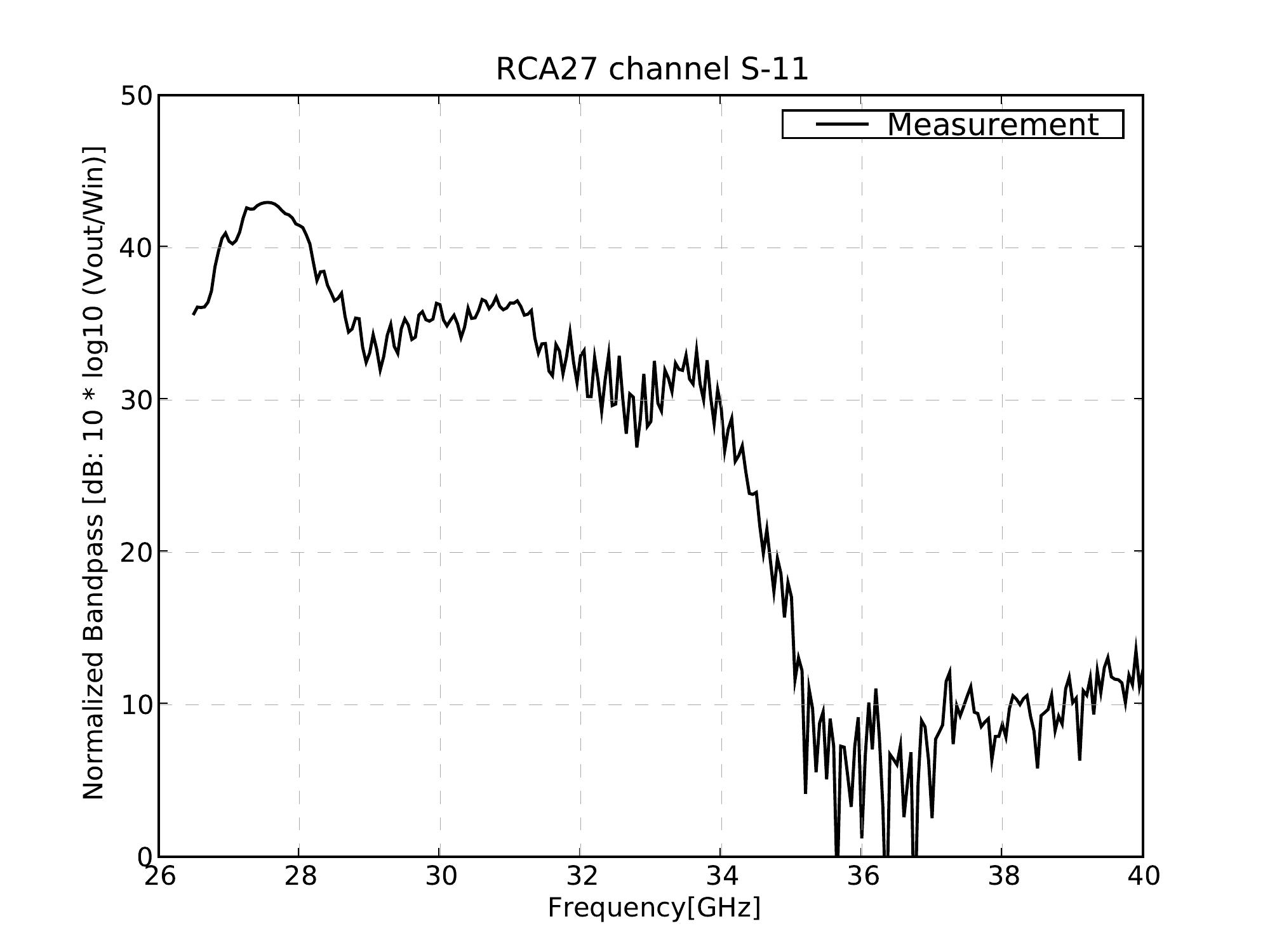}}
    \caption{Swept Source bandpass measurements of 30 GHz: RCA LFI28S-11 is an example of channels showing ripples due to injector orientation, while LFI27S-11 an example of a clean measurement, bandpasses measured are just relative, therefore the normalisation is arbitrary, in this case the integral was set equal to the corresponding modelled bandpasses - see section~\ref{sec:comparison}}
    \label{fig:spr3044}
\end{figure}

The measurement campaign gave poor results for 2 channels of the same polarisation axis of each 30 GHz RCA due the presence of 1 GHz spaced ripples, see upper figure~\ref{fig:spr3044}, and a very low and noisy Volt output. Investigation at warm temperature showed that the power delivered to the Feed Horn was highly dependent on injector orientation; the ripples disappeared by connecting the injector with its polarisation plane at 45 degrees with respect to OMT Main arm and Side arm polarisation planes (refer to \cite{rca27_report}).
Unfortunately it was not possible to repeat the tests in cryogenic chamber, therefore frequency characterisation of these channels can only rely on simulations.

Among the eight 30 GHz channels, four of them show ripples as highlighted in the previous paragraph, while the remaining 4 are very clean, see lower figure~\ref{fig:spr3044}, and extends down to 26.5 GHz; this limit was established considering the nominal bandwidth of 20\% of the central frequency, e.g. 6 GHz [27-33 GHz] for 30 GHz and that 26.5-40 GHz is the nominal band of WR28 waveguides, used for injecting the signal into the sky load. However, the response at 26.5 GHz is still high and useful bandwidth is expected below this limit, but the shape of this low frequency cut was not measured.

\begin{figure}[h]
    \centering
    \includegraphics[width=\columnwidth]{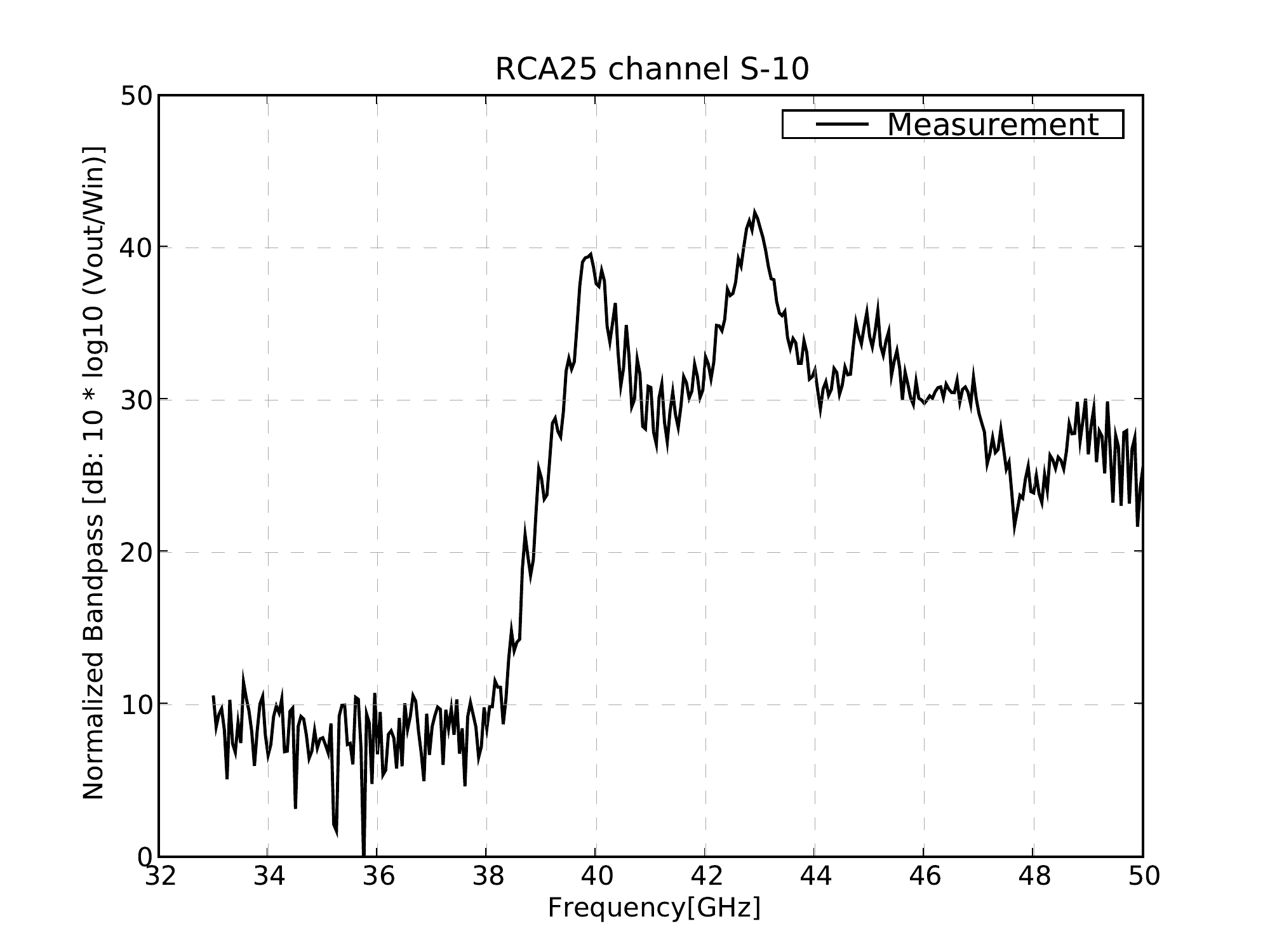}
    \caption{Swept Source bandpass measurements of LFI25S-10, 44 GHz, bandpasses measured are just relative, therefore the normalisation is arbitrary, in this case the integral was set equal to the corresponding modelled bandpasses - see section~\ref{sec:comparison}}
    \label{fig:spr44}
\end{figure}

All twelve 44 GHz channels measurements are very clean and frequency covered range is sufficient, see figure~\ref{fig:spr44}; the response at 50 GHz is still higher than background noise but already one order of magnitude less than central band gain.

\clearpage
\subsection{70 GHz Radiometers}
\label{sec:swept70}

\subsubsection{Experimental setup}
\label{sec:setup70}

\begin{figure}[h]
    \centering
    \includegraphics[width=\columnwidth]{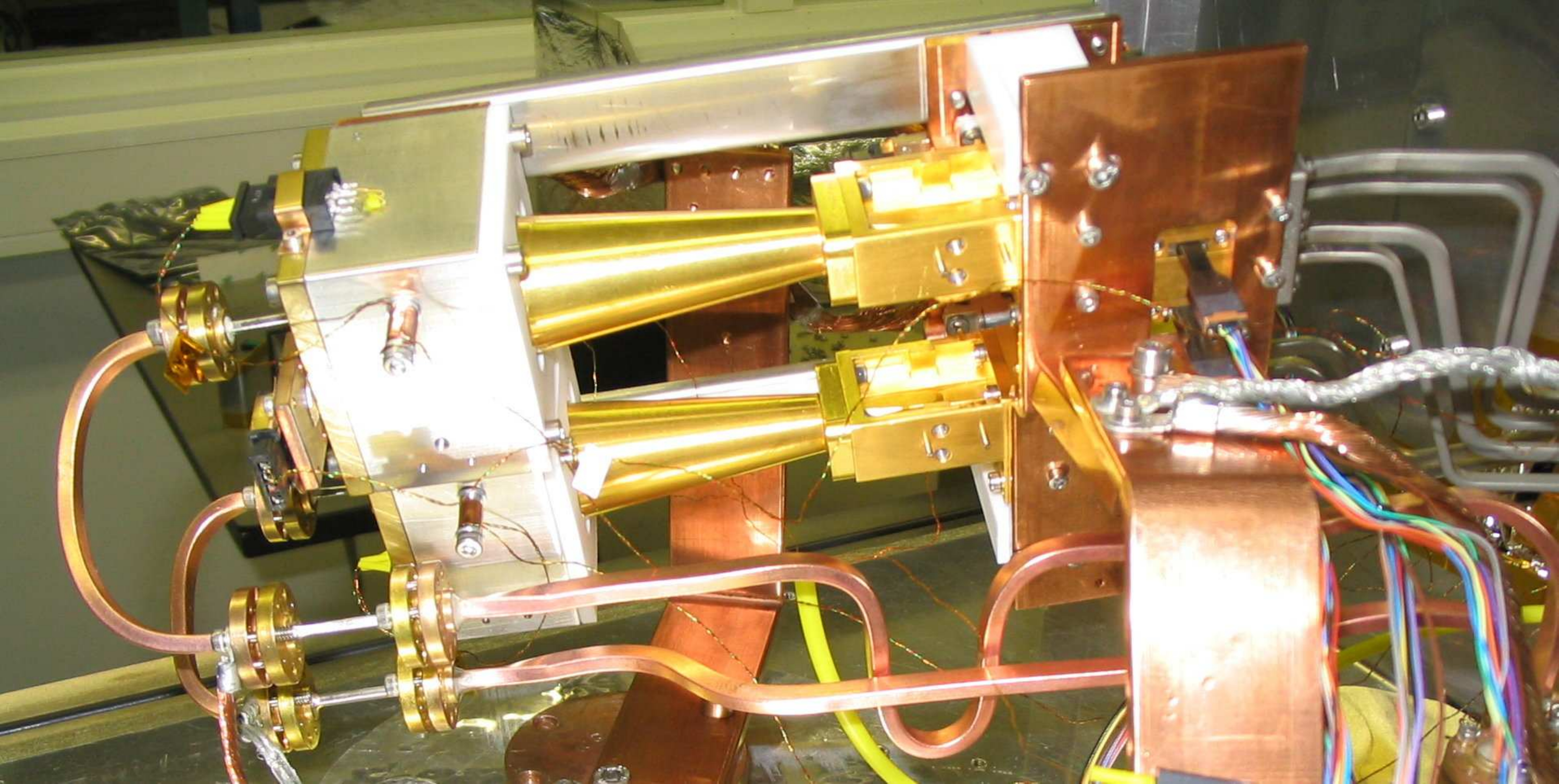}
        \caption{Picture of the two 70 GHz horns connected with the small sky loads provided by Ylinen, input signal waveguides and injectors are located on the left}
    \label{fig:skyload70}
\end{figure}

70 GHz radiometers were produced, assembled and tested at Elektrobit Ylinen in Finland, see \cite{Ylinen18}. The Swept Source test setup was different than for 30 and 44 GHz tests: the Microwave Generator was connected to a small sky load directly in front of the horn, see figure~\ref{fig:skyload70}.

Thanks to this design it is possible to estimate the power effectively delivered to the horn by using a calibrated input source and applying corrections due to the connection waveguides and the injector; in this conditions the skyload delivers all the input power to the feed horn.

\begin{figure}[h]
    \centering
    \includegraphics[width=\columnwidth]{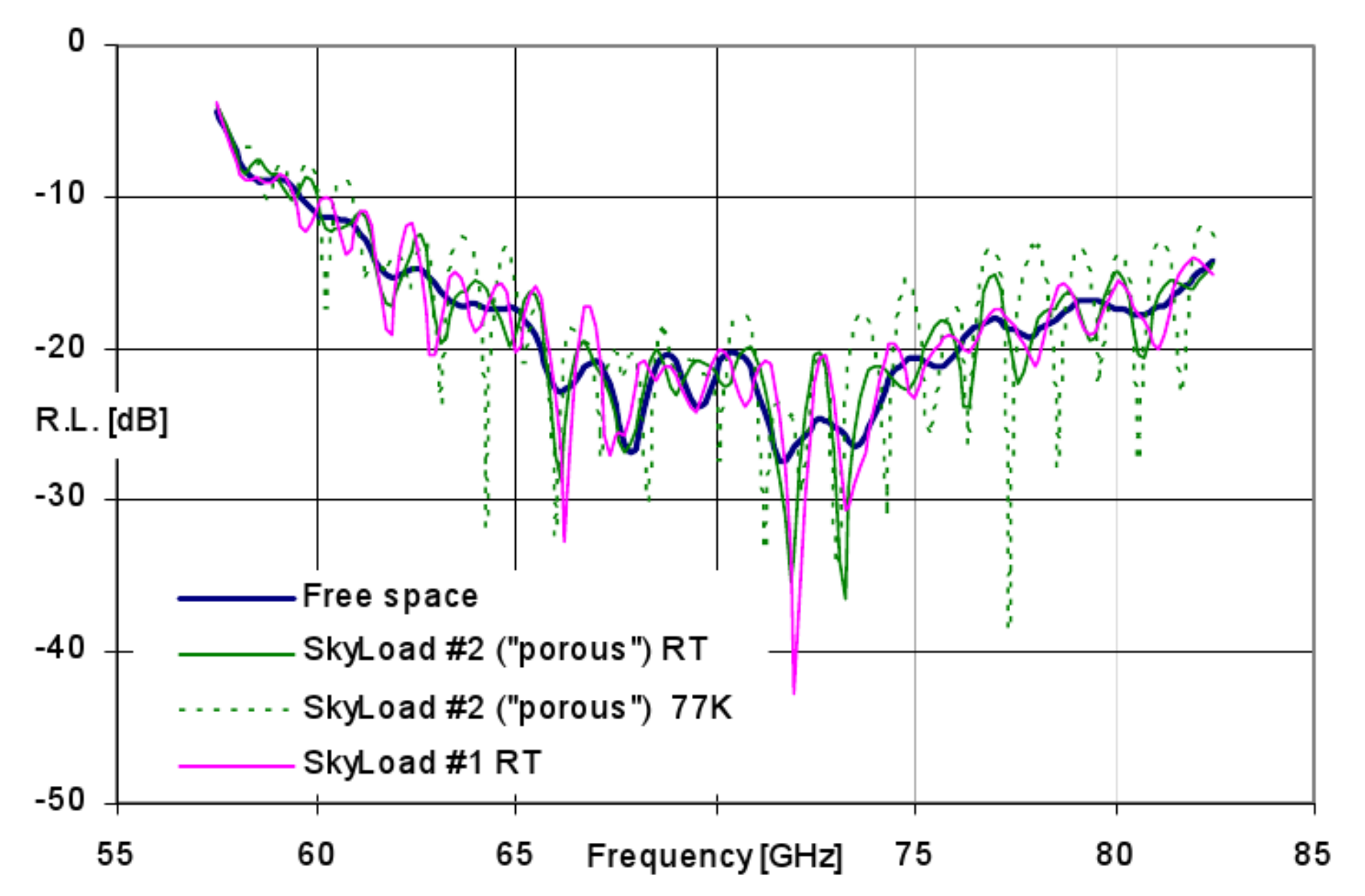}
        \caption{Measured return loss of the sky load used for testing the 70 GHz RCAs, the plot shows two different type of sky loads, \#2, which was used for the tests, at different temperatures.}
    \label{fig:skyload70rl}
\end{figure}

The main issue of this system is the return loss of the sky load; the requirement was -30 dB, but the measured performance didn't meet the requirement, see figure~\ref{fig:skyload70rl}. The favoured hypotheses is that the observed ripples in the bandpass measurement are caused by strong standing waves between the feed horn and the sky load due to its high return loss.

\subsubsection{Results}
\label{sec:res70}

The standing waves had a big impact on measured bandpasses, which suffered of regularly spaced ripples between 1 and 2 GHz wide and with a peak to peak amplitude between 5 and 15 dB.

\begin{figure}[h]
    \centering
    \includegraphics[width=\columnwidth]{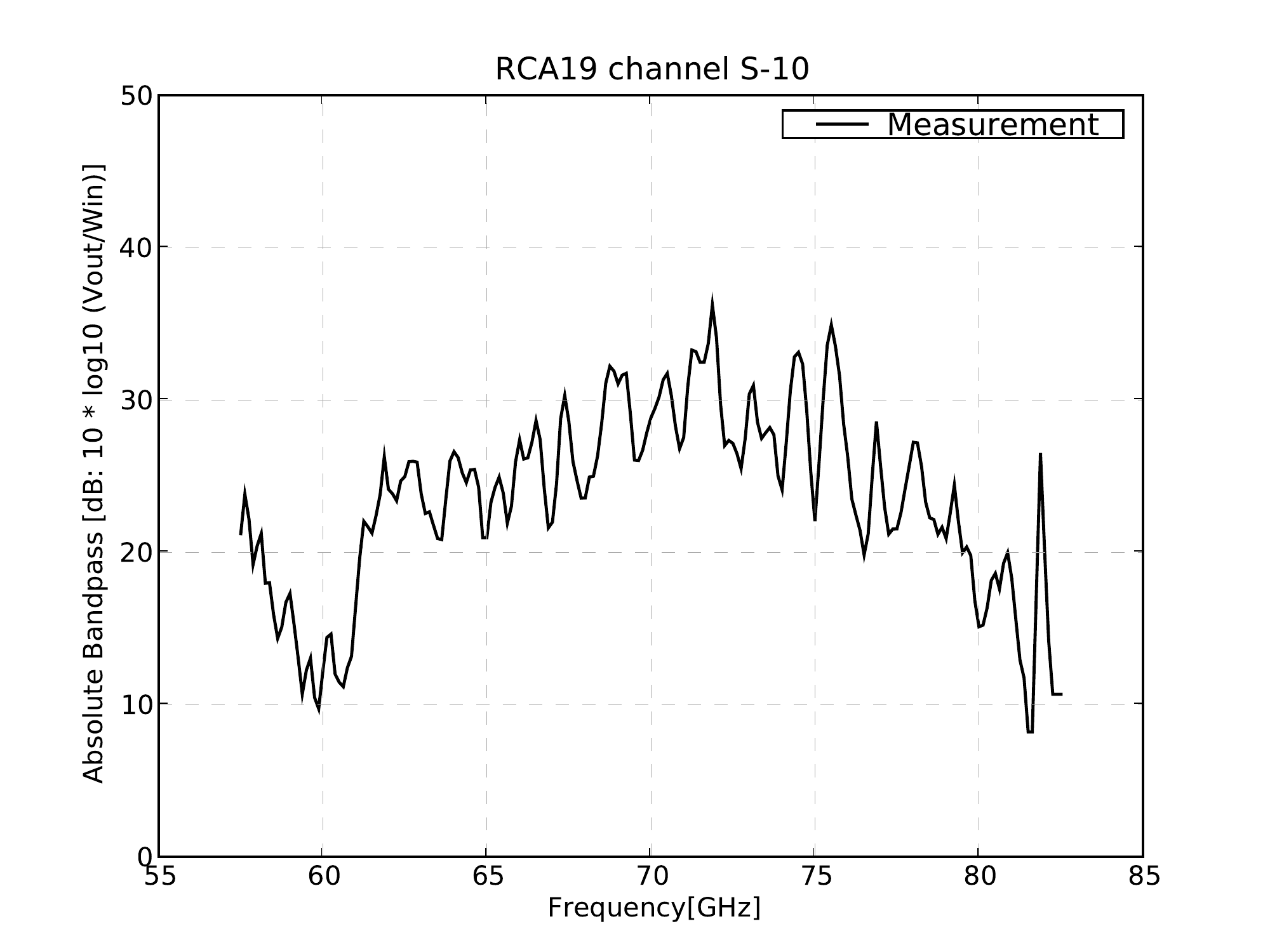}
    \caption{Swept Source bandpass measurements of LFI19S-10, one of the 70 GHz channels, bandpasses measured are absolute}
    \label{fig:spr70}
\end{figure}

The ripples are very similar in all 70 GHz channels, see an example in figure~\ref{fig:spr70}; moreover ripples pattern is not regular enough for removing it with software manipulation of the data without the risk of adding additional errors. 

\clearpage
\section{Comparison of measurements and simulations}
\label{sec:comparison}

As written in section \ref{sec:swept}, 30 and 44 GHz RCA Swept Source measurements provided only relative bandpasses while 70 GHz tests provided absolute bandpasses between input power and output voltage. 

Therefore, for 70 GHz the comparison process is straightforward, simulations bandpass is computed by:
\begin{equation}
G  = \dfrac{P_{high}-P_{low}}{(V_{high}-V_{low})}
\end{equation}

where the subscripts refer to simulations performed at 2 different temperatures. The dimensions of the result is $[W]/[V]$ and can be directly compared to Swept Source Tests results as explained in section \ref{sec:swept70}.

For 30 and 44 GHz channels the input power is very high compared to radiometer noise temperature, therefore $T_{noise}$ effect on the output is negligible and it is possible to compare the normalised bandpass directly with the normalised simulated gain.

In order to compare measurements to simulations measured data have been normalised as model results:
\begin{eqnarray}
\tilde{G}_{meas}(\nu) & =  & \lambda G_{meas} \\ 
\text{where:} & &\\
\lambda &  =  & \dfrac{G_{meas}(\nu)}{\int G_{meas} d\nu } \cdot \int G_{model} d\nu
\label{eq:normalization}
\end{eqnarray}

Comparisons of measurements and simulations for all LFI channels are plotted in figure~\ref{fig:comp70} and figure~\ref{fig:comp3044} at the end of the chapter.

  \subsection{30 GHz Radiometers}

One of the noticeable results of the QIMP simulations has been to provide information about the bandwidth below 26.5 GHz, down to 21.3 GHz. In fact RCA measurements were available only from 26.5 GHz to 40 GHz.

FEM response at lower frequencies has been obtained extrapolating the available data using as a reference VNA measurements performed on a larger bandwidth, see \ref{sec:fem_tests}. OMT response instead was extrapolated by exploiting its similarity with 44 GHz OMTs, which were measured on a much larger frequency span, see \ref{sec:omt_extra}.

\begin{figure}
    \centering
    \resizebox{\hsize}{!}{\includegraphics{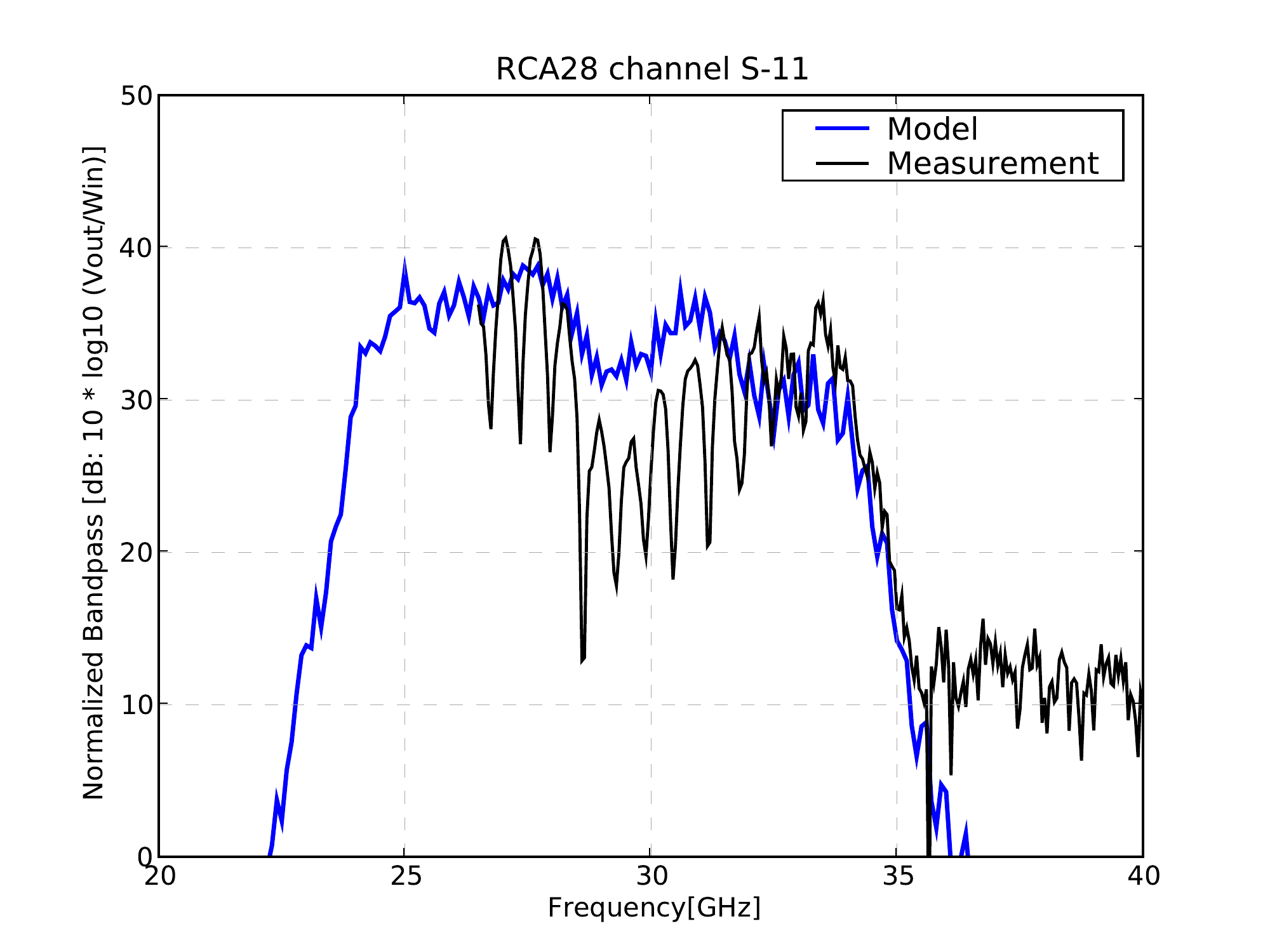}}
    
        \caption{30 GHz channel LFI28S-11, an example of the polarisation issues on half of the 30 GHz channels: measurements were normalised considering model output as a reference, bandpasses units are $dB: 10 \log10([V]/[W])$}
    \label{fig:30GHz-band}
\end{figure}

\begin{figure}
    \centering
    \resizebox{\hsize}{!}{\includegraphics{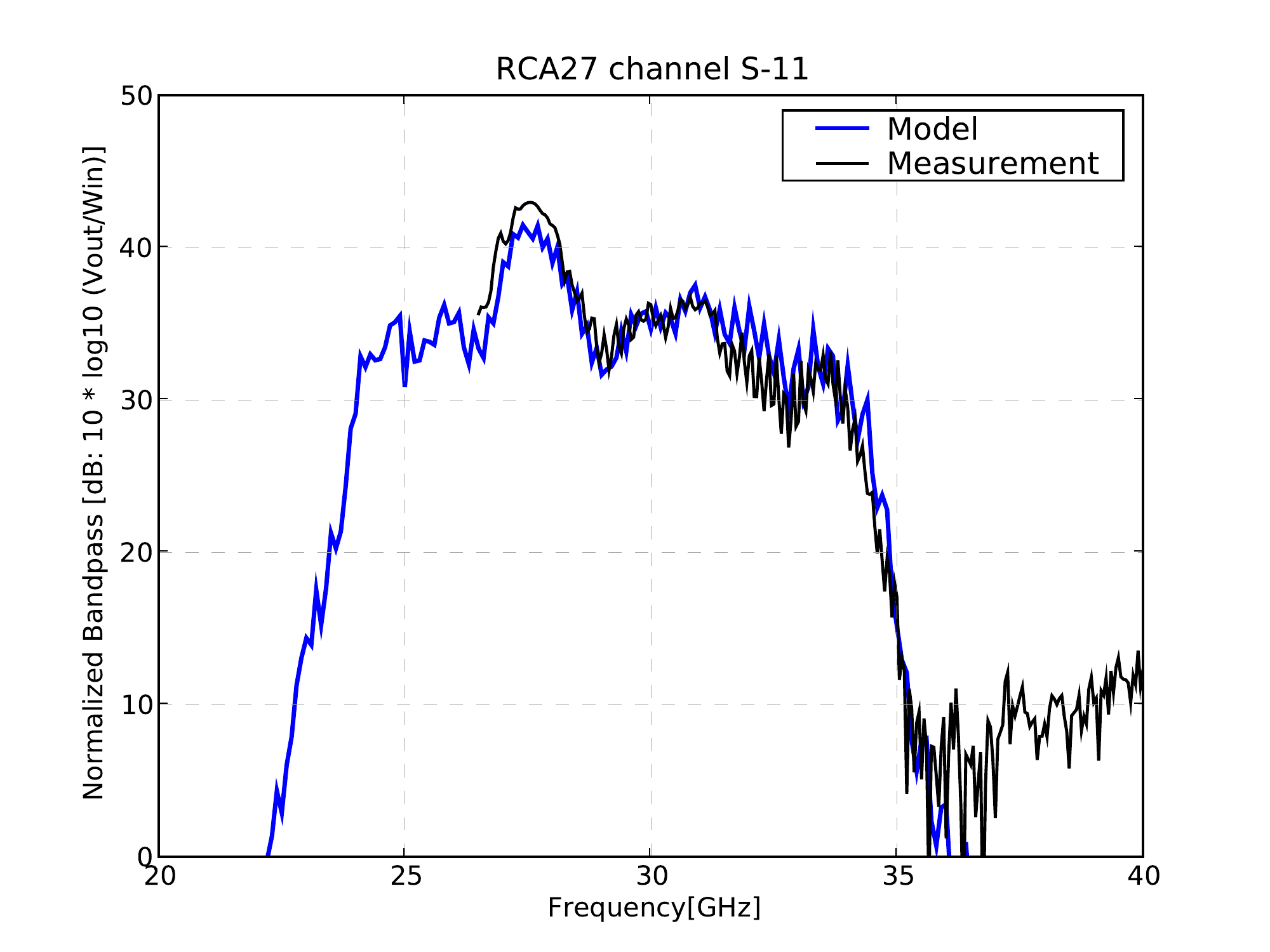}}
    \caption{30 GHz channel LFI27S-11, good agreement between measurements and simulations}
    \label{fig:30GHz-band27}
\end{figure}

\begin{figure}
    \centering
    \resizebox{\hsize}{!}{\includegraphics{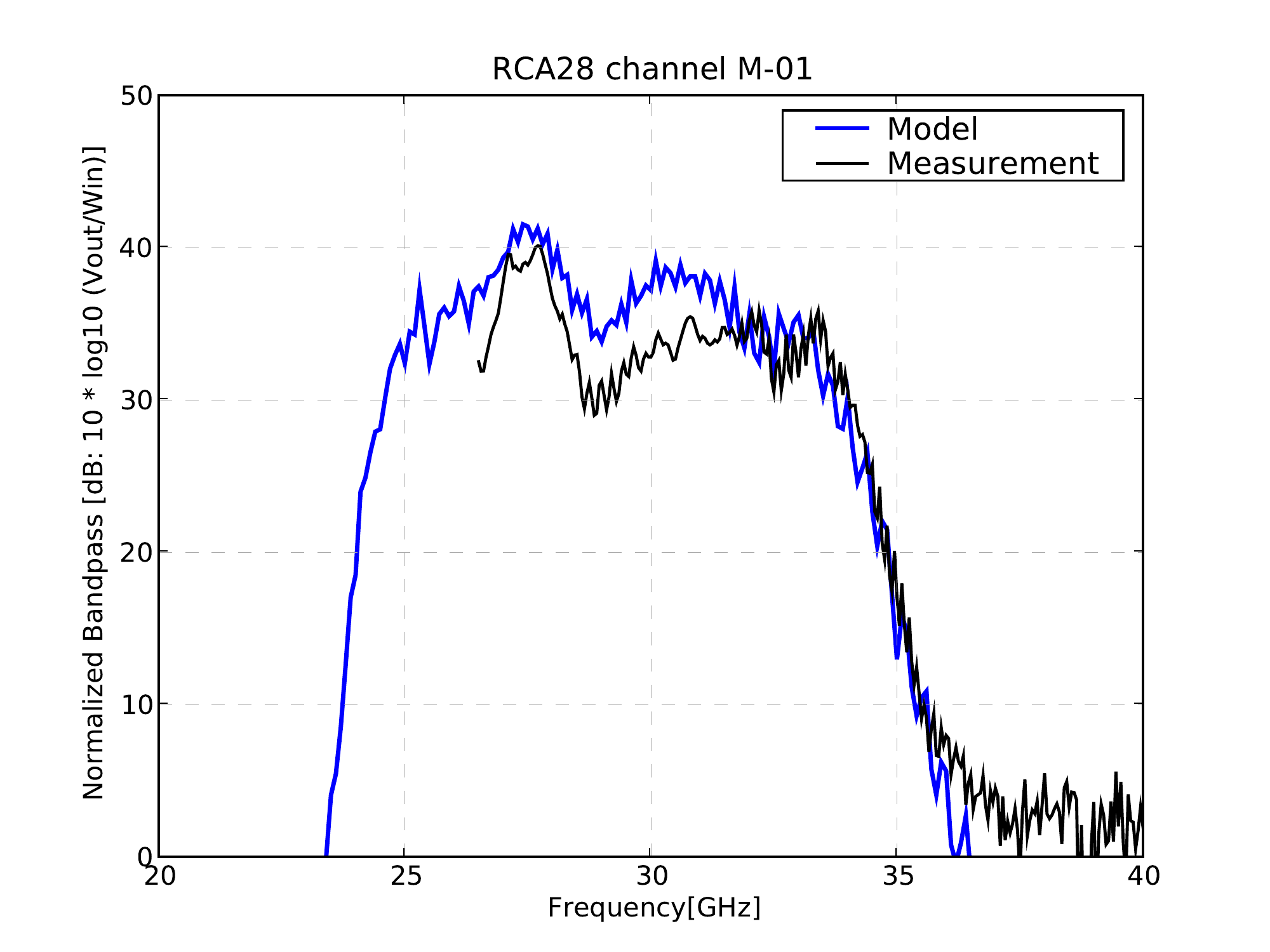}}
    \caption{30 GHz channel LFI28M-01, general good agreement beside a discrepancy in the central part of the band}
    \label{fig:30GHz-band28}
\end{figure}

\begin{figure*}
    \begin{center}
    \begin{tabular}{m{0.3cm} |c c c c} 
   & \textbf{M-00} & \textbf{M-01} & \textbf{S-10} & \textbf{S-11}\\

\input{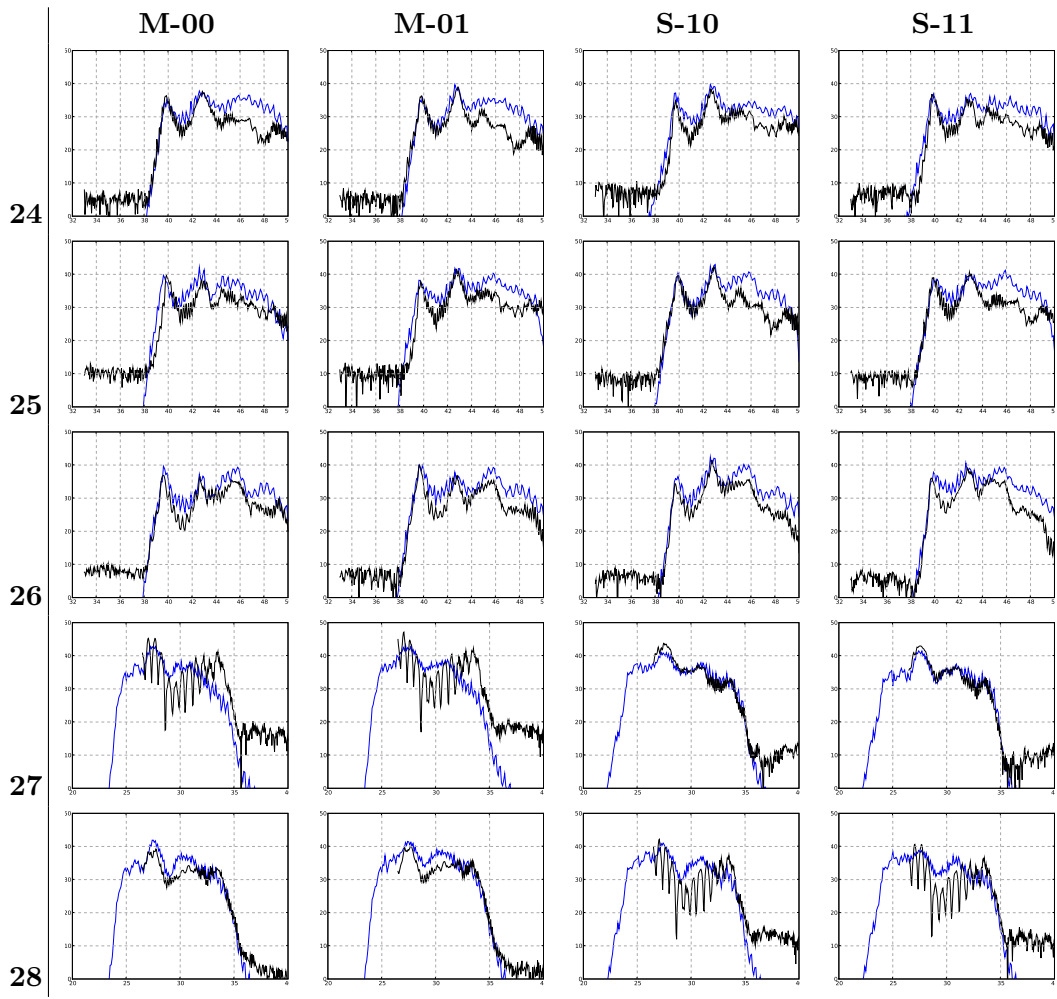}

    \end{tabular} 
    \caption{30 and 44 GHz RCAs comparison between measurements(black) and model(blue) results}
    \label{fig:comp3044}
    \end{center}
\end{figure*}

At the end of the chapter, in figure~\ref{fig:comp3044}, we show the comparison between simulations and measurements for all the 30 GHz channels.

It is important to note that measurements convey in this case, and for 44 GHz as well, only a relative information. They have a free normalisation factor which translates into an offset in logarithmic scale; therefore they were normalised using the model output as a reference.

In general there is good qualitative matching between measurements and simulations, which supports the hypotheses that there are no major influence among the components, i.e. the performance of radiometer devices doesn't change significantly when assembled into an RCA.
 
On the other side this is also a good indirect test of performance tests at component level, which have been effective in characterising radiometer frequency response.

Some channels show strong discrepancies due to issues on the measurements setup while others show smaller discrepancies still under investigation.

In particular Figure~\ref{fig:30GHz-band} shows an example of a channel affected by high ripples due to the measurements test setup, see section~\ref{sec:swept3044}, in this case comparison is useful just for fitting the high frequency gain drop slope. Channels affected by this issue are 4 on the total of 8 channels.

Figures \ref{fig:30GHz-band27} and  \ref{fig:30GHz-band28} shows comparison from RCA 27 and 28, these channels are very similar to their own twin channels; discrepancy above 36 GHz is due to the thermal noise of the BEMs, which is present in the measurements but not taken into account by the model.

RCA 27 comparison is very good while RCA 28 shows an important discrepancy in the central part of the band which is still under investigation.

  \subsection{70 GHz Radiometers}

Due to the different test setup, 70 GHz measurements gave absolute bandpasses, therefore the comparison with the model is both on the output level and on the band shape.

In this case measurements are slightly broader than simulations, 57.5-82.5 GHz versus 60-80 GHz, but they are highly affected by ripples very likely caused by standing waves.

As highlighted in section \ref{sec:swept70}, the main issue concerning 70 GHz Swept Source Tests is the strong presence of ripples in the bandpass; the modulation of the superimposed ripples pattern is even larger than the modulation of the radiometer bandpass, its removal, indeed, is not straightforward and may introduce uncontrolled systematics in the result.

\begin{figure}
    \centering
    \includegraphics[width=\textwidth]{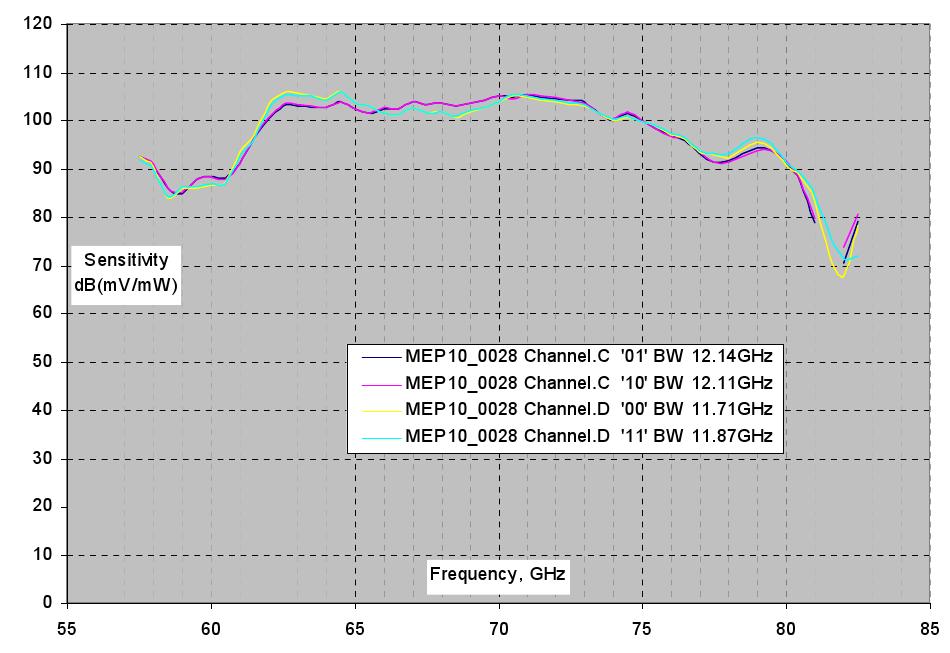}
    \caption{Ylinen Tests on FEM and BEM with signal injection in waveguide directly into FEM}
    \label{fig:fembem70}
\end{figure}

However, the difference between model and measurement looks really consistent with a standing wave pattern; a good evidence supporting this hypothesis is that Elektrobit performed a preliminary test of FEM and BEM alone with direct signal injection in waveguide and its result doesn't show traces of ripples, see figure~\ref{fig:fembem70}.

\begin{figure}
    \centering
    \resizebox{\hsize}{!}{\includegraphics{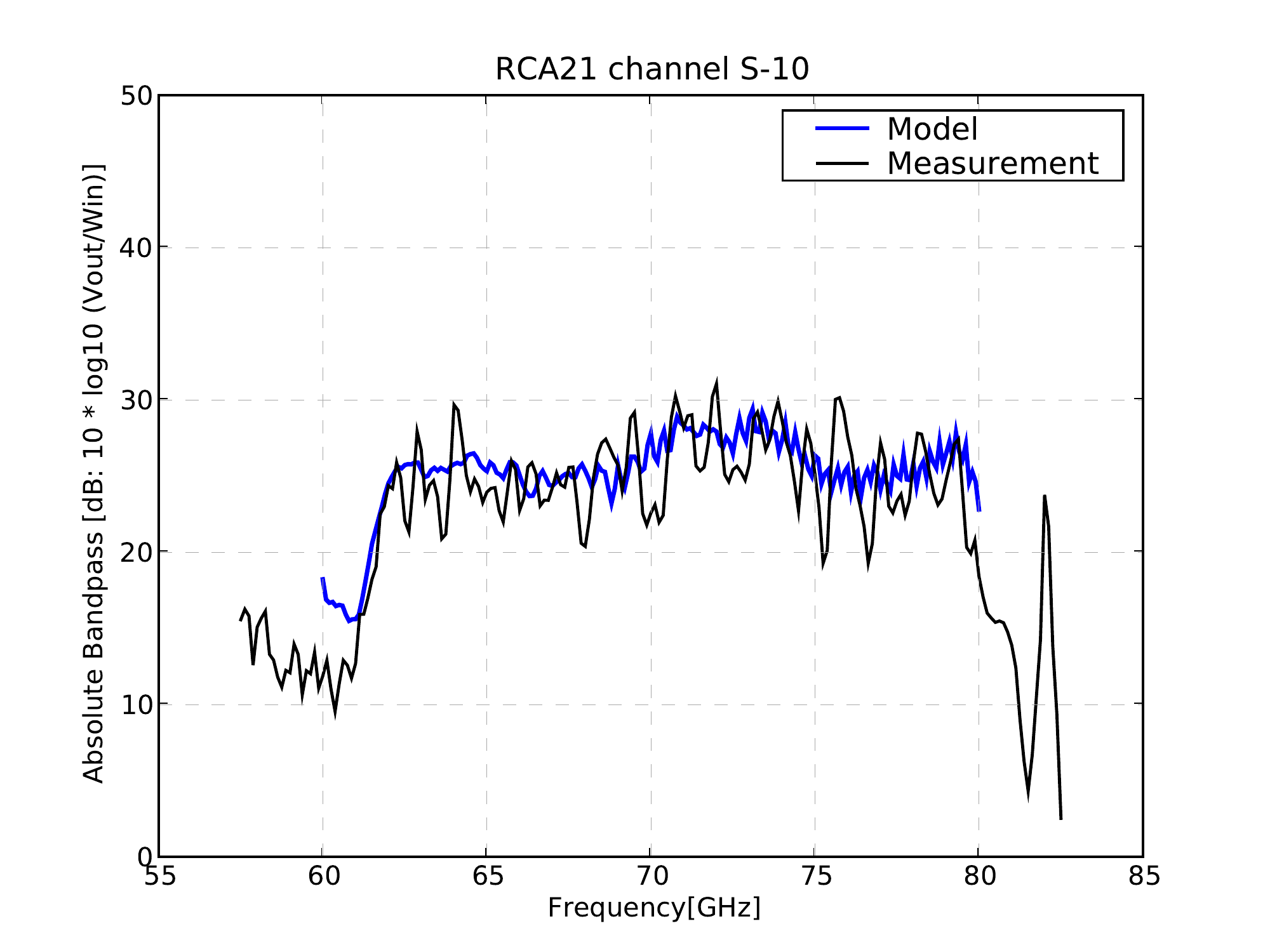}}
    \resizebox{\hsize}{!}{\includegraphics{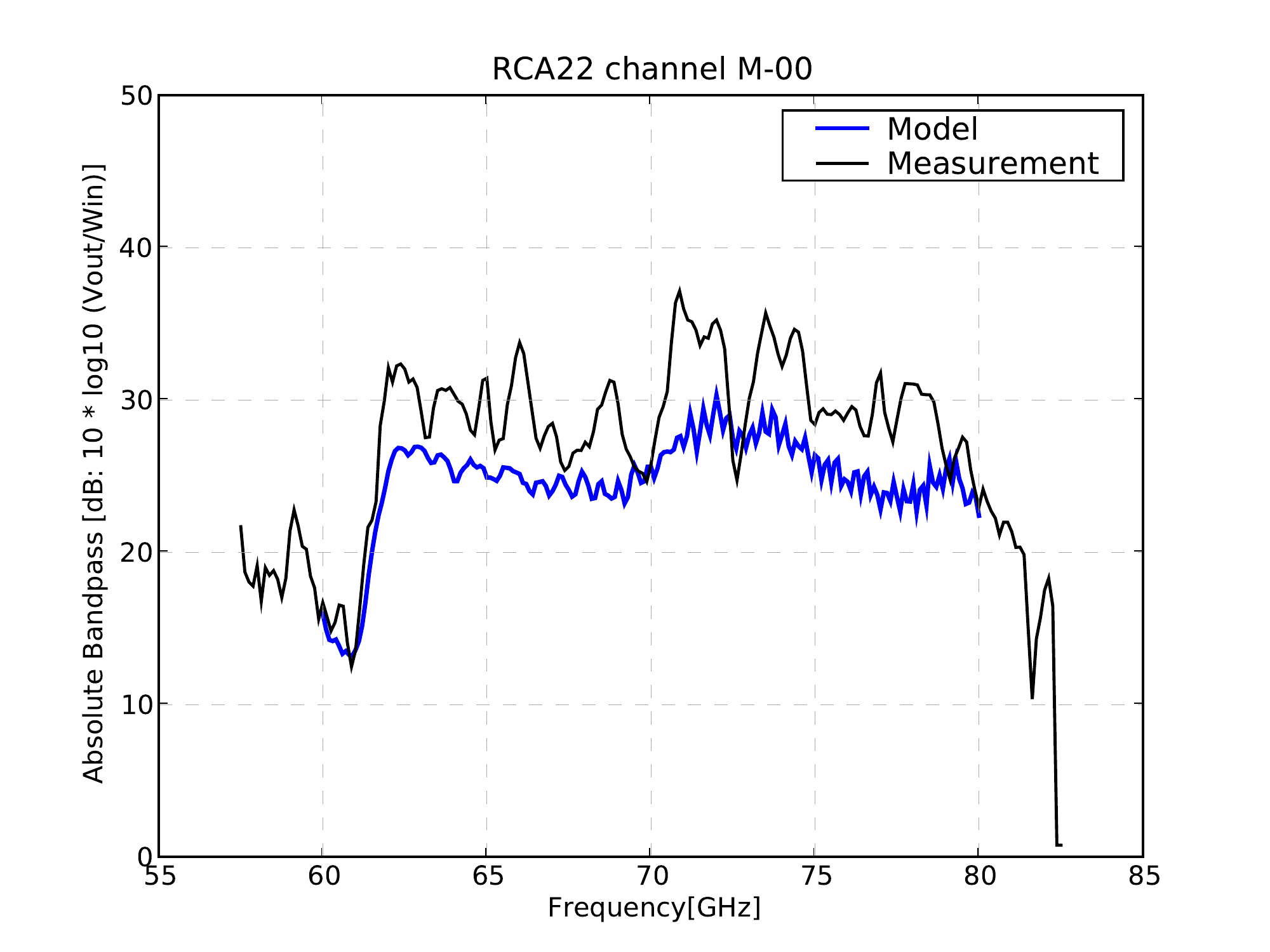}}
    \caption{70 GHz channels LFI21S-10 on top, LFI22M-00 on bottom, bandpasses measured and simulated, for 70 GHz channels the comparison is on absolute value and not only on normalised shapes}
    \label{fig:70GHz-band}
\end{figure}

Most of the channels show high consistency between model and measurements (see, e.g., top of figure~\ref{fig:70GHz-band}); few channels have an offset in logarithmic scale, which means they differ by a multiplicative factor, see middle of figure~\ref{fig:70GHz-band}. The cause of this effect is not clearly understood, but its impact on bandpasses modelling is low because the main objective of bandpass modelling and simulation is just to deliver relative bandpasses. Absolute level consistency is good for increasing confidence on radiometers modelling but it is not a required feature.

RCA 23 instead, see bottom of figure~\ref{fig:70GHz-band}, is a peculiar case, because all its 4 channels, in the centre of the band, show a model bandpass higher than measurements bandpass; this issue is under investigation, the favoured hypotheses is a problem on model input data.

\begin{figure}
    \centering
    \resizebox{\hsize}{!}{\includegraphics{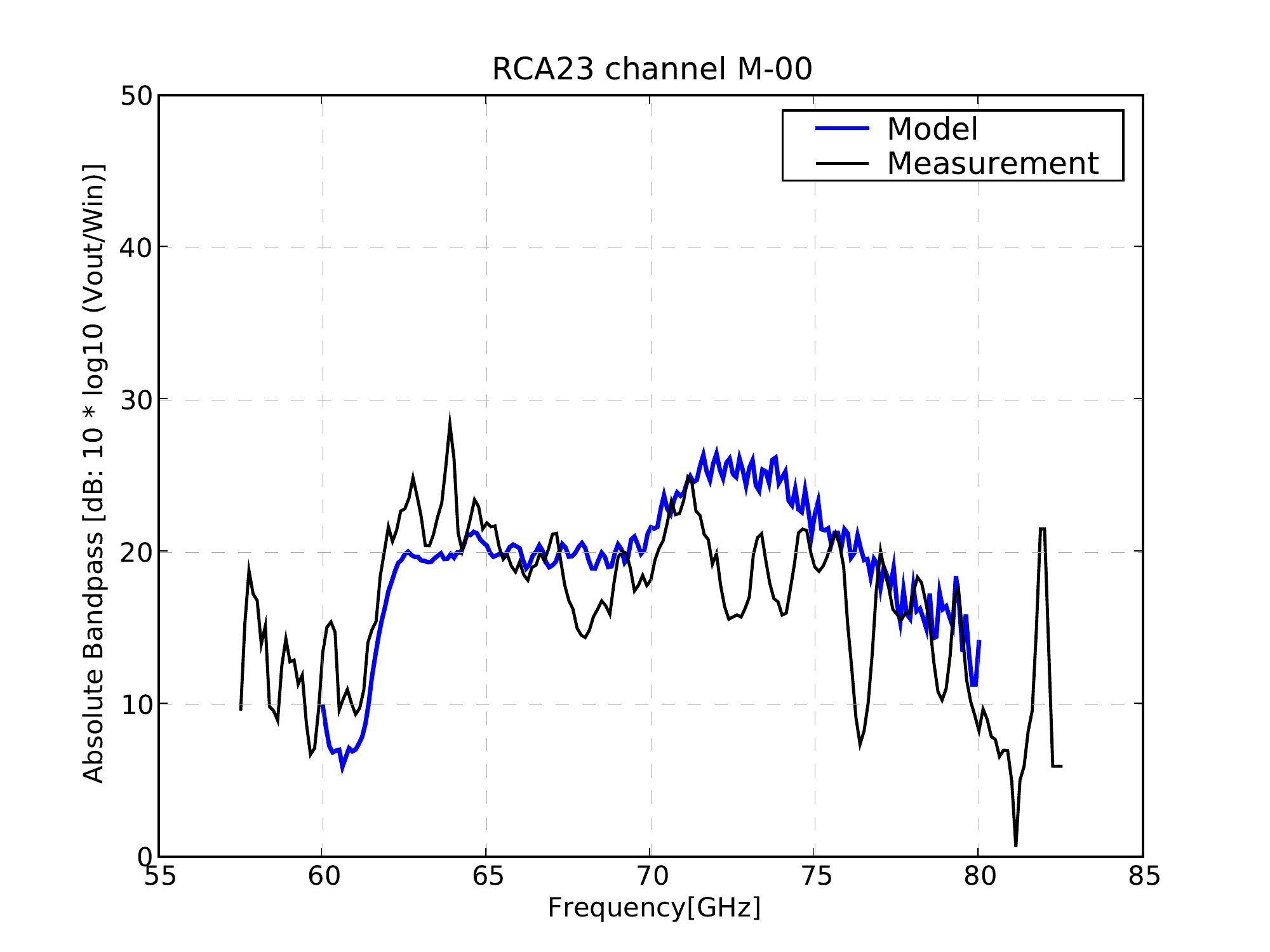}}
    \caption{70 GHz channel LFI23M-00, bandpasses measured and simulated, for 70 GHz channels the comparison is on absolute value and not only on normalised shapes}
    \label{fig:70GHz-band23}
\end{figure}
  \subsection{44 GHz Radiometers}

\begin{figure}
    \centering
    \includegraphics[width=\columnwidth]{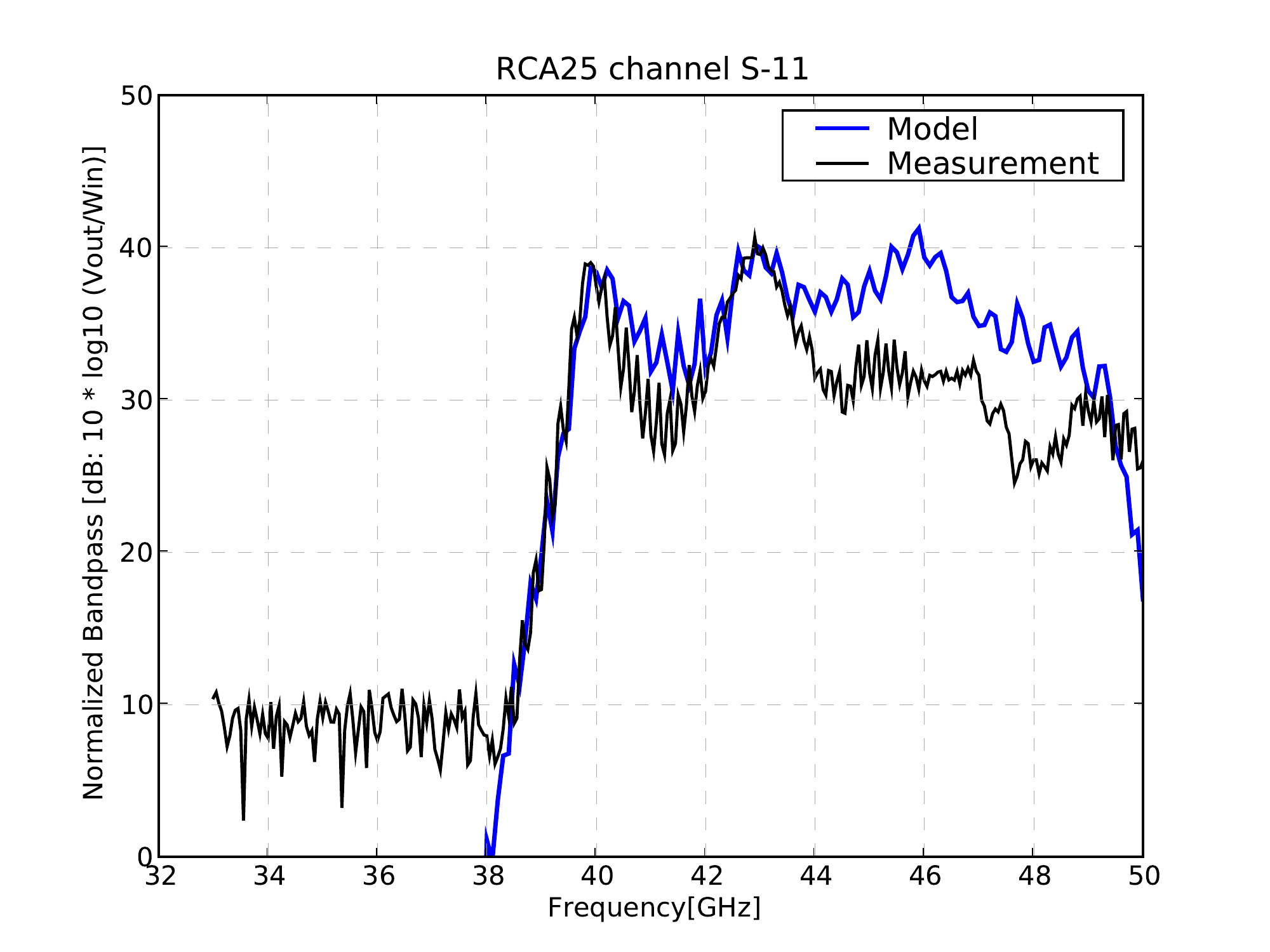}
    \caption{RCA 25 44 GHz channel S-11 bandpass measured and simulated: measurements were normalised in order to have the same integral as the model, a systematic discrepancy between 44 and 49 GHz is present in all 44 GHz channels and it is still not understood}
    \label{fig:rca25-band}
\end{figure}

Figure \ref{fig:rca25-band} shows an example of a 44 GHz channel; few notes about the comparison:
\begin{itemize}
    \item measurements show no systematic effects
    \item model and measurements have the same frequency coverage
    \item measurements were normalised to the same integral of simulations, using the same procedure applied on 30 GHz channels
    \item agreement is good apart from a systematic discrepancy in all 44 GHz channels from 44 to 49 GHz, where simulations have an higher response
\end{itemize}

The discrepancy is important and at the moment it is not clear what is the cause of this effect, among the hypotheses there are both model deficiencies:
\begin{itemize}
    \item systematic error in a subcomponent campaign test
    \item insufficient modelling of the FEM return loss on the back side (facing the waveguides)
\end{itemize}
or measurements issues:
\begin{itemize}
    \item sweeper input power loss
    \item error on the measurements data analysis process
\end{itemize}

The last possibility at present is the favourite because a preliminary reanalysis on the data acquired during swept source tests on just one channel gave a result much more similar to QIMP simulations that the first analysis. However, this is not true for all channels, therefore I plan to analyse all the data using a different software in order to investigate this issue further.

\subsection{30 GHz Spare radiometer}

In 2008 Stefania De Nardo, during her master thesis \cite{thesis_denardo}, performed new bandpass measurements in warm conditions on a 30 GHz spare radiometer. Spare units are equivalent to flight units and were built to be used for replacing flight units in case of damage.

These tests were performed on a much larger band, from 21.3 to 40 GHz, and were also performed both on the complete RCA and on different chain sections. In particular here is the list of tests performed, in which the signal was injected at different points:
    \begin{itemize}
        \item into the skyload, with 3 different input waveguide orientation
        \item into the OMT, connecting the injection waveguide directly to OMT main and side channels
        \item into the Waveguides
            \item into the BEM alone
    \end{itemize}
    
    In all cases the DC signal output from the BEM was sampled.
    
Tests with skyload injection suffer strong ripples probably due to the presence of standing waves in the skyload, tests with injection into the OMT, instead, are very clean.
The feed horn effect on bandpass is very low, because its losses in general are negligible, therefore tests with injection at OMT level can be regarded as representative of the RCA response.

\begin{figure}
    \centering
    \includegraphics[width=\textwidth]{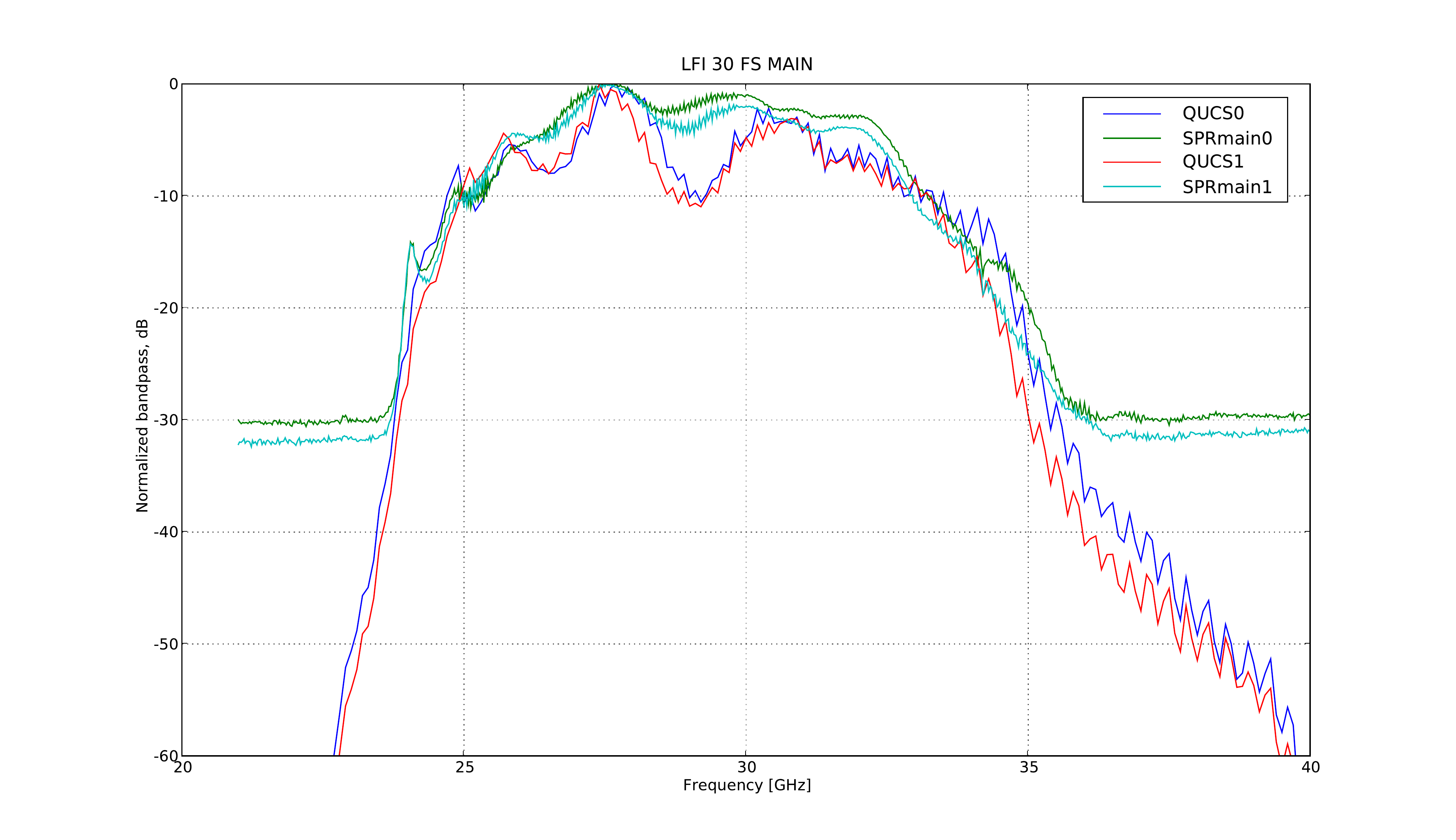}
    \caption{Comparison of measurements and QIMP simulations on 30 GHz spare MAIN channels}
    \label{fig:30fsmain}
\end{figure}

\begin{figure}
    \centering
    \includegraphics[width=\textwidth]{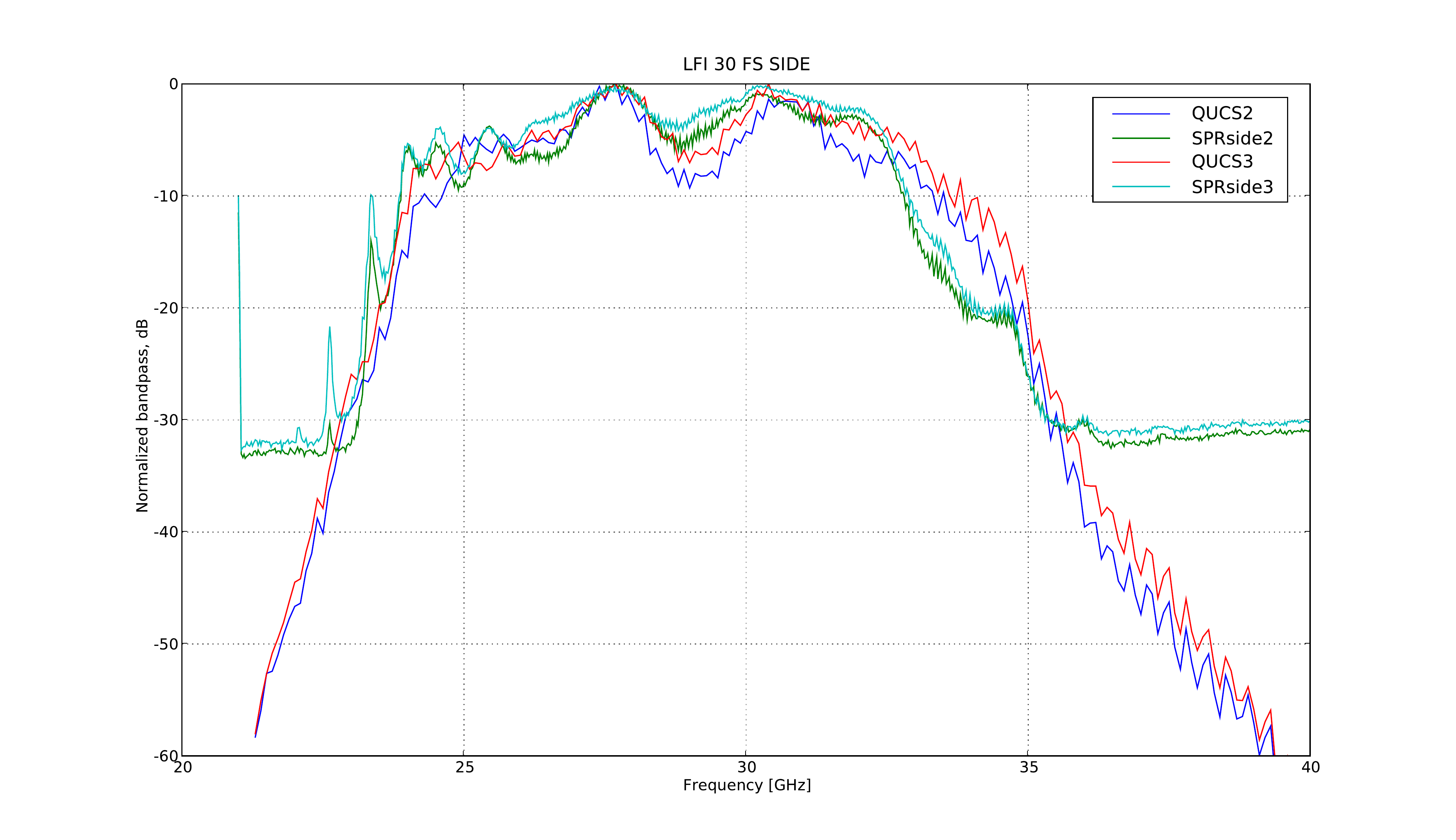}
    \caption{Comparison of measurements and QIMP simulations on 30 GHz spare SIDE channels}
    \label{fig:30fsside}
\end{figure}

Figures~\ref{fig:30fsmain} and \ref{fig:30fsside} show the comparisons between measurements and simulations for main and side channels:
    \begin{itemize}
        \item the agreement is very good, even below 25 GHz, where FEM and OMT data were extrapolated in the model, and this is an important validation of QIMP results
        \item in Main channels the discrepancy is mainly in the dips at the centre of the band; this could be due to the power reflected back by the FEM which is reflected by the test setup again into the FEM. Such an effect tends to flatten the band and in nominal configuration it is not present because the skyload absorbs the power reflected back by the FEM.
        \item in Side channels the effect of bandpass flattening is weaker, probably because it is already flatter than Main band, however there is a discrepancy between 33 and 35 GHz which is still under investigation.
    \end{itemize}

\section{Conclusion}
\label{sec:conclusion}
Swept Source Measurements suffer many systematic effects, mainly standing waves on 70 GHz (section \ref{sec:swept70}) and polarisation issues plus low frequency coverage on 30 GHz (section \ref{sec:swept3044}).
These effects are zero order effects and are so important that prevent using SPR as best bandpass estimator, and make the mathematical difference between measurements and simulations unmeaning for judging their likelihood.

On the other hand single components performance tests (which are less prone to systematic effects because of simpler setups) and QIMP simulations provide the best knowledge of LFI radiometers frequency response.

The QIMP model has been successfully validated against the bandpass measurements. Some discrepancies remain at 44 GHz, this case is still under investigation.

The best estimation of Planck LFI radiometers bandshapes up to now is represented by the QIMP simulations, obtained by exploiting single components performance tests. Simulations have still margin for improvement, thanks to the analysis of spare components tests. For the first time they included SPR tests at ambient temperature of sections of RCA in order to better understand how the microwave signal is transmitted by each component.

\subsection{Use of bandpasses into data analysis}

    \subsubsection{Temperature measurements}
    \label{sec:temperature}

CMB anisotropies signal is highly contaminated by foregrounds emission and data analysis strategies rely on their different frequency dependence in order to distinguish them from the CMB.

Figure \ref{fig:foregrounds} in chapter \ref{ch:cosmology} shows the different frequency dependence of dust, synchrotron, free-free emissions and CMB; thank to its wide frequency coverage, {\sc Planck} will be able to separate the different components by combining maps from different channels.

Both radiometers and bolometers, however, have a broadband response, therefore their output is a convolution of the input spectra with their own frequency response.\\
A fine correction can be applied at map level by an iterative process which separates the different components thanks to their spectral indexes, convolve them one by one with the instrument frequency response, and compares their sum with the detected map.

\subsubsection{Polarization measurements}
  \label{sec:polarization}
The new frontier of precision cosmology is CMB polarization measurements; several HFI channels at 100, 143, 217 and 353 GHz are sensitive to polarization thanks to Polarization Sensitive Bolometers, \cite{PSB}.

Each LFI Feed Horn, instead, is connected to an OrthoMode Transducer which splits the orthogonal components to be detected by two twin radiometers. Each radiometer is calibrated separately against the CMB dipole signal and polarization measurements are performed differencing the signals of the 2 radiometers connected to the same Horn.

The main issue is that foregrounds have different spectral indexes with respect to the CMB dipole which determined the calibration, therefore spurious polarization signal is generated when differencing channels with a different bandpass, producing leakage of foregrounds temperature signal to polarization.

Using the radiometers frequency response, it is possible to estimate the impact of bandpass mismatch on CMB maps and power spectra, this work was performed for 70 GHz radiometers by Torsti Poutanen in \cite{bandpass_torsti}.

\subsection{Delivered bandpasses}

QIMP bandpasses convey information both on relative bandshape and absolute gain; however, there is a more reliable technique for measuring the absolute gain, which consists on dedicated tests which measure the Volt output as a function of input power and then fit radiometer response with an suitable model, details in \cite{prelaunch_receivers}.

Therefore bandpasses need just to characterise a relative response, and they are normalised: in detail a weight $w_i$ is computed by integrating the bandpass $G(\nu)$ over 0.1 GHz, they are then normalised to have a sum of 1.

Absolute frequency response can then be computed by multiplying QIMP bandshapes by the absolute gain, called photometric constant.

\begin{figure}
    \centering
    \includegraphics[height=\textwidth,angle=270]{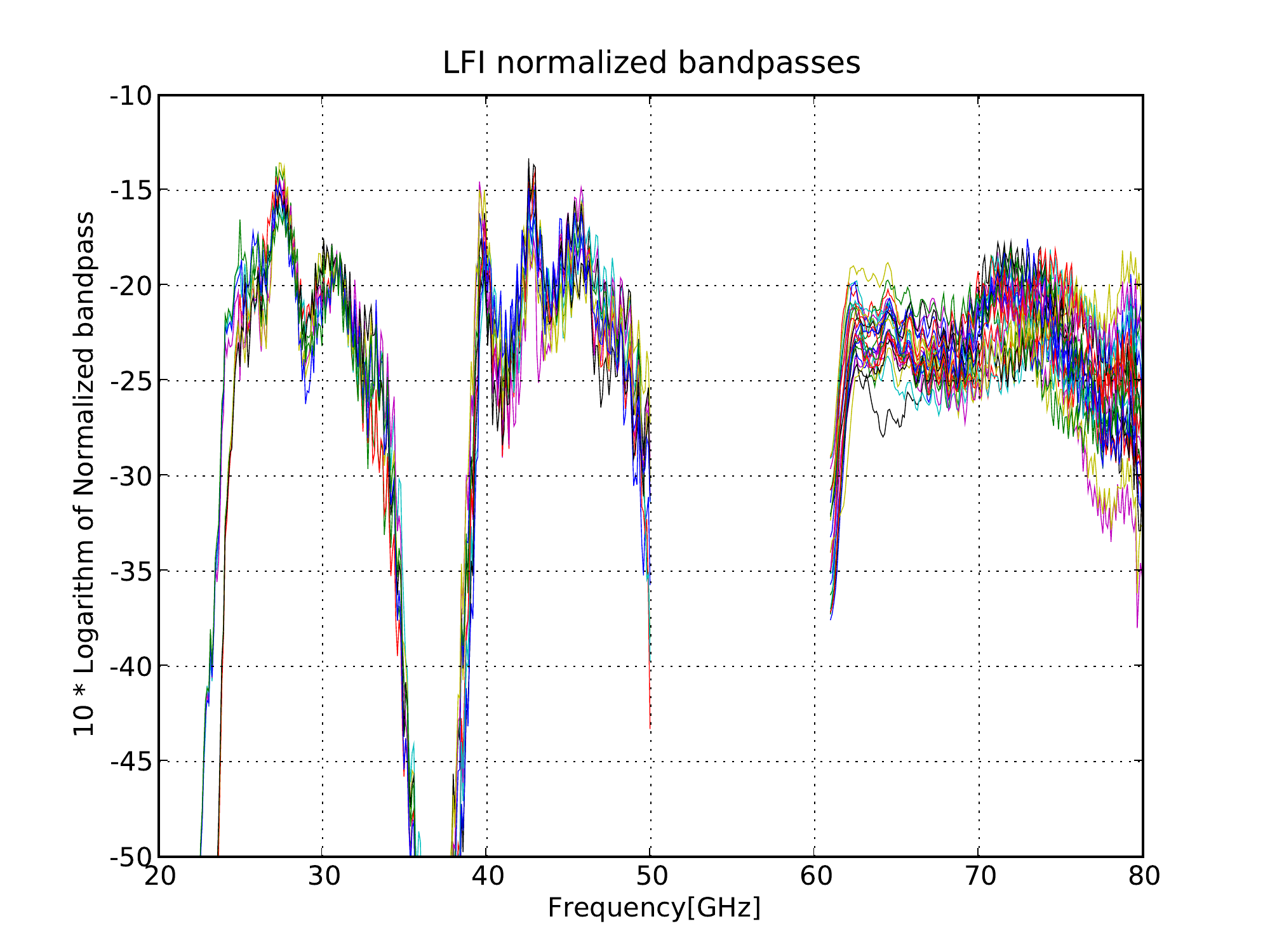}
    \caption{All the 44 LFI channels bandpasses released on September, the 18th, 2007, bandpasses are normalised to unit integral and plotted in logarithmic format computed as $10 \log w_i $ where each $w_i$ is the radiometer gain $G(\nu)$ integrated over 0.1 GHz, the separation in frequency, and normalised to sum equal to 1.}
    \label{fig:lfibands}
\end{figure}

Tabulated bandpass files are available at LFI Data Processing Centre in Trieste or from Andrea Zonca at request upon approval of the LFI Instrument Team. Figure \ref{fig:lfibands} shows all the 44 channels normalised bandpasses on the same plot in logarithmic format.

\newpage
\section{Comparison of all the channels bandpasses}
\label{sec:all_plots}

The following figures plot channel by channel all the comparisons between the measured and QIMP bandpasses.


 \begin{figure*}[h]
    \begin{center}
    \begin{tabular}{m{0.3cm} |c c c c} 
      & \textbf{M-00} & \textbf{M-01} & \textbf{S-10} & \textbf{S-11}\\

\input{compare_70.tex}

    \end{tabular} 
    \caption{70 GHz RCAs comparison between measurements(black) and model(blue) results}
    \label{fig:comp70}
    \end{center}
\end{figure*}


\chapter{Thermal aspects}
\label{ch:thermal}

\section{Introduction}
\label{sec:plancktherm}

Cryogenics and thermal issues were a strong driver in the design of the Planck Satellite, HFI bolometers need a temperature of 0.1 K to work, while LFI radiometers need a Front End temperature of $\sim 20K$ to achieve the required sensitivity.

\begin{figure}
    \centering
    \includegraphics[width=0.5\textwidth]{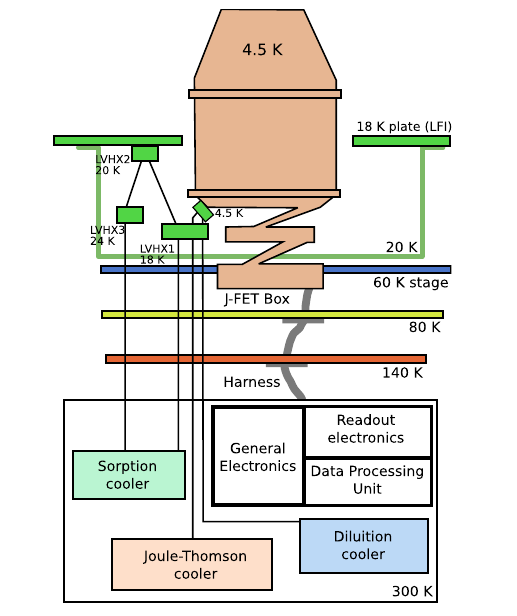}
    \caption{Schematic of Planck cooling chain}
    \label{fig:planckthermal}
\end{figure}

Planck cooling system is composed by, see figure~\ref{fig:planckthermal}:
    \begin{itemize}
        \item three passive radiators at 140, 80 and 60 K which decrease gradually the temperature from the 300 K of the Service Module, to the interface with the first cooling stage
            \item the Sorption cooler is the most important cooling system onboard Planck because it cools the complete focal plane, including both LFI and HFI, from 60 to about 20 K. Technically it is a vibrationless hydrogen cooler through which the hydrogen is pumped exploiting a chemical sorption process. It is composed by six compressors, each compressor is periodically cycled between heating and cooling phases, each phase lasts 667 seconds and a complete cycle $667 \text{s} \times 6 = 4002 \text{s}$.
            It is capable of providing $\sim 1 W$ of cooling power at 20 K to cool the LFI radiometers and as precooling stage of the 4K.
            \item the Joule-Thomson cooler is a mechanical helium compressor, it cools the HFI external shell to 4K. The 4K shell is very important because the it includes both the HFI focal plane, where its horns are located, and the LFI reference loads. In order to improve the stability of the 4K stage top plate, where horns are located, a PID controller commands a heater. LFI 70 GHz reference loads are very near to the top plate and they benefit from the PID action, 30 and 44 GHz are located further and their stability is lower.
            \item the Diluition cooler is used by HFI to keep the bolometric detectors at 0.1 K
    \end{itemize}

\section{4K box thermal transfer functions}
\label{sec:4kboxtf}

\subsection{Thermal transfer functions}

Before discussing the work performed on data analysis of the measurements, I'd like to introduce the concept of thermal transfer functions.

If a solid body has a boundary imposed temperature which is sinusoidally fluctuating:
    \begin{equation}
        T(\mathbf{x_0},t) = T_0 + T_s sin (2\pi \nu t)
    \end{equation}
where $T_0$ is the background temperature, $T_s$ the amplitude of the sinusoidal fluctuation and $\nu$ its frequency, the generic solution at any other point of the solid body is still a sinusoid:
       \begin{equation}
            T(\mathbf{x},t) = T_0 + T_s \gamma(\mathbf{x},\nu) sin (2\pi \nu t + \phi(\mathbf{x},\nu)
    \end{equation} 

where $\gamma(\mathbf{x},\nu)$ is the amplitude transfer function and  $\phi(\mathbf{x},\nu)$ is the phase transfer function.
This derives from the fact that the equation of heat propagation is linear.

The knowledge of the transfer functions for each frequency $\nu$ allows to compute the fluctuations at a point $\mathbf{x}$ due to any boundary condition by working in the Fourier domain: any smooth source fluctuation can expanded as a series of sinusoids, each sinusoid is transformed to the output by adding the phase and multiplying by the amplitude transfer function and the fluctuation at the new point is reconstructed by merging again the transformed sinudoids.

\subsection{Introduction}
This section presents the analysis performed on the data acquired during the thermal transfer functions measurements, performed at the Liege Space Centre in cryogenic chamber on the Planck Cryogenic Model in 2005. The Planck Cryogenic Model consisted in:

     \begin{itemize}
         \item the qualification model of the satellite, built for qualification and testing purposes
         \item HFI qualification model (CQM), the complete HFI instrument built for testing and qualification, just some redundancies were missing with respect to the flight model
         \item LFI instrument was not involved in the test, it was replaced by a dummy structure needed for simulating its passive thermal behaviour
     \end{itemize}

During the test periodic thermal fluctuations with periods of 100, 666 and 4000 seconds were forced at LVHX1 and the resulting effects measured with the HFI thermometers on lower part of the 4K box (Cernox), on the side, where LFI 30 and 44 GHz reference loads are located (Ther4KL1,2) and on the top 4K bolometers plate (PID4C,R,N).

The tests for 100 and 666 seconds were performed both with PIDs controllers on and off while the 4000 seconds only in PID on condition, due to the more important drift effect.
In order to estimate the thermal transfer functions, an analysis has been performed in the time domain by computing the ratio between the fluctuation amplitudes at the input and at the output.

This method is straightforward but it is not very reliable when the input and output signals are not perfect sine waves, therefore it was applied, for validation purposes, only in analysing data of the 666 PID off and 100 PID off in order to compare its results with an analysis in the frequency domain.

\subsubsection{Analysis using the Fourier Transform}
The thermal transfer function versus input frequency was evaluated by applying the Fast Fourier Transform both to the input and to the output time streams, after having removed the transient.
The results were divided by their own sampling frequency in order to compare different type of sensors (for example LVHX1 has 1 Hz sampling while PID4* thermometers have 160 Hz).

Finally the amplitude transfer function was computed by taking the ratio between the peaks of the output on the input FFTs, at the stimulus frequency; while the phase was computed making the difference between the phase of the output and the input peaks.

\subsection{Test setup}
\subsubsection{Description}
The HFI CQM was integrated on the Planck CQM satellite. The Sorption cooler Pipe Assembly Cold End (PACE), the 4K cooler CQM and the dilution cooler CQM were also integrated on the satellite and connected to the instrument.

The whole satellite was integrated in a vacuum chamber cooled at 20 Kelvins.

The PACE was fed by Hydrogen bottles instead of compressor. 

The thermal transfer function measurement principle for the HFI at CSL was to generate various thermal stimuli at the 18K cold ends of the PACE by modulating the hydrogen flow in order to
quantify the impact of in-flight Sorption Cooler fluctuations on
HFI stages stability. 

Various pressure changes sequences in the
PACE allowed to generate different periodic fluctuations of LVHX1
and LVHX2 with various amplitudes. From the result of the various
PACE fluctuation tests it is then possible to deduce the thermal
transfer function between LVHX1 and the other HFI stages.

The complete test report is available from \cite{CSLreport}.
 
\subsubsection{Thermometers location and performance}

\begin{figure}[ht!]
\begin{center}
\includegraphics[width=12cm]{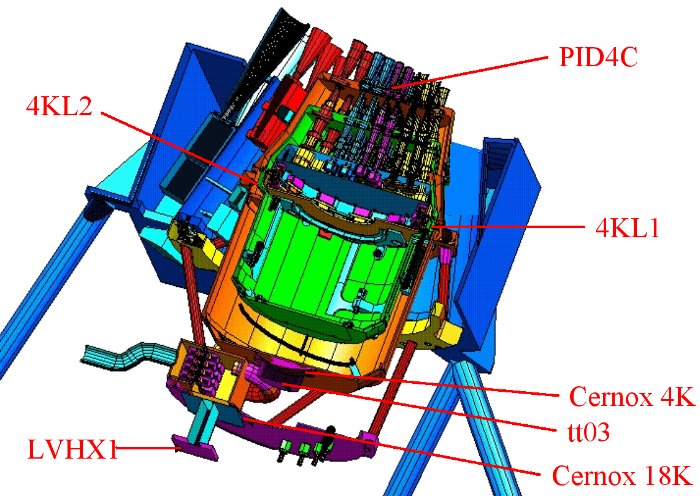}
\caption{CSL-CQM thermometers location} \label{CSL_CQMthermos}
\begin{tabular}{|c|l|c|c|}
\hline \textbf{Sensor}&\textbf{Location}&\textbf{Sampling[\small{$Hz$]}}\\
\hline PID4N& 4K stage horns&  160\\
\hline 4KL2& Side of the 4K box&  160\\
\hline 4KL1& Side of the 4K box&  160 \\
\hline Cernox\_4K& Back of the 4K box&  1/70\\
\hline tt03& 4K Cooler cold end&  160\\
\hline Di842& LFI support&  160\\
\hline Cernox\_18K& HFI 18K plate&  1/70\\
\hline tt04& 4K Cooler pre-cooling temperature&  160\\
\hline LVHX2& Sorption Cooler cold end on LFI&  1\\
\hline LVHX1& Sorption Cooler cold end on HFI&  1\\
\hline
\end{tabular}
\end{center}
\end{figure}

Figure~\ref{CSL_CQMthermos} shows the location of the thermometers on the HFI 4K shell and includes a table with the sampling frequency of all the thermometers. Thermometers are divided in three groups:
    \begin{itemize}
        \item high precision and high frequency thermometers share the same acquisition electronics of the bolometers, due to their key role for the processing of HFI data, i.e. bolometers are highly sensitive to any temperature signal, not only to microwaves, therefore any drift needs to be removed with high accuracy
        \item low precision thermometers (Cernox) which serves just for housekeeping
        \item Sorption cooler sensors, used to monitor the performance of the Sorption cooler
    \end{itemize}

For the rest of the chapter I will be often referring to the 4KL1 and 4KL2 sensors as LFI reference load sensor; it is understood that I'm referring to the 30 and 44 GHz reference loads, because the 70 GHz loads are located near to the top of the 4K box.

\pagebreak

\subsection{TIME domain analysis}

The amplitude transfer function was computed in the time domain by taking the ratio of the peak-to-peak amplitude of the input and output fluctuations; this method has strong contra:
    \begin{itemize}
        \item the transfer functions are defined on sinusoids, taking the ratio of input square waves amplitudes versus output triangular waves is a strong simplification
        \item a slow thermal drift or high frequency noise can worsen the estimation
    \end{itemize}

However, this method has the advantage of being very simple and fully under control, i.e. it is possible to check its results by eye, therefore is a good method for having a rough but robust estimate of the amplitude transfer functions.

\subsubsection{666 seconds sinusoidal input, PID Off}

\begin{figure}[htbp]
\begin{minipage}{0.5\linewidth}
\centering
\includegraphics[width=\textwidth]{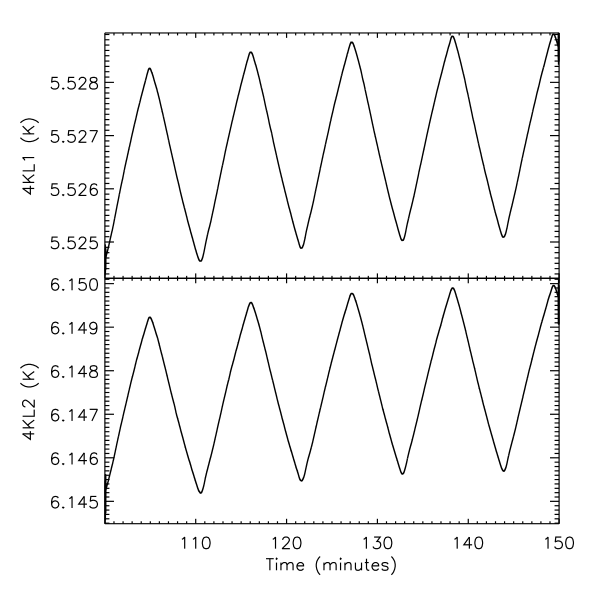}
\caption{4K LFI reference loads fluctuations}
\label{4ka}
\end{minipage}
\begin{minipage}{0.5\linewidth}
\centering
\includegraphics[width=\textwidth]{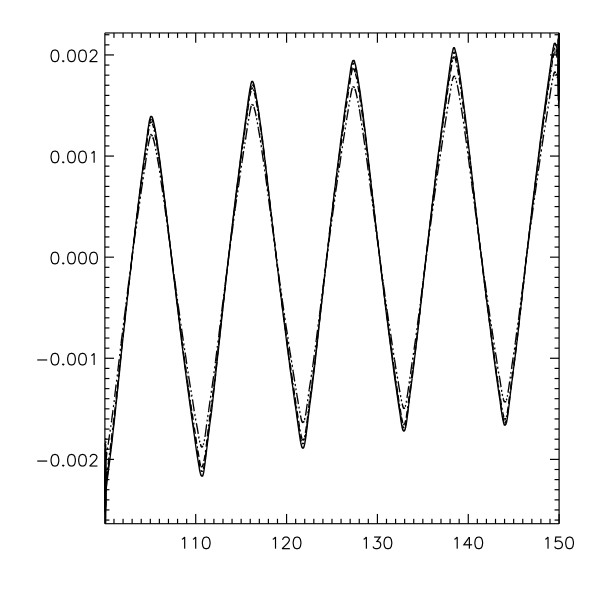}
\caption{4K plate fluctuations (mean removed)}
\label{4kb}
\end{minipage}
\end{figure}

\begin{figure}[htbp]
\centering
\includegraphics[width=\textwidth]{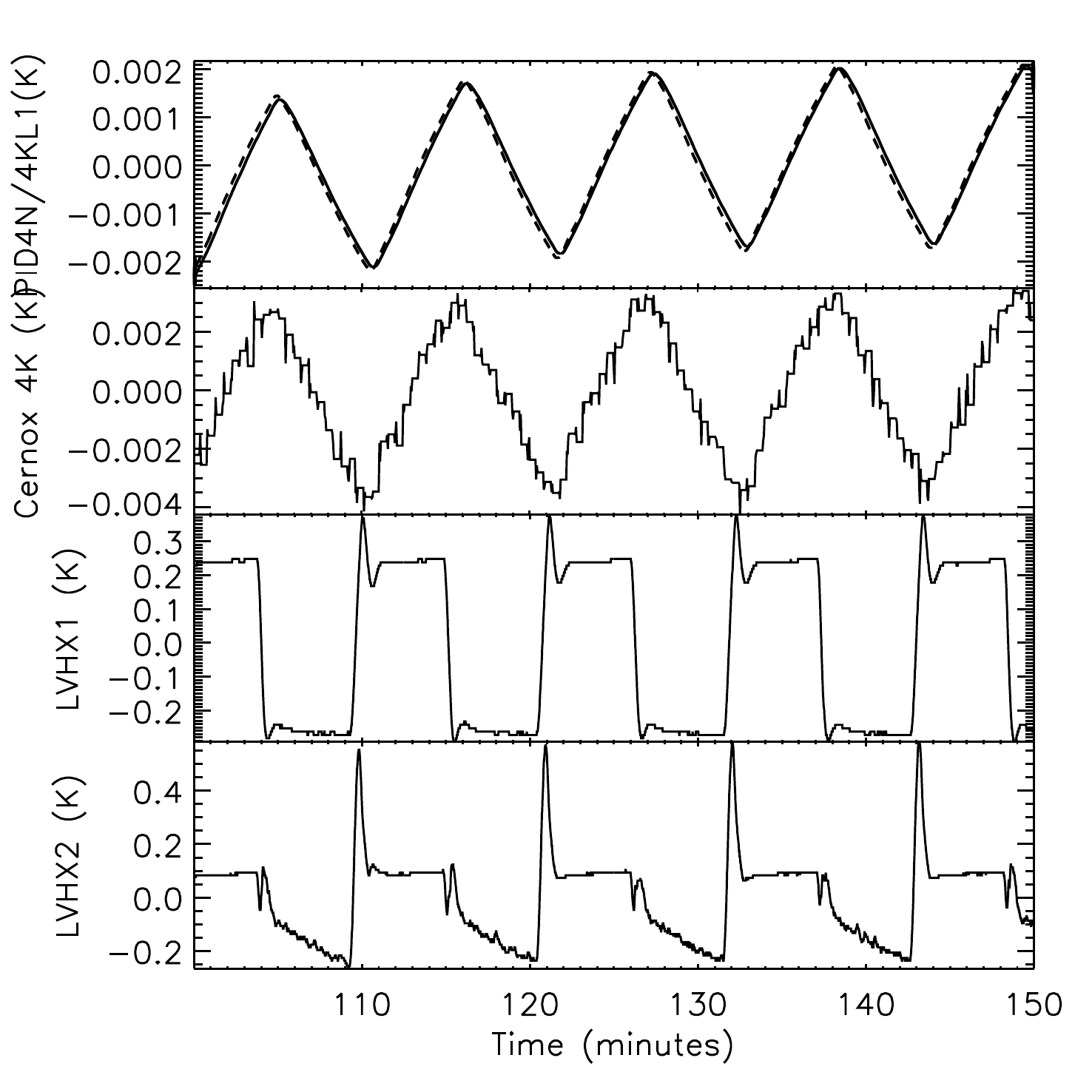}
\caption{666 sec, PID OFF: comparison between LVHX, Cernox, 4K plate and 4K load}
\label{4kc}
\end{figure}

This test was performed with an input square wave at LVHX1 of an amplitude of $\sim 0.5 K$ and a frequency of 1.5 mHz.
Figures~\ref{4ka} and \ref{4kb} show the resulting fluctuations measured by the high precision thermometers; figure~\ref{4kc} shows the comparison between the square input fluctuation at LVHX1 and LVHX2 and the outputs at the top, middle and bottom of the 4K box.

Table~\ref{FDT_666sec} lists the measured outputs and the computed amplitude transfer functions.
\begin{table}[!h]
\begin{center}
\begin{tabular}{|c|l|l|}
\hline \textbf{Sensor}&\textbf{Fluctuation amplitude (K)}&\textbf{Transfer function from LVHX1(K/K)}\\
\hline PID4N&     0.00383&    0.00719\\
\hline PID4R&     0.00370&    0.00695\\
\hline PID4C&     0.00339&    0.00637\\
\hline 4KL1&     0.00393&    0.00739\\
\hline 4KL2&     0.00438&    0.00823\\
\hline LVHX2&       0.356&      0.669\\
\hline LVHX1&       0.532&       1\\
\hline
\end{tabular}
\end{center}
\caption{Transfer function gains deduced from PACE 666sec
fluctuations measurement}\label{FDT_666sec}
\end{table}
\pagebreak

\subsubsection{100 seconds sinusoidal input, PID Off}

\begin{figure}[htbp]
\begin{minipage}{0.5\linewidth}
\centering
\includegraphics[width=\textwidth]{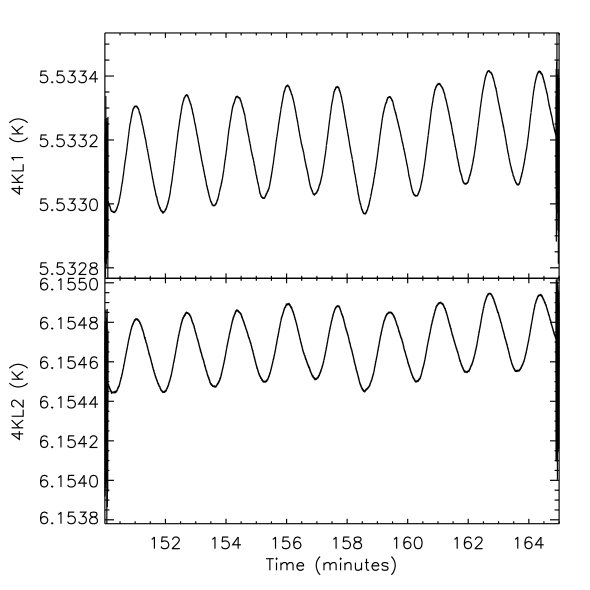}
\caption{4K LFI reference loads fluctuations}
\label{4ka1}
\end{minipage}
\begin{minipage}{0.5\linewidth}
\centering
\includegraphics[width=\textwidth]{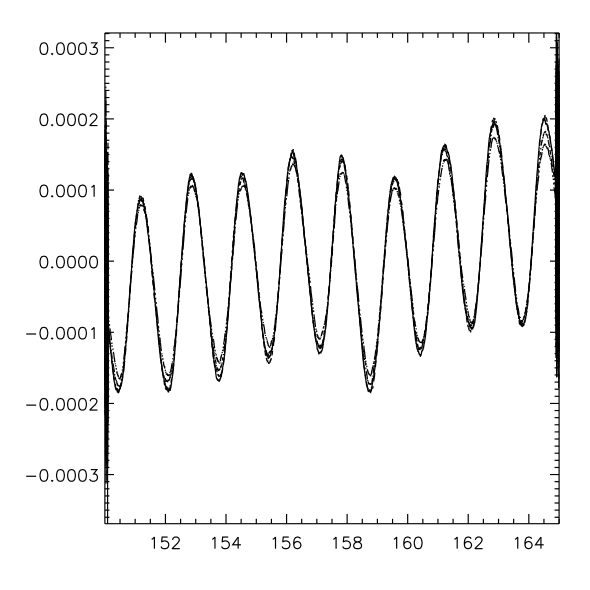}
\caption{4K plate fluctuations (mean removed)}
\label{4kb1}

\end{minipage}
\end{figure}

\begin{figure}[htbp]
\centering
\includegraphics[width=\textwidth]{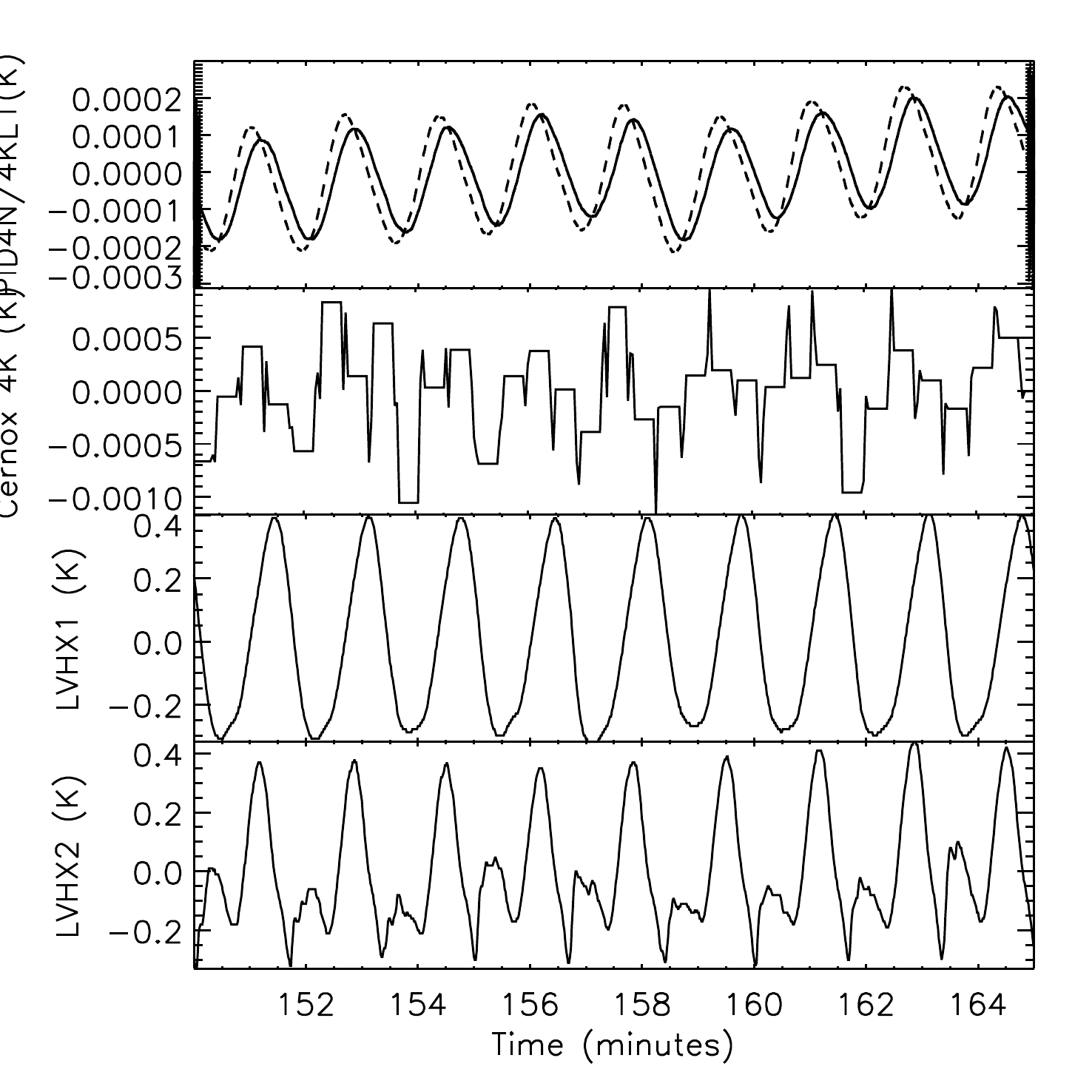}
\caption{100 sec,PID OFF: comparison between LVHX, Cernox, 4K plate and 4K load}
\label{4kc1}
\end{figure}

This test was performed with an input square wave at LVHX1 of an amplitude of $\sim 0.5 K$ at a frequency of 10 mHz.
Figures~\ref{4ka1} and \ref{4kb1} show the resulting fluctuations measured by the high precision thermometers; figure~\ref{4kc1} shows the comparison between the square input fluctuation at LVHX1 and LVHX2 and the outputs at the top, middle and bottom of the 4K box.

\begin{table}[!h]
\begin{center}
\begin{tabular}{|c|l|l|}
\hline \textbf{Sensor}&\textbf{Fluctuation amplitude (K)}&\textbf{Transfer function from LVHX1(K/K)}\\
\hline PID4N&    0.000269&   0.000384\\
\hline PID4R&    0.000294&   0.000420\\
\hline PID4C&    0.000241&   0.000345\\
\hline 4KL1&    0.000336&   0.000480\\
\hline 4KL2&    0.000376&   0.000537\\
\hline LVHX2&       0.690&      0.986\\
\hline LVHX1&       0.700&       1\\
\hline
\end{tabular}
\end{center}
\caption{Transfer function gains deduced from PACE 100sec
fluctuations measurement}\label{FDT_100sec}
\end{table}
\clearpage

\subsection{FREQUENCY domain analysis}

Here we discuss the method to extract the thermal transfer functions based on frequency domain analysis. The procedure is described step by step taking as example the 100 seconds PID off test.

\subsubsection{100 sec PID off}

\begin{figure}[htbp]
\centering
\includegraphics[width=\textwidth]{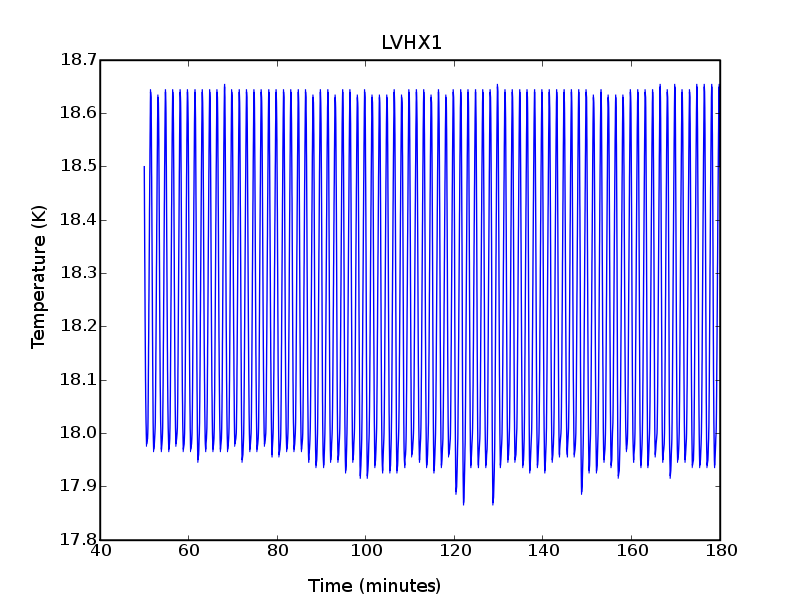}
\caption{LVHX1 100 sec fluctuations}
\label{100lvhx1}
\end{figure}

\paragraph{Selection of the stable part of the timestream}
The first step is to remove transients and drifts which could impact to the FFT. In this case the timestream is very clean and the drift on the thermometers is only of few mK, see figure~\ref{100lvhx1}

In each case a trade-off was performed between removing the drift and having enough samples for a sufficient frequency resolution

\begin{figure}[htbp]
\begin{minipage}{0.5\linewidth}
\centering
\includegraphics[width=\textwidth]{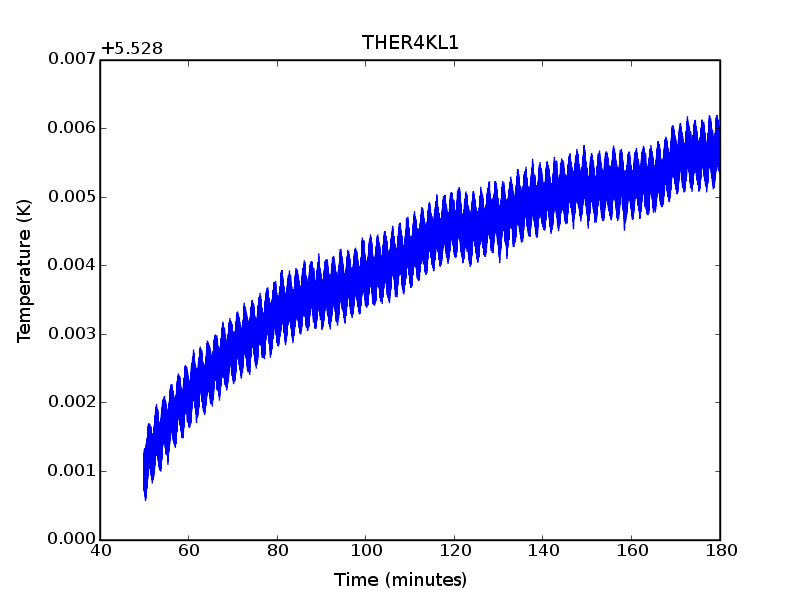}
\caption{4K LFI reference loads fluctuations}
\end{minipage}
\begin{minipage}{0.5\linewidth}
\centering
\includegraphics[width=\textwidth]{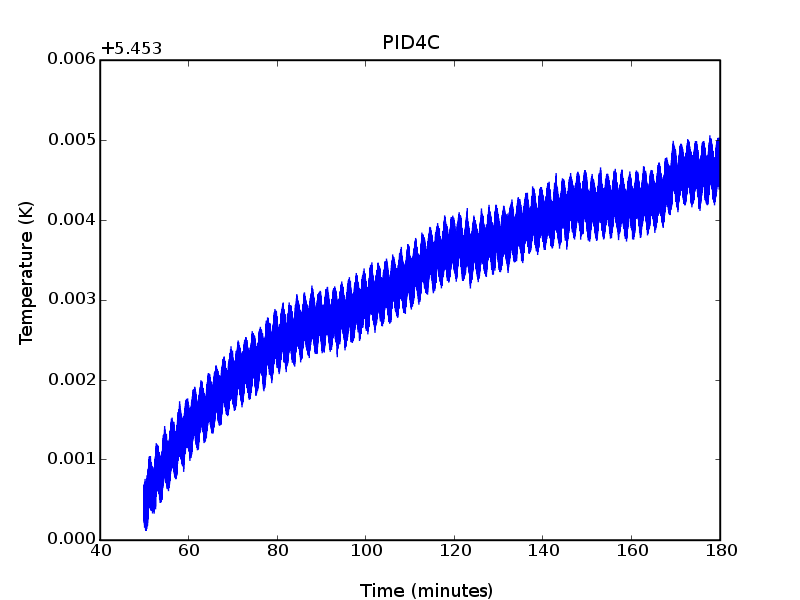}
\caption{4K plate fluctuations}
\end{minipage}
\end{figure}

\paragraph{FFT}
The Fast Fourier Transform module used is the FFTPACK included in Numpy 1.0 \footnote{Numpy is a multi dimensinal array and mathematical package for Python, see \url{http://numpy.scipy.org}}, the output is then divided by the sampling frequency, no other normalisation is performed because the objective is just to compute ratios.

\begin{figure}[htbp]
\centering
\includegraphics[width=\textwidth]{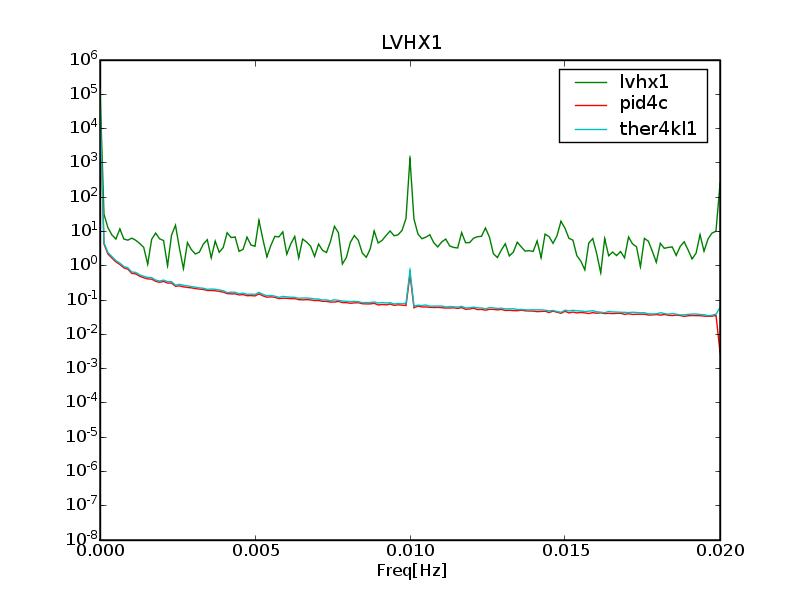}
\caption{FFT of the input LVHX1 signal and the thermometer measurement on the top of the 4K box, PID4C and in the middle 4KL1. The FFT is the raw output of the Numpy FFTPACK, no normalisation was performed}
\label{fft}
\end{figure}

In the 100 seconds case the input fluctuations are sine waves while for the 666 and 4000 case they are square waves, therefore the frequency peak is not always as clear as in this example, see figure~\ref{fft},but it is always defined.

\paragraph{Computation of the transfer function}
The final step is to identify the peak in the frequency domain around the input signal frequency and take the ratio between the peaks.

It was also taken into account to consider not just the peak point but also the surrounding points, by making an integral over 3 or 5 points and taking the ratio of this integral; however, even the nearest points are at least one order of magnitude lower, and already at the level of the mean of the signal, therefore they are not significant for the measurement; for this reason it was judged more reliable to consider just the peak point for taking the ratio.

\begin{figure}[htbp]
\centering
\includegraphics[width=\textwidth]{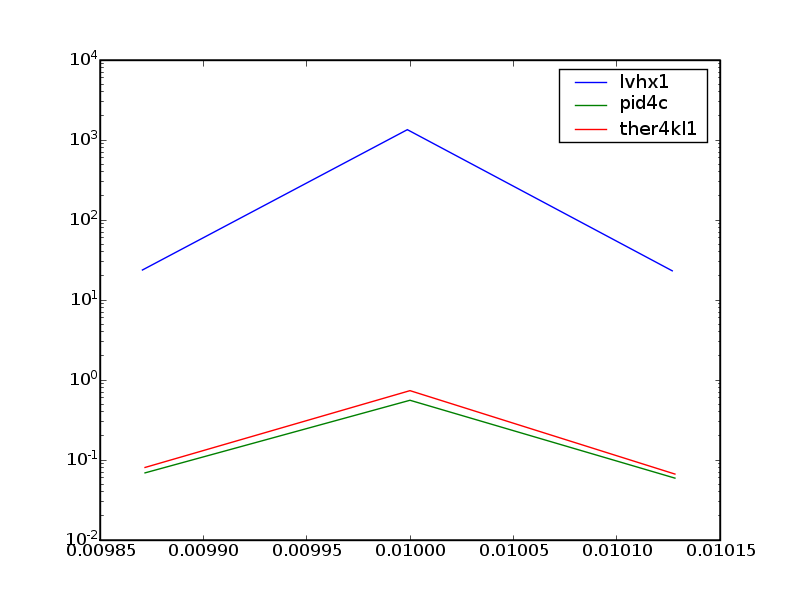}
\caption{FFT of the 3 signals - peaks on a logarithmic scale}
\label{peaklog}
\end{figure}

Figure~\ref{peaklog} shows the peaks in a logarithmic scale.

The results, compared to the frequency domain analyses, are displayed in table \\

\begin{table}[!h]
\begin{center}
\begin{tabular}{|c|l|l|l|}
\hline \textbf{Sensor}&\textbf{FFT} &\textbf{Time domain} & \textbf{Difference (\%)}\\
\hline PID4C&   0.000414& 0.000345 & 17\%\\
\hline 4KL1&  0.000547 &  0.000480 & 12 \%\\
\hline
\end{tabular}
\end{center}
\caption{Comparison between time and frequency domain results for 100 sec PID OFF, the percentage difference was computed as (FFT-Time)/FFT}

\end{table}

%


\clearpage
\subsubsection{Results}
The next table and the figures gather all the results obtained in the Fourier analysis:
\begin{table}[!h]
\begin{center}
\begin{tabular}{|c|l|l|}
\hline

\textbf{Sensor}&\textbf{100s PID off} &\textbf{666s PID off}\\
\hline

THER4KL1  & 0.0005466 & 0.005048 \\
THER4KL2  & 0.0006102 & 0.00562 \\
PID4C  & 0.0004145 & 0.004423 \\
PID4N  & 0.0003987 & 0.005011 \\
PID4R  & 0.0004556 & 0.004853 \\
THER4KH1  & 0.0004585 & 0.00489 \\
THER4KH2  & 0.0004788 & 0.005106 \\
\hline
\end{tabular}
\end{center}
\caption{Amplitude transfer functions, PID off}
\end{table}


\begin{table}[!h]
\begin{center}
\begin{tabular}{|c|l|l|l|}
\hline \textbf{Sensor}&\textbf{100s PID on}&\textbf{666s PID on}&\textbf{4000 PID on}\\

THER4KL1  & 0.0002789 & 0.001344 & 0.003143 \\
THER4KL2  & 0.0003105 & 0.001484 & 0.003451 \\
PID4C  & 4.898e-05 & 1.512e-05 & 8.703e-06 \\
PID4N  & 5.197e-05 & 1.371e-05 & 5.764e-06 \\
PID4R  & 5.499e-05 & 3.204e-05 & 6.475e-05 \\
THER4KH1  & 5.3e-05 & 1.225e-05 & 8.208e-06 \\
THER4KH2  & 5.609e-05 & 1.843e-05 & 2.583e-05 \\
\hline

\end{tabular}
\end{center}
\caption{Amplitude transfer functions, PID on}
\end{table}


\pagebreak
%
%

\subsection{4K timelines simulations}

The experimental results just showed were obtained with a set of PID settings still preliminary; the HFI team performed an optimisation of the PID settings in order to reduce their power consumption while still meeting the stability requirements.
The optimisation was performed the HFI thermal software model.

The new PID settings were then used as input to their thermal model in order to produce a long duration simulation (several months) of the expected fluctuations on the 4K stage. These simulations used as input a realistic datastream of temperatures at LVHX1 simulated by the Sorption cooler team.

I have used the same Fourier method for analysing this dataset; however, in this case I could compute the transfer function for all the range of frequencies permitted by the signals sampling frequencies. In fact in measurements it is necessary to give a strong source signal in order to disentangle effectively what component of a sensor measurement is due to the input source and what part instead has a different cause. Simulations instead are simpler and there is no doubt that all the signal we see on each 4K sensor is actually the effect of the Sorption cooler fluctuations.

\begin{figure}
    \centering
    \includegraphics[width=\textwidth]{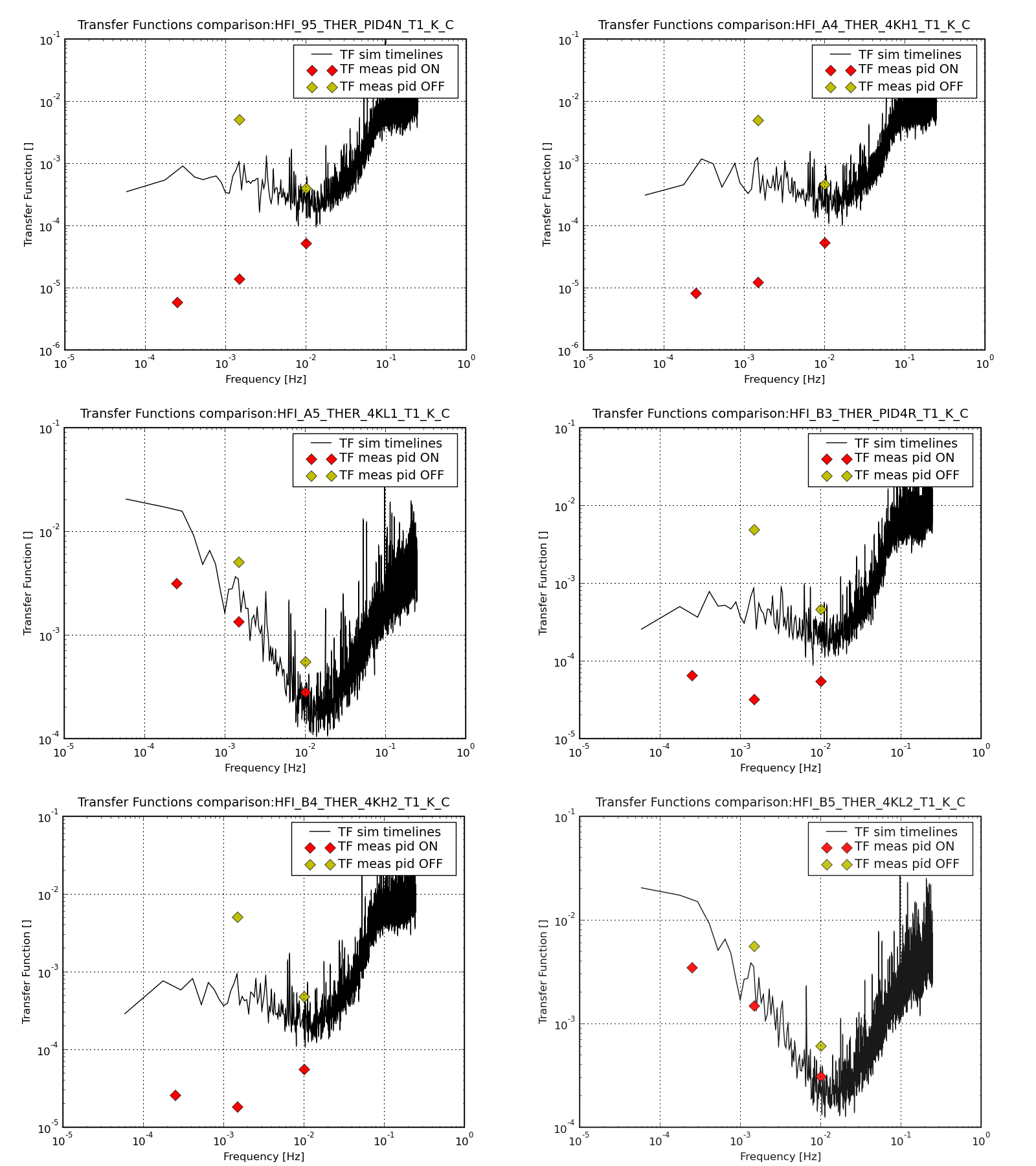}
    \caption{Comparison of the transfer functions computed from the simulations (black curve) and measurements at CQM with PID ON and PID OFF}
    \label{fig:alltf}
\end{figure}

Figure~\ref{fig:alltf} shows the results of the comparison between the transfer functions computed from the simulations and the transfer functions from measurements. As expected the simulated transfer functions lie between the PID OFF and the PID ON measurements, which is the effect of the reduction of PIDs power consumption.

\section{Conclusion}

In this chapter I have presented the estimation of the amplitude thermal transfer functions of the HFI 4K box with respect to the Sorption cooler cold end LVHX1, with and without PID regulation. They have been estimated from the qualification tests held in CSL in 2005 using an analysis method in the Fourier domain which proved to be consistent with a simpler time domain method at the level of 17\%.

However, the expected behaviour in flight is different, because the setting of the PID thermal regulators has been optimised for reducing power consumption. The expected performance was simulated by the HFI team which provided a long duration simulation of the temperature measurements of all 4K sensors considering a realistic Sorption cooler input. I have then computed the expected transfer functions from the simulations: their comparison with the measurements is consistent; as expected, the expected transfer functions give a low pass filtering effect weaker than PID ON measurements but better then PID OFF measurements.

The next step is the analysis of the measurements acquired during the flight model cryogenic testing campaign held in CSL in summer 2007. During this campaign there were not dedicated tests to characterise transfer functions; however, computing the amplitude thermal transfer functions during a long stable acquisition test we can check whether the prediction of the simulations were accurate.


\pagebreak
\section{Annex}

The following figures~\ref{fig:comp1} and \ref{fig:comp2} show the complete timelines of the PID OFF tests at 100 and 666 seconds.

\begin{figure}[h]
    \centering
\includegraphics[width=12cm]{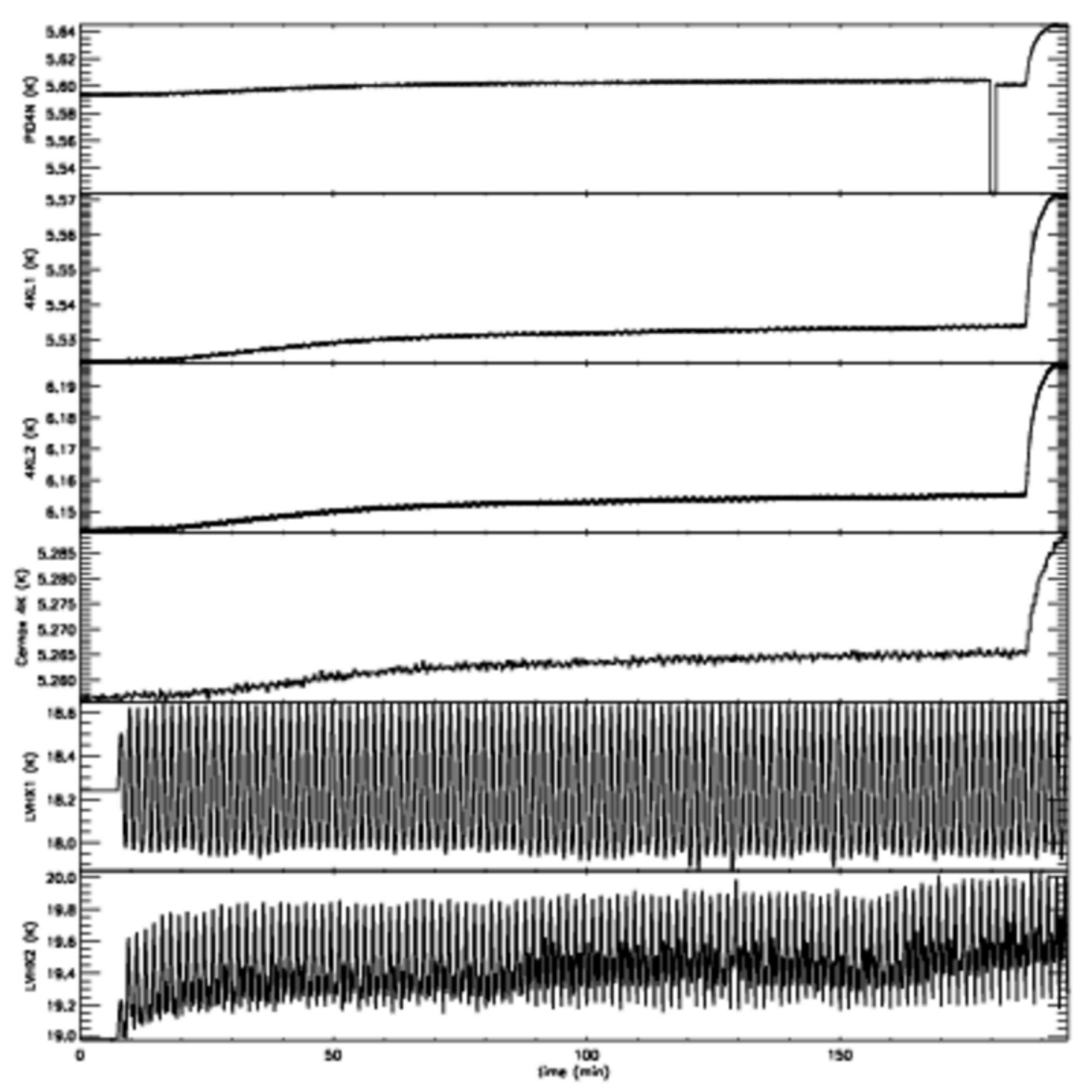}
    \caption{\textbf{100 seconds PID OFF} test, complete timelines}
    \label{fig:comp2}
\end{figure}

\begin{figure}[h]
    \centering
\includegraphics[width=12cm]{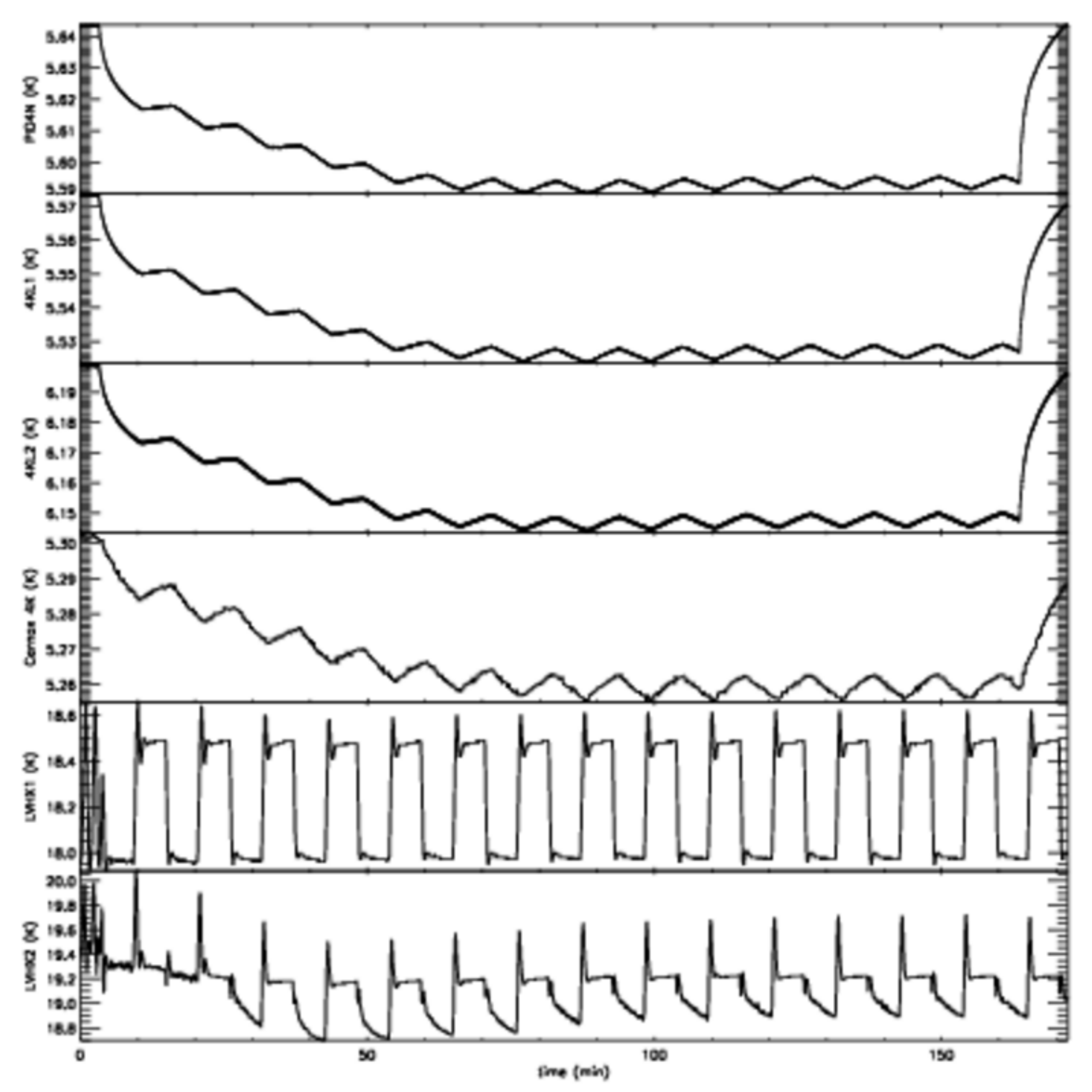}
    \caption{\textbf{666 seconds PID OFF} test, complete timelines}
    \label{fig:comp1}
\end{figure}

\clearpage

\chapter{Combined quick look analysis}
\label{ch:combined_qla}

\section{Introduction}

Quick look analysis represents the first check of instruments health and performance; precisely it refers to the operation of visualising the data in real-time and performing simple analysis tasks, like producing spectral densities, change data binning and so on.

This analysis has been regularly performed during testing campaigns to check in real-time the output of the scientific channels and of housekeeping, i.e. thermometers, instruments supply and control biases, power supply parameters and other.

The HFI team used KST for this purpose, while the LFI team used Rachel just for tests on single RCAs and TQL (Telemetry Quick Look) for tests of the integrated instrument.

The same tools used during the testing campaigns will be used during the commissioning phase (CPV), i.e. the first 90 days of the flight dedicated to check functionality and to set the instruments in the best configuration before starting the CMB survey, and during the rest of the mission.

Its use during the tests, both on ground and during CPV, it fundamental in order to monitor in real-time what is the instrument response to the commands sent and the testing procedures require that the test operator checks specific scientific or housekeeping channels during test execution.

During the surveys the objective of the quick look analysis is different, mainly focused on monitoring that no anomalies are present and that each instrument keeps its nominal configuration, which assures the best performance.

Since Mission planning it was clear that also a quick look analysis involving simultaneously both instruments would be very useful in order to identify possible systematic effects on the cross correlations; this exploits one of the big strenghts of Planck, which is having instruments with completely different technologies in the same focal plane; cross correlation can highlight issues that are not visible analysing a single instrument or can allow an immediate cross check on the other instrument if something unexpected is seen in the data stream and gives more chances of identifying its source.

During Planck Mission, LFI and HFI scientific and housekeeping data will be exchanged daily using the Exchange Format interface (EFDD), see \cite{efdd}. Exchange Format is a specific fits file format designed to be easily transferred and managed by the other DPC; it is already planned that the exchange will be organised in 2 subsequent phases, see \cite{efddicd}:
    \begin{enumerate}
        \item \textbf{\texttt{converted}} timelines will be created daily and available the day after data reception from the satellite, \texttt{converted} involves a minimal systematic effect removal and calibration to $V$ for scientific channels and $K$ for thermometers
        \item \textbf{\texttt{reduced}} instead includes also a deeper systematic removal and a calibration of scientific channels to $K$ or to $W$; they will be created once a week for the data of the previous week.
    \end{enumerate}
In detail:
    \begin{itemize}
        \item The HFI \texttt{converted} science timeline data is composed by the data, demodulated, low-pass filtered and converted to
volts using the appropriate gain
\item The HFI \texttt{reduced} science timeline data is composed by the \texttt{converted} data after the systematic effect
removal, (the headers of the files will include conversions to sky units for the bolometer and to
Kelvin for the thermometers).
\item The LFI \texttt{converted} timeline is composed by differenced radiometer data after the best gain modulation factor ($r$, see section~\ref{sec:erre})
application and calibrated in Volts.
\item The LFI \texttt{reduced} science timeline is composed by differenced radiometric data after the $r$
factor application, after calibration to Kelvin and after the systematic effect removal
    \end{itemize}

Following this concept, as soon as the data of the other instrument in Exchange Format arrives the day after reception, an operator should manually inspect them with the support of a handy graphical tool.

Moreover, this tool can be useful in later phases when a poorly understood effect for example on LFI data could be easily checked against HFI scientific or housekeeping data.

\section{KST for data crosscorrelation}

\subsection{KST}
\label{sec:kst}

KST software was created by Barth Netterfield \cite{kstinterview}, an astrophysicist of the University of Toronto. He needed a tool to look at the data in real time during acquisition and to inspect quickly large datasets.

The software is coded in C++ and based on the QT graphical libraries and the KDE libraries, it was released as Free Software under GPL and is now released together with the K Desktop Environment.

It is a GUI oriented plotting and analysis tool focused on working with large datasets or directly interfaced with streaming acquisition electronics; the implementation is focused on the GUI being quick and easy to use. It also supports plugins in order to extend its features.

\begin{figure}
    \centering
    \includegraphics[width=.9\textwidth]{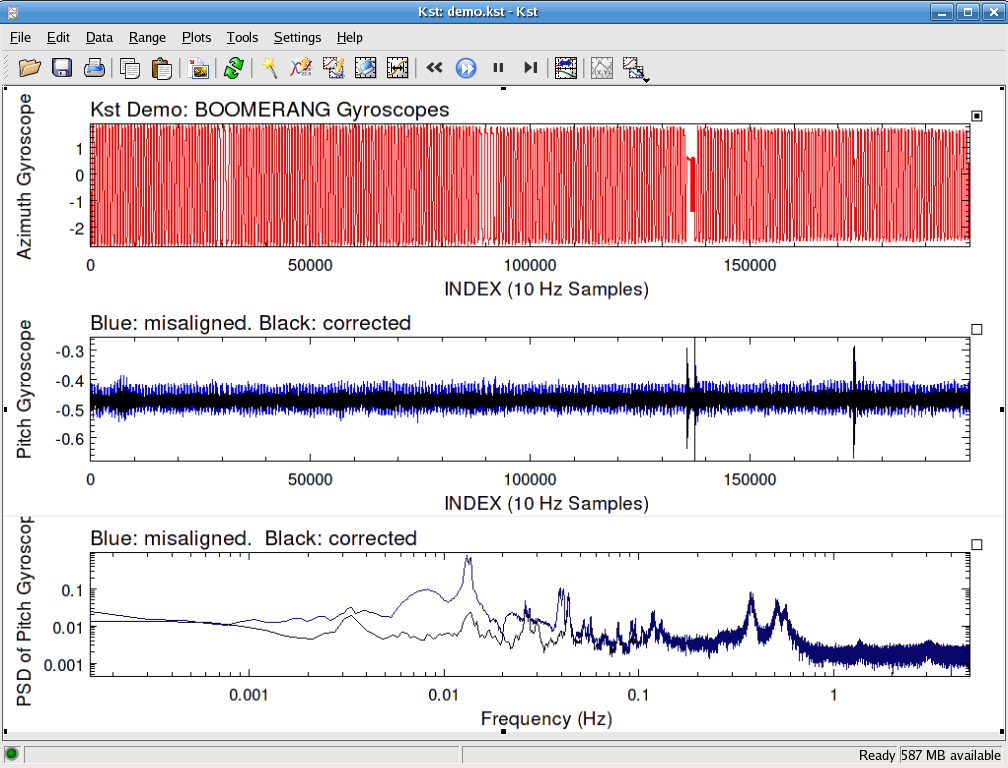}
    \caption{Screenshot of KST plotting BOOMERANG data}
    \label{fig:kstdemo}
\end{figure}

It was since then used by the Cosmic Microwave detection balloon experiment Boomerang (\cite{bernardis99},\cite{boomerangWWW}) as a tool for graphical plotting and navigating the data, see an example in figure~\ref{fig:kstdemo}.

The Canadian Space Agency funded the project in order to support the ESA Planck Mission with 2 main objectives:
    \begin{enumerate}
        \item Being the HFI quick look analysis software used for plotting and inspecting live the data coming from the satellite and the tool for quickly inspecting data afterwards.
        \item Being the designated tool for data cross correlation between HFI and LFI to be used to manage data in Exchange Format
    \end{enumerate}

\subsection{Concept of quick look analysis sessions}

KST is very efficient for loading data and producing plots, however manually loading data, adjusting legends and labels is quite tedious to be performed routinely.

Therefore the need of automating this tasks has been fulfilled by preparing a configurable set of scripts which will manage the repetitive part of data visualisation.

Beside this, having predefined analysis sessions helps standardising the way different operators look at the data, improving robustness and reliability of data analysis.

Another interesting KST feature is that once a part of a dataset, for example two hours of data, including amplitude spectral densities or histograms, it is visualised, it is easy to switch to the next hours back and forth just by pressing buttons; this allows to set up even a complex analysis environment and then apply it sequentially to the complete duration of the dataset.

\subsection{Implementation}

KST supports a scripting language and an interactive console which is available at the bottom of KST GUI; the implementation is based on javascript and binds most of the C++ objects which form the KST window.

It is possible to run command line functions interactively on KSTscript console, and prepare scripts offline that can be run and take arguments from this console.

KSTscript allows to load data, create windows, make plots, modify labels, create power spectra and so on.

The first objective was to create an abstraction layer for accessing the data, in order to be able to build analysis sessions without worrying about how the data are stored.

This abstraction layer is a javascript library named PlanckData I wrote using an object oriented structure, it allows the access to different data format with the same syntax.

For now the available dataset interfaces are:
    \begin{itemize}
        \item \texttt{Level S} 24 hour simulation: HFI and LFI simulated timestreams produced by the Planck simulator, called \texttt{Level S}
        \item Exchange format: ability to read fits files in Exchange format structured in predefined folders
    \end{itemize}
It is important to note that data access itself is already built in into KST, what this library allows is to call the data from javascript with a very simple syntax, typically, once defined what data source we are using, the syntax is:

\begin{lstlisting}
source = new PlanckData("24hLevelS");
time = source.ch["LFI23a"].getTime(start,stop,binning);
signal = source.ch["LFI23a"].getSig(start,stop,binning);
\end{lstlisting}

This example is based on the 24hlevelS dataset which makes use of the old LFI naming conventions; this dataset contains data already differenced, therefore each RCA has an $a$ and a $b$ streams.

The library also provides consistent labelling to all sessions, sampling frequencies computations and data binning.

Scripts do not directly manage the data, they just control the behaviour of the C++ objects which do the hard work.

\subsection{Base sessions}
Base sessions perform atomic actions on KST, and they are the basis for building advanced sessions.

Each base session has an inline help and a test function in order to improve user friendliness.

Sessions already implemented includes data loading and plotting functions for:
    \begin{itemize}
        \item data streams
        \item amplitude spectral densities
        \item histograms
        \item simple correlation
        \item correlation function
        \item autocorrelation
        \item cross correlation
        \item cross power spectrum
    \end{itemize}

User interface is managed in 3 different ways:
    \begin{itemize}
        \item command line interface, i.e. call a function with arguments:
            \begin{lstlisting}
tod({ch:['2700','2410','1911','2201'], t0:0, dt:2, bin:0, over:0});
            \end{lstlisting}
            where ch is channel list, t0 begin time, dt time interval, bin binning and over defines whether each curve should be on its own plot or overplotted.
            \item a pop up dialogue interface, which asks the user the needed arguments with hints
            \item a complete GUI, with a table of all available channels to be selected by clicking on tickboxes
    \end{itemize}

Basic sessions have been already developed and tested on Level S test data and on an Exchange format dataset and are fully functional; they are suitable for supporting the activity of an operator looking at the data, but they are not proper analysis sessions by themselves.

\begin{figure}
    \centering
    \includegraphics[width=\textwidth]{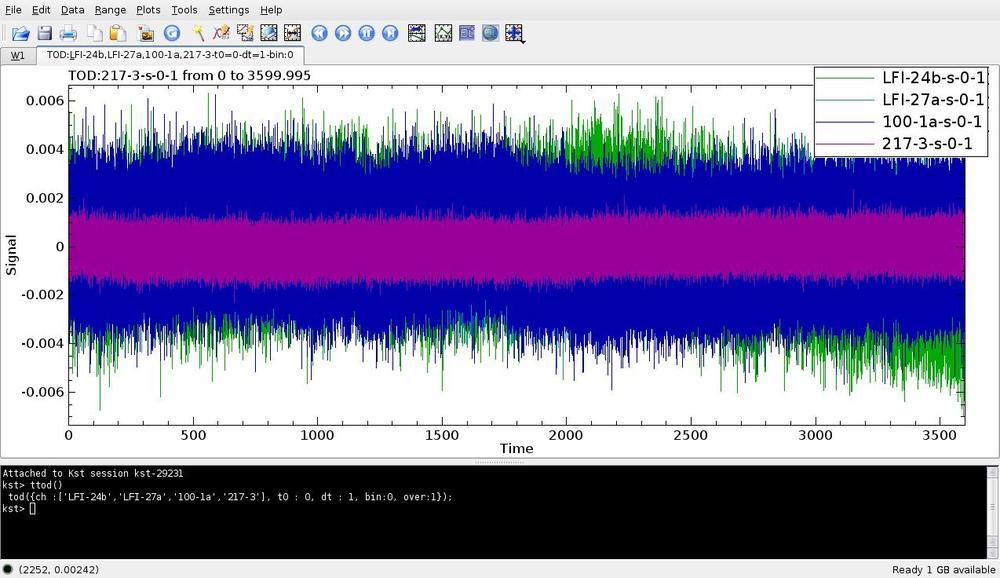}
    \caption{Overplotting of HFI and LFI simulated data streams produced with \texttt{tod} session, in the console there is the function call}
    \label{fig:tod}
\end{figure}

\begin{figure}
    \centering
    \includegraphics[width=\textwidth]{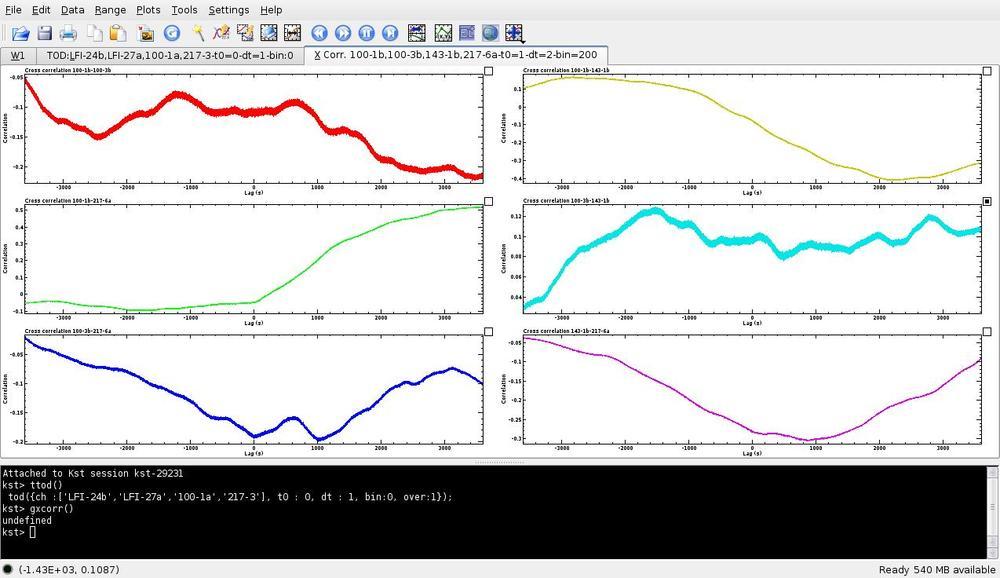}
    \caption{All possible crosscorrelation function plots of the previously selected channels produced by \texttt{xcorr}}
    \label{fig:xcorr}
\end{figure}

\begin{figure}
    \centering
    \includegraphics[width=\textwidth]{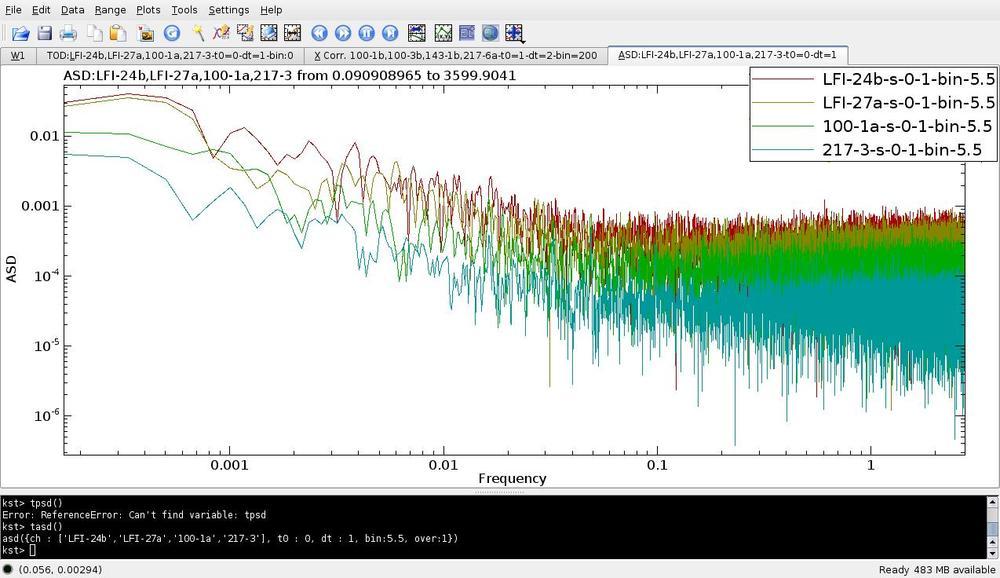}
    \caption{Overplotted amplitude spectral densities produced by \texttt{asd}}
    \label{fig:asd}
\end{figure}

\subsection{Advanced sessions}

The role of advanced sessions is different, their aim is to be a good, even if partial, view of a possible specific issue; therefore they should be self standing and should contain all the information needed for understanding if an issue is present or not.

This type of session could also be used as a starting point for reporting, i.e. the operator, after having looked through the plots, could log that there is no evidence of unexpected signatures and mark that the data range is clean.

In this context I'm evaluating the possibility of exploiting the available interface with the ELOG software, which is an electronic logbook which was used by HFI during test campaigns to log all the operations performed on the instrument directly from within KST, being also able to save a .kst file or a KST screenshot.

This could be handy to keep track of all the analysis performed on data cross correlation and having the possibility to easily search through the electronic logbook to go back to a previously identified issue.

\subsection{List of proposed advanced analysis sessions}

Before discussing HFI and LFI data cross correlation, we first discuss simpler sessions, based on data from a single instrument. This will help understanding the sessions structure and the main feature present in the data.

\subsubsection{LFI-only sessions}
\begin{itemize}
    \item Powergroup based:

LFI is biased thanks to four different powergroups; one of the first analysis tasks is to check if there are common effects on the same powergroup.
Here follows a list showing the correspondence between the RCAs and the powergroups:

    \begin{itemize}
        \item powergroup 1:    18,26
        \item powergroup 2: 19,20,28
        \item powergroup 3: 21,22,24,27
        \item powergroup 4: 23,25 
    \end{itemize}

In order to simplify the analysis it would be better to choose the same frequency, therefore 70 GHz, whose RCA are spread on all powergroups.

Proposed sessions:
\begin{enumerate}
\item  18,19,21,23: crosscorrelation between different powergroups
\item 19,20: crosscorrelation on the same powergroup
\item  21,22: crosscorrelation on the same powergroup
\end{enumerate}

Total of \textbf{8} plots, which could be accommodated on a single KST window, we can choose to plot just one channel for each RCA listed using differenced data

\item Frequency based:

26,23,28 different frequencies and different powergroups in order not to see effects in the same powergroup

\textbf{3} plots

\item RCA internal correlation

The correlation between the channels of the same RCA should show the strong correlation of the channels of the same radiometer due to the $1/f$ amplifiers gain fluctuations; while the correlation with the channels of the other radiometer is just due to the sky signal.
The number of plots for checking all the correlations for each RCA are:
\begin{itemize}
    \item \textbf{6} single channel;
    \item \textbf{1} combined diodes. 
\end{itemize}

\end{itemize}

\subsubsection{HFI-only sessions}

            Considering a channel for frequency and type (i.e. spider web or polarisation sensitive bolometers), we have 9 channels:

\begin{enumerate}
 \item 100-1b
 \item 143-1a
 \item 143-5 
 \item  217-5a
 \item  217-2 
 \item  353-3a
 \item  353-7 
 \item  545-3 
 \item  857-3 
\end{enumerate}

HFI channels readout electronics is organised in groups of six channels; the channels on the same group share the same FPGA (Field-programmable gate array) controls the channel settings and performs the AC demodulation.

A possible consistent choice of sessions, which was discussed with HFI experts, can be:
    \begin{itemize}
        \item 2 channels at different frequency on the same belt, in order to identify possible electrical effect on the belts

\item 2 nearby channels  in order to check for possible correlations due to thermal coupling

\item both bolometers of a Polarisation Sensitive Bolometer

\item a channel for each frequency

    \end{itemize}

The objective of the KST sessions is to be quickly analysed by a human operator, therefore it is necessary to choose among these possible cross correlations a consistent subset with no more than two windows and less than nine plots each. 

\subsubsection{HFI-LFI cross correlation sessions}

\begin{itemize}
    \item Satellite power supply monitoring
    
    A very simple correlation analysis involves monitoring the electrical power supply line connecting the scientific instruments to the satellite.
    
    In case of anomalies in the instrument outputs, it could be useful to have a session ready to cross correlate voltage and current measurements of the supply line.

Tables~\ref{tab:hfisup} and \ref{tab:lfisup} summarise the HFI and LFI housekeeping parameters that can be used in this correlation analysis.
    \begin{center}
        \begin{table}[h]
\caption{HFI supply line parameters}

\begin{tabular}{ll}
Parameter Code & Description \\
\hline
HM013250 	& 	DPU + 5V (V)\\
HM014250 	& 	DPU - 15V (V)\\
HM015250 	& 	DPU +15V (V)\\
HM016250 	& 	DPU +2.5V (V)\\
HM006250 	& 	DPU Intensity of +5V (mA)\\
HM007250 	& 	DPU Intensity of -15V (mA)\\
HM008250 	& 	DPU Intensity of +15V (mA)\\
HM009250 	& 	DPU Intensity of +25V (mA)\\
    \end{tabular}
    \label{tab:hfisup}
            \end{table}

    \end{center}
    
        \begin{center}
        \begin{table}

        \caption{LFI supply line parameters}

\begin{tabular}{ll}
Parameter Code & Description \\
\hline
LM214342 & VCC Current   (A) \\
LM215342 & +15V Current  (mA)\\
LM216342 & -15V Current  (mA)\\
LM218342 & VCC Voltage   (V) \\
LM219342 & +15V Voltage  (V) \\
LM220342 & -15V Voltage  (V) \\

    \end{tabular}
    \label{tab:lfisup}
            \end{table}

    \end{center}

    \item Crosscorrelation between 70 \& 100 GHz data
    
    This analysis can be performed by correlating data from two LFI 70 GHz and two HFI 100 GHz horns, on symmetric positions on the focal plane; cross correlating for example a 70 GHz channel with a 100 GHz very near and another on the opposite side of the focal plane can highlight correlations due to their proximity, e.g. something related to cross talk or to non idealities in a region of the telescope reflectors.
    
    All the possible crosscorrelations are just six and can fit in a single KST window or being overplotted.
    
    \item Large frequency coverage cross correlation
    
A cross correlation involving a wider set of Planck channels could check for signatures of unexpected correlated effects coming, for example, from the telescope.

In order not to increase too much the time the operator needs to go through this session, a subset of channels per frequency can be:
    \begin{itemize}
        \item a LFI 44 GHz channel differenced
        \item a LFI 70 GHz channel differenced
        \item a HFI 100 GHz channel
        \item a HFI 143 GHz channel
        \item a HFI 217 GHz channel
    \end{itemize}
    All the possible cross correlations, which could be either performed using the correlation function or using cross spectra, use 10 plots.

\item Polarisation dedicated sessions
    
    Possible polarisation dependent effects could be investigated by correlating the polarisation measurement from LFI 70 GHz and HFI 100 GHz.
    
    For LFI it is necessary to consider two radiometers connected to the orthogonal polarisations of the same horn, for each of them taking the difference of sky versus load and calibrate to sky units; at that point polarisation is measured by taking the difference of the two streams. This procedure gives a first estimation of the polarised signal, for a deeper analysis it is necessary to combine the data of the two horns symmetric in the focal plane.
    
    The corresponding procedure should be performed on HFI 100GHz, exploiting the data of the orthogonal Polarisation Sensitive Bolometers.
    
    The \texttt{reduced} exchange data are calibrated to Kelvin units, LFI data are already differenced, therefore this analysis can give a rough estimation of the polarised signal as a function of time.
\end{itemize}

The previous sessions mainly examined issues that shouldn't be present in nominal conditions; the only correlations should come from the sky signal, which will be removed by a dedicated algorithm already implemented into the data analysis pipeline.

The next session instead is fully dedicated to thermal issues that are already expected to be present; therefore the aim is to check that they comply with requirements and identify any anomaly.

\begin{itemize}
    \item sessions dedicated to the 4K stage
    
    LFI reference loads are attached to the HFI 4K shield; correlations between the LFI scientific signal and 4K loads fluctuations are already top priority for correcting the observation results: it is already planned that the data analysis pipeline will ingest the 4K thermometers data in order to clean the LFI scientific signal from these fluctuations.
    
    The objective of an analysis with KST instead is focused on involving also HFI scientific channels, because the focal plane, where feed horns are located, is part of the 4K stage, and its fluctuations could impact the bolometric output.
    
    Thermal fluctuations are very different depending on the position on the 4K shell; the interface to the 4K cooler is at the bottom of the structure, where the fluctuations are stronger; on the top plate, which is the HFI focal plane, instead, there is a strict thermal regulation performed by PID controlled heaters, therefore the temperature is very stable; in the middle, where LFI 30 and 44 GHz reference loads are located, there is an intermediate situation which is difficult to model due to the complex design.
    
    Therefore there are two possible scenarios:
        \begin{itemize}
            \item top thermal plate fluctuations: in this case the it would be useful to correlate a 70 GHz channel differenced data stream, an HFI thermal sensor located on the top plate and a 100 GHz channel.
            \item LFI 30/44 GHz load fluctuations: in this case a 30 GHz channel, a 44 GHz one, a 100 GHz and one of the HFI 4K belt thermometers (4KL1/2) should be used.
        \end{itemize}

%

\end{itemize}

\section{Preliminary tests on real data}

Planck satellite was tested in Liege Space Centre (CSL) during Summer 2008; this was the first time HFI and LFI worked together integrated into the satellite and in flight-like cryogenic conditions.

It was not in the test plans to produce exchange format and perform cross correlations between the two instruments data,


but the long duration test, where HFI and LFI acquired data in stable conditions for 24h was a really good opportunity for testing both exchange format production and cross correlation analysis.

Unfortunately, due to bugs in the exchange format conversion software, exchange format data were made available after the conclusion of CSL campaign, after much data were already been analysed with other tools.

However it is very interesting to use this dataset for testing the concept of sessions and also further analyse this dataset which still shows few unclear effects.

The main problem we are facing is that the exchange format data available now are just the \texttt{converted} streams, which means they are calibrated to volts, while most of the analysis sessions described in this chapter relies on scientific data calibrated in Kelvin units. This is the only way we can really compare data from different channels.

In the next future probably CSL data will be made available also as \texttt{reduced} timestreams, until then we can start testing the environment and few sessions.

\begin{figure}
    \centering
    \includegraphics[angle=-90,width=\textwidth]{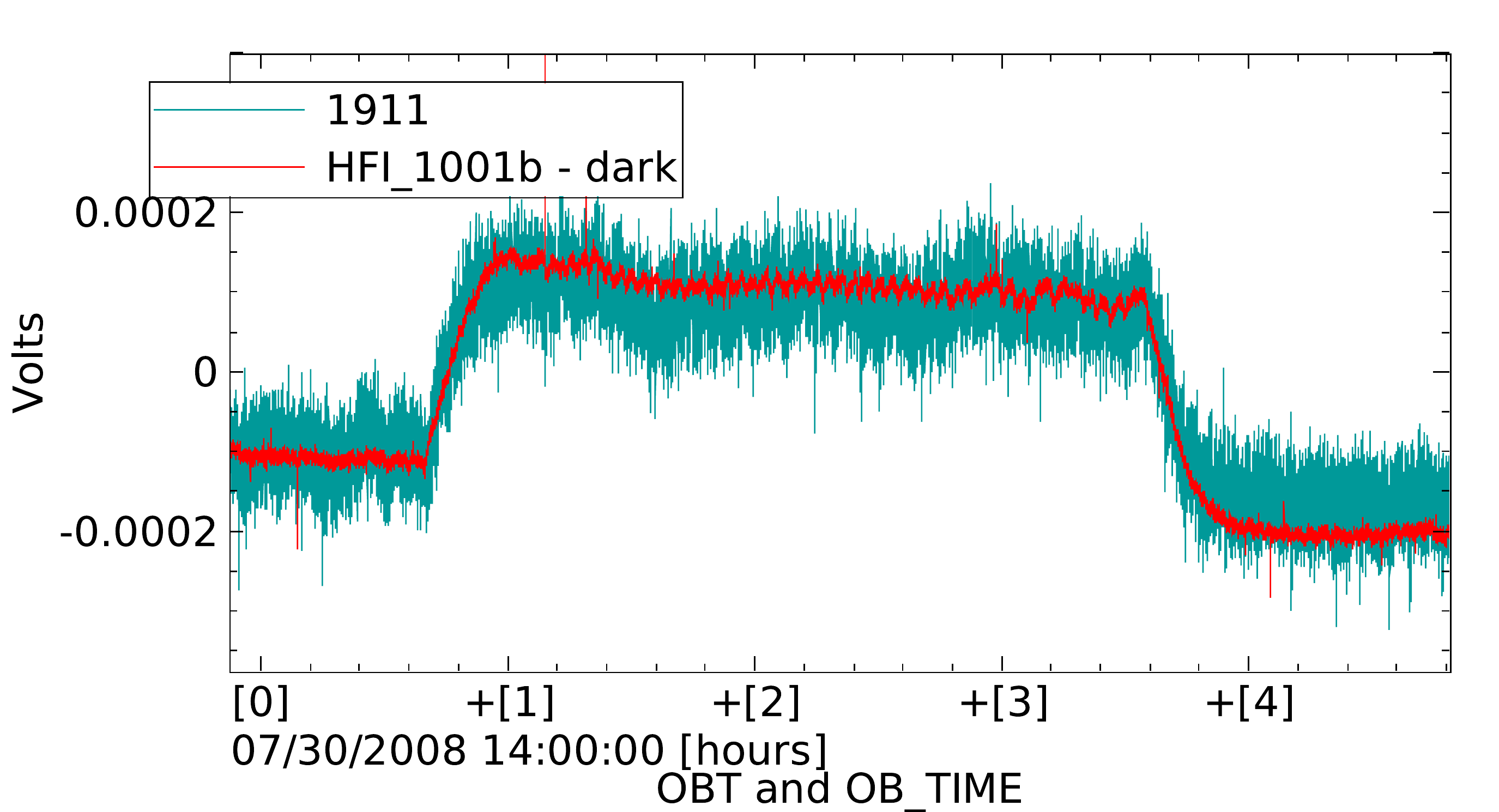}
    \caption{LFI19S-11 and HFI 100-1B channels overplotted during CSL long duration test}
    \label{fig:step_24h}
\end{figure}

In figure~\ref{fig:step_24h} HFI 100 GHz 1b is overplotted to LFI 19 70 GHz S-11, HFI bolometers show a strong drift, probably due to a thermal drift in the cold HFI stages. Therefore I have subtracted this drift by differencing the HFI 100 GHz 1b channel with one of the dark bolometers, which are bolometers equivalent to the others but that are not connected to any horn and are therefore very useful to distinguish between thermal effects and sky signal.

It shows a few hours period when the thermal shield was refilled with helium, which caused is a distinct temperature change recorded by all HFI and LFI channels. Then I have removed the mean and applied an arbitrary scale factor in order to mimic the effect of a calibration to Kelvin and obtain a similar step height.

\begin{figure}
    \centering
    \includegraphics[angle=-90,width=\textwidth]{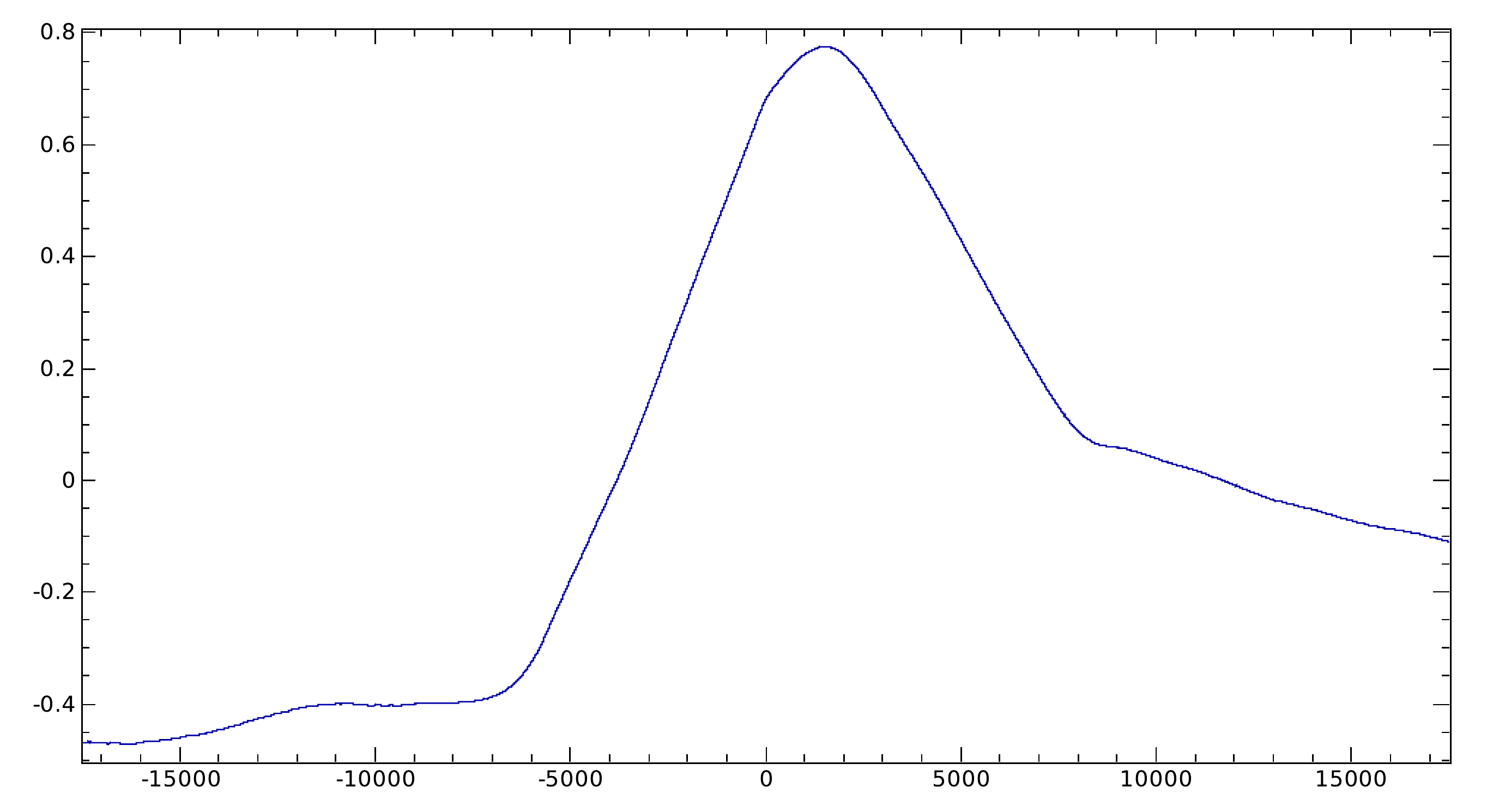}
    \caption{Crosscorrelation of LFI19S-11 and HFI 100-1B channels}
    \label{fig:step_24h_corr}
\end{figure}

Figure~\ref{fig:step_24h_corr} shows the cross correlation function between the data from these channels; as expected there is a very strong correlation around 0 offset, which means that the signal is very synchronous on both channels.

\section{Conclusion}

This project \texttt{kst4planck}, developed by Rodrigo Leonardi (UCSB), Luis Mendes (ESA), Andrew Walker (UBC) and myself, is focused on delivering a set of predefined quick analysis sessions in order to simplify the process of cross correlating HFI and LFI data; the objective is to support the removal of systematics and to look for unexpected issues during the survey.

At this point there is already a set of working base sessions, its development showed that the session concept is really applicable to KST and technically successful.

The next important step is to demonstrate its validity also on the field, by concentrating on the CSL long duration test and improve the knowledge about the environmental conditions by exploiting the crosscorrelation of scientific data and housekeeping of both instruments.

This is a fundamental step on the path for delivering a flight-ready tool both to the LFI and the HFI DPCs; kst4planck will be available among other tools and will be specifically dedicated (but not only) to exchange format data crosscorrelation.

\chapter{Conclusion}
\label{ch:conclusion}

The work I have discussed in this PhD thesis is focused on three main topics: LFI radiometers software modelling, 4K stage thermal transfer functions and HFI-LFI cross-correlation.

This work has been performed in the context of the ESA Planck Mission, which has the objective of providing the definitive measurement of the Cosmic Microwave anisotropies temperature down to the smallest scales, a good measurement of anisotropies E-mode polarisation and possibly the first detection of B-mode polarisation. 

\paragraph{QIMP model and radiometers bandpasses}

During my PhD I have developed QUCS Integrator of Measured Performances (QIMP), a software model which relies on measured frequency response of the LFI radiometers component and on an analytical waveguide simulator in order to simulate the LFI bandpasses channel by channel.

Precise knowledge of radiometer bandpasses is fundamental for effective foregrounds removal. Foregrounds removal algorithms work on maps at different frequencies by exploiting the different spectral indexes of the foreground components. However, each HFI and LFI channel has a broadband response, and each map is the convolution of the spectra of foregrounds and CMB with the instruments frequency response. Thus, an iterative process involving component separation and maps rescanning with realistic bandpasses is needed to take this effect into account.
Moreover, detailed bandpasses mismatch characterisation of the two radiometers of the same horn allows the study of the leakage of foregrounds temperature signal into polarisation signal, due to the fact that the calibration is tailored to the spectrum of the CMB, while foregrounds have different spectra.

LFI bandpasses were measured on dedicated tests during the integrated RCAs campaigns; however a part of these measurements showed strong systematic effects due to the test setup (30 GHz and 70 GHz) or had an insufficient frequency coverage (30 GHz).
The QIMP model can compensate for this issue by providing bandpass estimations based on frequency response tests on the single RCA components and an analytic waveguide simulator. QIMP results were validated against measurements and showed a good agreement, with the exception of a discrepancy on 44 GHz channels which is still under investigation.

Single components performance tests (which are less prone to systematic effects because of simpler setups) and QIMP simulations provide the best knowledge of LFI radiometers frequency response.

\paragraph{4K stage thermal transfer functions}
I have analysed the thermal transfer functions between the Sorption cooler cold end LVHX1 and all thermometers on the 4K stage, see chapter~\ref{ch:thermal}.

I had the opportunity of computing the transfer functions on data acquired during the HFI CQM cryogenic tests in Liege Space Centre with and without PID thermal regulation and compare these results with the thermal functions computed from the simulations performed with the HFI thermal model after PID settings optimisation; the results are consistent considering that the optimisation aimed at reducing PIDs power consumption.

4K stage thermal transfer functions are fundamental for LFI because they allows to compute what is the effect of Sorption cooler and 4K cooler fluctuations on the 4K reference loads. The requirement on the spectrum of the 4K load thermometers is $10 \micro K/\sqrt(Hz)$ for frequencies over $10 mHz$ and $100 \micro K/\sqrt(Hz)$ on lower frequencies; the last estimation of this effect, dated June 2006, showed that for 30 and 44 GHz RCAs the requisite was not met, while for  70 GHz it was met.  In the next future the fluctuations will be analysed both by using the latest transfer functions and by directly computing the spectrum of the thermometers, in order to check what is the impact on maps of these fluctuations.

\paragraph{HFI-LFI cross-correlation}

Starting from September 2007 I have led the development of the \texttt{kst4planck} project, with the objective of designing and implementing quick look analysis sessions for the cross-correlation of HFI and LFI scientific and housekeeping data in Exchange format, see chapter~\ref{ch:combined_qla}.

Analysis sessions are implemented in KST, a data visualisation software; we have developed and tested the basic sessions for loading, displaying and performing simple analysis, like spectra, cross-correlation, cross power spectra.
 
Basic sessions are the elements for building high level sessions, i.e. a integrated set of windows and plots focusing on a specific issue, that will allow an operator to quickly navigate the data for identifying unexpected systematic effects in the cross-correlations.
Among the huge amount of data and possible cross-correlations we have identified a set of useful cross-correlation sessions that we are now ready to implement.

HFI-LFI cross-correlation is a topic of growing interest in the Planck community; Planck peculiarity of having two very different detection technologies, and therefore different systematic effects and different signatures in the data of external systematics, is a strong opportunity for identifying systematic effects and remove their effect from the scientific data.

\section{Future work}

\paragraph{QIMP} QIMP still has strong margins of improvement: first of all thanks to the dedicated large band and section by section bandpass tests which were performed by Stefania De Nardo in her master thesis (\cite{thesis_denardo}) on a 30 GHz spare RCA. A detailed comparison between the test results and the model simulations will improve the model bandpass characterisation; the improved model will then be used to run again simulations on the flight bandpasses.
Moreover, 44 GHz test data will be analysed again in order to figure out if this could have caused the bandpass discrepancy.

QIMP can also be used to model other instrument performances, its use is very interesting because we can have a deeper insight on what component inside the RCA is the reason of a specific behaviour. Follows a list of interesting analysis topics that can be covered by QIMP:
    \begin{itemize}
    \item implement the non linear model for 30 and 44 GHz RCA by including into the BEM model the compression factor measured during the test campaigns
    \item use the results of the measurements of the radiometer output as a function of input temperature (LIS) in order to validate the absolute output of the 30 and 44 GHz radiometers
            \item estimate LFI radiative emission: an analysis on LFI radiative emission was performed by Cristian Franceschet in his master thesis \cite{thesis_franceschet}, he used a radiometer model based on LFI nominal performance and showed that LFI radiative emission towards HFI is about two orders of magnitude below requirements; it would be interesting to analyse now LFI emission with the real radiometers performance
        \item improve the modelling of the bandpass response for the reference load including second order effects like spillover, return loss and eventually straylight (in collaboration with Francesco Cuttaia)
    \end{itemize}
    
\paragraph{Thermal transfer functions} 

The next important step for improving the knowledge of the thermal behaviour of the 4K stage is the estimation of the thermal transfer functions in the integrated flight model satellite cryogenic tests performed at the Liege Space Centre in 2008. Thanks to this we can check the consistency with the transfer functions predicted by the simulations. The most suitable test for this analysis is the long duration test, where the instruments were left in stable nominal acquisition mode for more than 24 hours.

For concluding this study it would be interesting to fit the transfer function with a simple low pass filter in order to have a rough estimation of the thermal time constants of the system and to have a simple representation of the thermal behaviour of the 4K shell, that could be used in place of the full response when a quick estimation is needed.

\paragraph{HFI-LFI cross-correlation}

\texttt{kst4planck} is still in a early stage of development, the first task is the implementation of the high level sessions, but this is just the beginning of a deep test work based on the CSL long duration test data aimed at refining the design of the high level sessions in order to be ready for the operations.

The main focus, however, is not on the tool development, but on training people on understanding what we can learn by cross-correlation, where to search for systematics and how to better exploit this combined analysis.
In fact, KST is just one of the tools for engaging cross-correlation, in the near future it will be important to widen the point of view and understand how LIFE (i.e. the LFI radiometers data analysis software), KST and the DPC pipeline can be used to take the most out of HFI-LFI cross-correlation.
  
\bibliographystyle{plain}
\bibliography{arxiv}

\begin{thebibliography}{10}

\bibitem{efddicd}
C.~Mercier A.~Zacchei.
\newblock Icd 030 dpc dpc timelines exchange.
\newblock Technical report, OATS, 2008.

\bibitem{kstinterview}
Tom Acrewoods.
\newblock Interview to barth netterfield about kst.
\newblock \url{http://tom.acrewoods.net/writing/kstinterview}, 2007.

\bibitem{thesis_battaglia}
Paola Battaglia.
\newblock Advanced software model of the planck low frequency instrument
  radiometer.
\newblock Master's thesis, Universita' degli Studi di Milano, 2002.

\bibitem{PlanckLFIScientificReq}
M.~Bersanelli, M.~Seiffert, R.~Hoyland, and A.~Mennella.
\newblock Planck-lfi scientific requirements.
\newblock Technical report, IASF/Universit\`a{} degli Studi di Milano, Mar
  2002.

\bibitem{bhandari2004}
P.~Bhandari, M.~Prina, R.~C. Bowman, C.~Paine, D.~Pearson, and A.~Nash.
\newblock Sorption coolers using a continuous cycle to produce 20 k for the
  planck flight mission.
\newblock {\em Cryogenics}, 44(6-8):395--401, June-August 2004.
\newblock 2003 Space Cryogenics Workshop.

\bibitem{QUcsschema}
M.~E. Brinson and S.~Jahn.
\newblock Qucs: A gpl software package for circuit simulation, compact device
  modeling and circuit macromodeling from dc to rf and beyond.
\newblock In {\em {MOS-AK} Meeting}, MiPlaza, High Tech Campus Eindhoven, 2008.

\bibitem{2003A&A...409..375C}
B.~{Cappellini}, D.~{Maino}, G.~{Albetti}, P.~{Platania}, R.~{Paladini},
  A.~{Mennella}, and M.~{Bersanelli}.
\newblock {Optimized in-flight absolute calibration for extended CMB surveys}.
\newblock {\em aap}, 409:375--385, October 2003.

\bibitem{2001Carroll}
Sean~M. {Carroll}.
\newblock {The Cosmological Constant}.
\newblock {\em Living Reviews in Relativity}, 4:1, February 2001.

\bibitem{bbn}
A.~{Coc}, E.~{Vangioni-Flam}, P.~{Descouvemont}, A.~{Adahchour}, and
  C.~{Angulo}.
\newblock {Updated Big Bang Nucleosynthesis Compared with Wilkinson Microwave
  Anisotropy Probe Observations and the Abundance of Light Elements}.
\newblock {\em apj}, 600:544--552, January 2004.

\bibitem{prelaunch_wg}
O~et~al D'Arcangelo.
\newblock The planck-lfi ﬂight model composite waveguides.
\newblock {\em AA}, 2009.

\bibitem{bernardis99}
P.~de~Bernardis, P.~A.~R. Ade, R.~Artusa, J.~J. Bock, A.~Boscaleri, B.~P.
  Crill, G.~De~Troia, P.~C. Farese, M.~Giacometti, V.~V. Hristov,
  A.~Iacoangeli, A.~E. Lange, A.~T. Lee, S.~Masi, L.~Martinis, P.~V. Mason,
  P.~D. Mauskopf, F.~Melchiorri, L.~Miglio, T.~Montroy, C.~B. Netterfield,
  E.~Pascale, F.~Piacentini, P.~L. Richards, J.~E. Ruhl, and F.~Scaramuzzi.
\newblock Mapping the {CMB} sky: The {BOOMERanG} experiment.
\newblock {\em New Astronomy Review}, 43:289--296, July 1999.

\bibitem{thesis_denardo}
Stefania DeNardo.
\newblock Measurements and analysis of the planck lfi receivers spectral
  response.
\newblock Master's thesis, Universita' degli Studi di Milano, 2008.

\bibitem{planckbluebook}
G.~{Efstathiou}, C.~{Lawrence}, and I.~{Tauber}, editors.
\newblock {\em Planck: the Scientific Programme}.
\newblock European Space Agency, 2 edition, 2005.

\bibitem{1931Einstein}
Albert Einstein.
\newblock Zum kosmologischen problem der allgemeinen relativit\"atstheorie.
\newblock {\em Sitzungsberichte der Pr\"u\ss{}ischen Akademie der
  Wissenschaften}, pages 235--237, Apr 1931.

\bibitem{prelaunch_receivers}
Mennella et~al.
\newblock The response of the planck-lfi fm receivers.
\newblock In preparation for Astronomy and Astrophysics.

\bibitem{thesis_franceschet}
Cristian Franceschet.
\newblock Application of the planck-lfi advanced simulator to performances
  assessments of the qualification models.
\newblock Master's thesis, Universita' degli Studi di Milano, 2004.

\bibitem{1981Guth}
A.~H. {Guth}.
\newblock {Inflationary universe: A possible solution to the horizon and
  flatness problems}.
\newblock {\em Physical Review D}, 23:347--356, January 1981.

\bibitem{CSLreport}
Guy Guyot.
\newblock Hfi csl cryotest report.
\newblock Technical report, IAS, Univ. Paris Sud/CNRS (France), 10 2006.

\bibitem{1929Hubble}
Edwin~Powell Hubble.
\newblock A relation between distance and radial velocity among extra-galactic
  nebulae.
\newblock {\em Proceedings of the National Academy of Sciences}, 15:168, 1929.

\bibitem{Ylinen18}
N~Hughes.
\newblock Planck 70ghz lfi pfm.5 - rca 018 acceptance test report.
\newblock Technical report, Elektrobit Microwave Ltd, 02 2006.

\bibitem{efdd}
C.~Mercier J.~Sternberg, A.~Zacchei.
\newblock Planck idis dmc exchange format design document.
\newblock Technical report, ESA, 2008.

\bibitem{PSB}
B.~{Jones}, R.S. {Bhatia}, J.J. {Bock}, and A.E. {Lange}.
\newblock A polarization sensitive bolometric detector for observations of the
  cosmic microwave background.
\newblock In {\em SPIE}, 2003.

\bibitem{2003Lamarre}
J.-M. {Lamarre}, J.~L. {Puget}, M.~{Piat}, P.~A.~R. {Ade}, A.~E. {Lange},
  A.~{Benoit}, P.~{De Bernardis}, F.~R. {Bouchet}, J.~J. {Bock}, F.~X.
  {Desert}, R.~J. {Emery}, M.~{Giard}, B.~{Maffei}, J.~A. {Murphy}, J.-P.
  {Torre}, R.~{Bhatia}, R.~V. {Sudiwala}, and V.~{Yourchenko}.
\newblock {Planck high-frequency instrument}.
\newblock In J.~C. {Mather}, editor, {\em IR Space Telescopes and Instruments.
  Edited by John C. Mather . Proceedings of the SPIE, Volume 4850, pp. 730-739
  (2003).}, volume 4850 of {\em Presented at the Society of Photo-Optical
  Instrumentation Engineers (SPIE) Conference}, pages 730--739, March 2003.

\bibitem{liddle}
Liddle.
\newblock {\em An Introduction To Modern Cosmology}.
\newblock Wiley, 2003.

\bibitem{1996ApJ...470...38L}
C.~H. {Lineweaver}, L.~{Tenorio}, G.~F. {Smoot}, P.~{Keegstra}, A.~J. {Banday},
  and P.~{Lubin}.
\newblock {The Dipole Observed in the COBE DMR 4 Year Data}.
\newblock {\em apj}, 470:38--+, October 1996.

\bibitem{imagingfirstlight}
A.~{Mennella}, C.~{Baccigalupi}, A.~{Balbi}, M.~{Bersanelli}, C.~{Burigana},
  C.~{Butler}, B.~{Cappellini}, G.~{De Gasperis}, F.~{Hansen}, D.~{Maino},
  N.~{Mandolesi}, M.~{Maris}, G.~{Morgante}, P.~{Natoli}, F.~{Pasian},
  F.~{Perrotta}, P.~{Platania}, L.~{Valenziano}, F.~{Villa}, and A.~{Zacchei}.
\newblock {Imaging the first light: experimental challenges and future
  perspectives in the observation of the Cosmic Microwave Background
  Anisotropy}.
\newblock {\em ArXiv Astrophysics e-prints}, February 2004.

\bibitem{2008arXiv0803.0593N}
M.~R. {Nolta}, J.~{Dunkley}, R.~S. {Hill}, G.~{Hinshaw}, E.~{Komatsu},
  D.~{Larson}, L.~{Page}, D.~N. {Spergel}, C.~L. {Bennett}, B.~{Gold},
  N.~{Jarosik}, N.~{Odegard}, J.~L. {Weiland}, E.~{Wollack}, M.~{Halpern},
  A.~{Kogut}, M.~{Limon}, S.~S. {Meyer}, G.~S. {Tucker}, and E.~L. {Wright}.
\newblock {Five-Year Wilkinson Microwave Anisotropy Probe (WMAP) Observations:
  Angular Power Spectra}.
\newblock {\em ArXiv e-prints}, March 2008.

\bibitem{1999Perlmutter}
S.~{Perlmutter}, G.~{Aldering}, G.~{Goldhaber}, R.~A. {Knop}, P.~{Nugent},
  P.~G. {Castro}, S.~{Deustua}, S.~{Fabbro}, A.~{Goobar}, D.~E. {Groom}, I.~M.
  {Hook}, A.~G. {Kim}, M.~Y. {Kim}, J.~C. {Lee}, N.~J. {Nunes}, R.~{Pain},
  C.~R. {Pennypacker}, R.~{Quimby}, C.~{Lidman}, R.~S. {Ellis}, M.~{Irwin},
  R.~G. {McMahon}, P.~{Ruiz-Lapuente}, N.~{Walton}, B.~{Schaefer}, B.~J.
  {Boyle}, A.~V. {Filippenko}, T.~{Matheson}, A.~S. {Fruchter}, N.~{Panagia},
  H.~J.~M. {Newberg}, W.~J. {Couch}, and {The Supernova Cosmology Project}.
\newblock {Measurements of Omega and Lambda from 42 High-Redshift Supernovae}.
\newblock {\em The Astrophysical Journal}, 517:565--586, June 1999.

\bibitem{bandpass_torsti}
Torsti Poutanen.
\newblock Effects of bandpass mismatch in 70 ghz maps.
\newblock Technical report, Helsinky University, 2007.

\bibitem{microwave-pozar:1998}
David~M. Pozar.
\newblock {\em Microwave engineering}.
\newblock John Wiley \& Sons, Inc, 2 edition, 1998.

\bibitem{1998Riess}
A.~G. {Riess}, A.~V. {Filippenko}, P.~{Challis}, A.~{Clocchiatti},
  A.~{Diercks}, P.~M. {Garnavich}, R.~L. {Gilliland}, C.~J. {Hogan}, S.~{Jha},
  R.~P. {Kirshner}, B.~{Leibundgut}, M.~M. {Phillips}, D.~{Reiss}, B.~P.
  {Schmidt}, R.~A. {Schommer}, R.~C. {Smith}, J.~{Spyromilio}, C.~{Stubbs},
  N.~B. {Suntzeff}, and J.~{Tonry}.
\newblock {Observational Evidence from Supernovae for an Accelerating Universe
  and a Cosmological Constant}.
\newblock {\em The Astronomical Journal}, 116:1009--1038, September 1998.

\bibitem{bigbang}
M.~{Roos}.
\newblock {Expansion of the Universe - Standard Big Bang Model}.
\newblock {\em ArXiv e-prints}, February 2008.

\bibitem{2002A&A...391.1185S}
M.~{Seiffert}, A.~{Mennella}, C.~{Burigana}, N.~{Mandolesi}, M.~{Bersanelli},
  P.~{Meinhold}, and P.~{Lubin}.
\newblock {1/f noise and other systematic effects in the Planck-LFI
  radiometers}.
\newblock {\em aap}, 391:1185--1197, September 2002.

\bibitem{boomerangWWW}
WWW Page, 2002.

\bibitem{gainmodel}
F.~Villa and L.~Terenzi.
\newblock Radiometer gain model.
\newblock Technical report, IASFBO, May 2006.

\bibitem{rca27_report}
Fabrizio Villa.
\newblock Fm 30ghz rca27 data analysis report.
\newblock Technical report, IASFBO, 2006.

\bibitem{wald1984}
Robert~M. Wald.
\newblock {\em General Relativity}.
\newblock {University Of Chicago Press}, June 1984.

\bibitem{2006ApJ...650....1W}
Y.~{Wang} and P.~{Mukherjee}.
\newblock {Robust Dark Energy Constraints from Supernovae, Galaxy Clustering,
  and 3 yr Wilkinson Microwave Anisotropy Probe Observations}.
\newblock {\em apj}, 650:1--6, October 2006.

\bibitem{Wedge1991}
Scott~W. Wedge, , and David~B. Rutledge.
\newblock Noise waves and passive linear multiports.
\newblock {\em Microwave and Guided Wave Letters, IEEE}, 1(5):117--119, May
  1991.

\end{thebibliography}

\chapter{Acknowledgements}

First of all I want to thank Marco Bersanelli and Aniello Mennella: Marco gave me the possibility of starting the adventure of my PhD, following my interest for cosmology, a change which was fundamental for my life; Aniello followed and encouraged me continously during these years. Both have a strong positive attitude and the ability of placing confidence on collaborators that I wouldn't have hoped to find anywhere.

The whole experimental cosmology group in Milan is a great place for enjoying working on research with helpful people, including in casual order Maurizio Tomasi, Simona Donzelli, Simona Pezzati, Benedetta Cappellini, Davide Maino.

In the larger Planck community, I found support and enjoyed working with people from IASF-Bologna, like Fabrizio Villa, Luca Terenzi, Francesco Cuttaia, Gianluca Morgante, Enrico Franceschi; from Trieste, like Andrea Zacchei, Samuele Galeotta, Marco Frailis, Anna Gregorio; from UCSB, Peter Meinhold and Rodrigo Leonardi; from ESA Luis Mendes; from IFP, Ocleto D'Arcangelo, from Alcatel, Cristian Franceschet, Paola Battaglia, Roberto Silvestri, Paolo Leutenegger, Maurizio Miccolis and Flavio Ferrari; from UBC Andrew Walker.

A special thank to Jean-Michèl Lamarre and Andrea Catalano, which made it possible a very interesting collaboration with the HFI team and hosted me in Paris for four months, sharing some of their knowledge of the HFI with me.

I'd like also to mention Lara Sidoli, Ada Paizis and all the people involved into the \href{http://www.iasf-milano.inaf.it/~InStrada/}{La ricerca in mezzo ad una strada} initiative, it was really important to participate in protesting against what the Italian government is doing against the Italian research.

Another thank to the Alcatel Alenia Space Milano, which supported the first year of my PhD and made me understand that working in industry would be really my last choice.

I've passed an important part of my PhD working on my laptop, a great thank to all the people who built and maintain the great \href{http://www.gnu.org/philosophy/free-sw.html}{Free Software} I've been using, like Ubuntu, vim, Python, Scipy, KST, Gnumeric, Kile and especially QUCS, which was fundamental for my work, and in particular Bastien Roucaries, which developed the waveguide component model, a long-time missing piece in my work.

On the personal side I want to thank my wife, that despite I caused her to suffer for being far away, busy and under stress, she've been on my side.
My family also was very supportive when I decided to change from Engineering to Physics, which was a quite difficult choice and also during my PhD.
A thank also to my brother which introduced me to GNU/Linux, and shares always my passion for Free Software.

\appendix

\chapter{Scattering parameters}

A very useful technique for modelling radiometric components is through an S-parameters 2-ports component. 

S-parameters, or Scattering parameters relate the amplitude and phase of reflected and output signal to the input signal as a function of frequency. In order to characterise the frequency response of a 2-ports component, four S-parameters need to be measured all over the bandwidth:
\begin{description}
\item[S11] reflected wave amplitude(port 1)/input wave amplitude(port 1) 
\item[S21] output wave(port 2)/input wave(into port 1)
\item[S12] output wave(port 1)/input wave(into port 2)
\item[S22] reflected wave(port 2)/input wave(port 2) 
\end{description}

Therefore the relation between input and output voltages can be expressed in matrix notation, see \cite{microwave-pozar:1998}:

\begin{equation}
 \begin{bmatrix}
    V^{out}_{1} \\
     V^{out}_{2}
   \end{bmatrix} = 
\begin{bmatrix}
    S_{11} & S_{12}  \\
    S_{21} & S_{22}
\end{bmatrix}
\begin{bmatrix}
    V^{in}_{1} \\
     V^{in}_{2}
   \end{bmatrix}
\end{equation}

\paragraph{Return and insertion losses}

Return loss (RL) is sometimes used instead of $S_{11}$, their relation is:
    \begin{equation}
        RL = 20 log_{10} (S_{11}) dB
    \end{equation}
Insertion loss is instead very different from $S_{21}$, Insertion loss refers just to the losses due to internal dissipation, while $S_{21}$ is the ratio between output and input amplitude, therefore it includes both the effect of reflection and the effect of internal dissipation. Their relation is:
    \begin{equation}
    IL = -10\log_{10}\frac{\left|S_{21}\right|^2}{1-\left|S_{11}\right|^2}
    \end{equation}

\chapter{Microwave Waveguides analytical treatment}
\label{sec:wg}

Electromagnetic wave guides are hollow cavity enclosed by metallic walls. The propagation of electromagnetic waves can be calculated by solving the Maxwell equations in the cavity:
\begin{align}
\nabla\times \overrightarrow{E} &= - j\omega \mu \overrightarrow{H} \\
\nabla\times\overrightarrow{H} &= j \omega \varepsilon \overrightarrow{E}
\end{align}
where $\overrightarrow{E}$ is the electric field, $\overrightarrow{H}$ the magnetic field, $\omega$ the 
angular frequency, $\mu$ is the magnetic permeability, $\varepsilon$ is electrical permittivity. 
These equations have multiple solutions, or modes, which are eigenfunctions of the equation system. 
Each mode is therefore characterised by an eigenvalue, which corresponds
to the axial propagation velocity of the wave in the guide.

\begin{figure}[h]
\hfill\includegraphics[width=0.6\columnwidth]{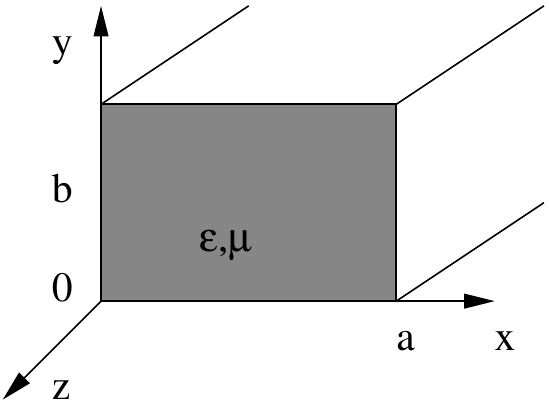}\hfill\hbox{}
\caption{Rectangular wave guide schematic}
\label{fig:rect}
\end{figure}

Let us consider the special case of rectangular waveguides(see figure~\ref{fig:rect}). 
Because this waveguide is invariant under translation along $z$ axis, the variation of the electric field can be written
as $\overrightarrow{E}(x,y,z)=\overrightarrow{E}(x,y) \left[ A e^{\gamma z} + B e^{\gamma^* z}\right]$, 
where $\gamma$ is the propagation constant, and $A$, $B$ two arbitrary complex constants. 
This waveguide structure allows propagation of two kinds of 
modes~\citep{microwave-pozar:1998}:
\begin{itemize}
\item $\text{TE}_{mn}$ mode (Transverse Electric) which have no electric field in the direction of propagation.
\item $\text{TM}_{mn}$ modes (Transverse Magnetic) which have no magnetic field in the direction of propagation.
\end{itemize}
Moreover it can be shown that modes exist only beyond a frequency called cut-off frequency. Below this frequency
propagation of this mode cannot occur apart from evanescent modes that decay exponentially. For a rectangular waveguide 
this frequency is defined by:
\begin{equation}
f_{c_{mn}} =  \frac{c}{2\pi \sqrt{\varepsilon\mu}} \sqrt{ \left(\frac{m\pi}{a}\right)^2 + \left(\frac{n\pi}{b}\right)^2}
\end{equation}
where $c$ is the speed of light in vacuum.
However, we will limit to the computation of the fundamental mode propagation, which is the only one necessary to characterise Planck LFI waveguides. Moreover, higher modes need the 
knowledge of excitation and are usually simulated using a full wave simulation (like FEM simulation or method of moments).

The propagation constant $\gamma_{n,m}$ can be decomposed in two 
terms one propagation term called $\beta$ and one loss 
term called $\alpha$: 
\begin{equation}
\gamma_{mn} = j \beta_{mn} + \alpha_{mn}
\end{equation}
Using the free space wavenumber $k=\omega\sqrt{\mu\varepsilon}$ and $k_{c_{mn}}=\frac{2\pi}{c} \sqrt{\varepsilon\mu} f_{c_{nm}}$, 
it could be show that:
\begin{equation}
\beta_{nm}=\sqrt{k^2-k_{c_{nm}}^2} 
\end{equation}
Loss term can be decomposed into a dielectric component and a resistive component ($\alpha=\alpha_c+\alpha_d$). 
The dielectric loss can be computed for all modes
and has the following form:
\begin{equation}
\alpha_d=\frac{k^2 \tan \delta}{2\beta}
\end{equation}
where $\delta$ is the material loss factor.

    The resistive loss, instead, is mode-dependent. For $\text{TE}_{10}$ mode we have:
\begin{equation}
\alpha_c=\frac{R_s}{a^3 b \beta k Z_0} \left( 2 b \pi^2 + a^3  k^2 \right)
\end{equation}
$R_s$ is the metal surface resistance which is linked to the metal conductivity, $\rho$, by:
\begin{equation}
R_s = \sqrt{ \pi f \mu \rho}
\end{equation}
    Another important parameter to compute is the waveguide impedance:
\begin{equation}
Z_\text{TE}=\frac{k\eta}{\beta}
\end{equation}
Where $\eta$ is the vacuum impedance ($\eta\simeq 120\pi \simeq 377 \Omega$)

Using this model a transmission line model (or S-parameters model) can be delined, where the S matrix is give by:
\begin{equation}of
S=\begin{pmatrix} 
  \frac{(\frac{Z_\text{TE}}{Z_c} - \frac{Z_c}{Z_\text{TE}}) \sinh (\gamma l)}{n} &  \frac{2}{n} \\
\frac{2}{n}  & \frac{(\frac{Z_\text{TE}}{Z_c} - \frac{Z_c}{Z_\text{TE}}) \sinh (\gamma l)}{n}                          
\end{pmatrix}
\end{equation}
where $l$ is waveguide length  and $n$ is given by:
\begin{equation}
n= 2 \cosh (\gamma l) + \left(z + \frac{1}{z}\right) \sinh (\gamma l)
\end{equation}
    $Z_c$ is an arbitrary chosen global circuit parameter called normalisation impedance. For ease of computation, 
it is usually taken for simulation purpose as $Z_C=\unit{1}{\ohm}$.

Noise is computed easily from S-parameters using Bosma's theorem~\cite{Wedge1991}.

\chapter{Plots of all the comparisons between measured and QIMP bandpasses}
\label{sec:compapp}

The following plots show the comparison between the bandpasses directly measured with a Swept source test and the simulated bandpasses obtained by QIMP integrating the frequency response measurements performed on the single components. 

On abscissa the frequency is expressed in GHz units, while in ordinate the bandapass consists of the ratio between output voltage in Volts and input power in Watts in the logarithmic units: 10 log 10([V]/[W]).

For 30 and 44 GHz channels, the bandpass measurements are just relative to the central frequency, therefore their integral was normalized to the integral of the model.

\section{70 GHz}
\subsection{RCA 18}

\begin{figure*}[h]
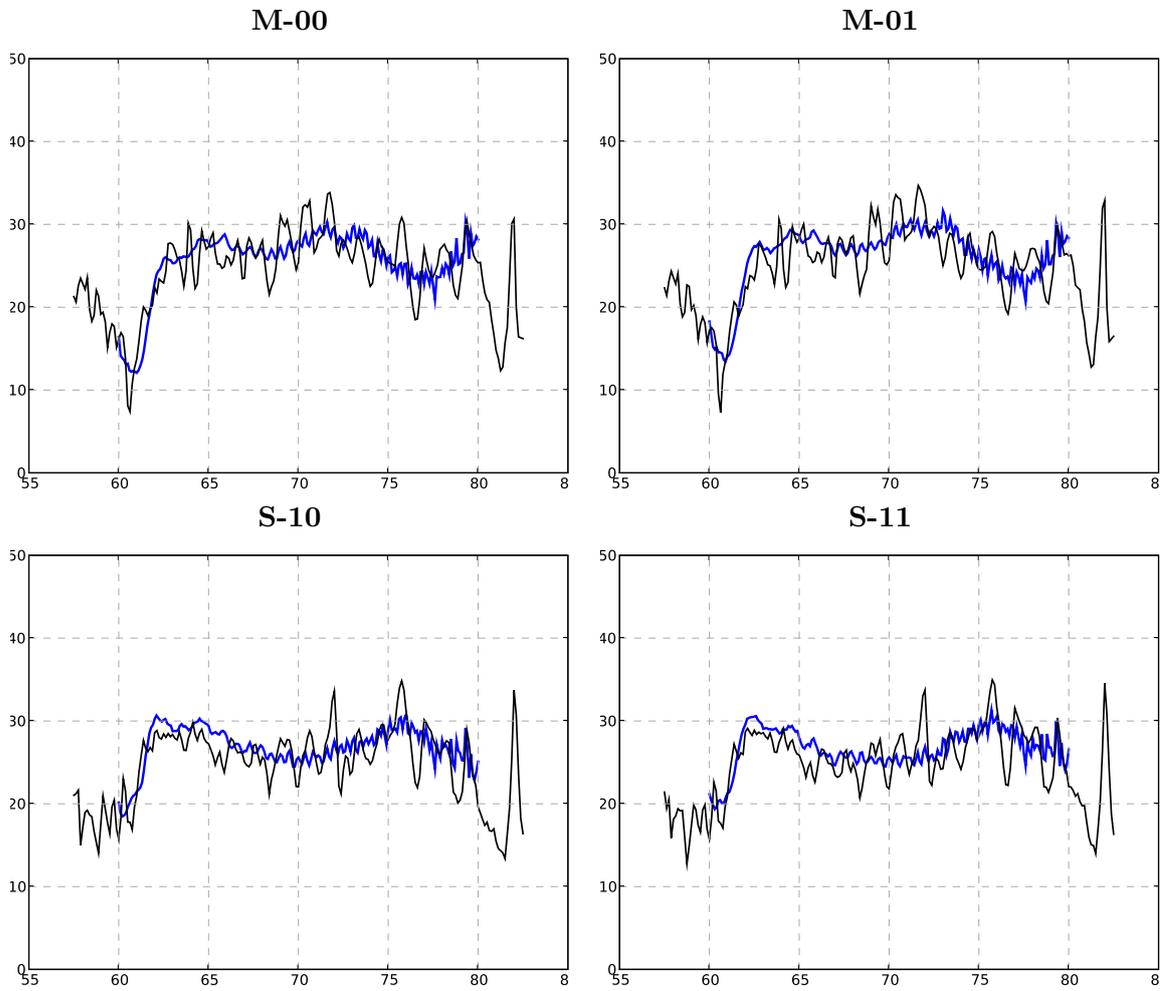

    \begin{center}
    \begin{tabular}{c c} 
    \textbf{M-00} & \textbf{M-01} \\
    \plota{compLFI18-M-00}  & \plota{compLFI18-M-01}\\
    \textbf{S-10} & \textbf{S-11}\\
    \plota{compLFI18-S-10}  & \plota{compLFI18-S-11} \\
    \end{tabular} 
    \caption{RCA \textbf{18} comparison between measurements(black) and model(blue) results}
    \end{center}
\end{figure*}
\clearpage
\subsection{RCA 19}

\begin{figure*}[h]
    \begin{center}
    \begin{tabular}{c c} 
    \textbf{M-00} & \textbf{M-01} \\
    \plota{compLFI19-M-00}  & \plota{compLFI19-M-01}\\
    \textbf{S-10} & \textbf{S-11}\\
    \plota{compLFI19-S-10}  & \plota{compLFI19-S-11} \\
    \end{tabular} 
    \caption{RCA \textbf{19} comparison between measurements(black) and model(blue) results}
    \end{center}
\end{figure*}
\clearpage
\subsection{RCA 20}

\begin{figure*}[h]
    \begin{center}
    \begin{tabular}{c c} 
    \textbf{M-00} & \textbf{M-01} \\
    \plota{compLFI20-M-00}  & \plota{compLFI20-M-01}\\
    \textbf{S-10} & \textbf{S-11}\\
    \plota{compLFI20-S-10}  & \plota{compLFI20-S-11} \\
    \end{tabular} 
    \caption{RCA \textbf{20} comparison between measurements(black) and model(blue) results}
    \end{center}
\end{figure*}
\clearpage
\subsection{RCA 21}

\begin{figure*}[h]
    \begin{center}
    \begin{tabular}{c c} 
    \textbf{M-00} & \textbf{M-01} \\
    \plota{compLFI21-M-00}  & \plota{compLFI21-M-01}\\
    \textbf{S-10} & \textbf{S-11}\\
    \plota{compLFI21-S-10}  & \plota{compLFI21-S-11} \\
    \end{tabular} 
    \caption{RCA \textbf{21} comparison between measurements(black) and model(blue) results}
    \end{center}
\end{figure*}
\clearpage
\subsection{RCA 22}

\begin{figure*}[h]
    \begin{center}
    \begin{tabular}{c c} 
    \textbf{M-00} & \textbf{M-01} \\
    \plota{compLFI22-M-00}  & \plota{compLFI22-M-01}\\
    \textbf{S-10} & \textbf{S-11}\\
    \plota{compLFI22-S-10}  & \plota{compLFI22-S-11} \\
    \end{tabular} 
    \caption{RCA \textbf{22} comparison between measurements(black) and model(blue) results}
    \end{center}
\end{figure*}
\clearpage
\subsection{RCA 23}

\begin{figure*}[h]
    \begin{center}
    \begin{tabular}{c c} 
    \textbf{M-00} & \textbf{M-01} \\
    \plota{compLFI23-M-00}  & \plota{compLFI23-M-01}\\
    \textbf{S-10} & \textbf{S-11}\\
    \plota{compLFI23-S-10}  & \plota{compLFI23-S-11} \\
    \end{tabular} 
    \caption{RCA \textbf{23} comparison between measurements(black) and model(blue) results}
    \end{center}
\end{figure*}
\clearpage
\section{44 GHz}
\subsection{RCA 24}

\begin{figure*}[h]
    \begin{center}
    \begin{tabular}{c c} 
    \textbf{M-00} & \textbf{M-01} \\
    \plota{compLFI24-M-00}  & \plota{compLFI24-M-01}\\
    \textbf{S-10} & \textbf{S-11}\\
    \plota{compLFI24-S-10}  & \plota{compLFI24-S-11} \\
    \end{tabular} 
    \caption{RCA \textbf{24} comparison between measurements(black) and model(blue) results}
    \end{center}
\end{figure*}
\clearpage
\subsection{RCA 25}

\begin{figure*}[h]
    \begin{center}
    \begin{tabular}{c c} 
    \textbf{M-00} & \textbf{M-01} \\
    \plota{compLFI25-M-00}  & \plota{compLFI25-M-01}\\
    \textbf{S-10} & \textbf{S-11}\\
    \plota{compLFI25-S-10}  & \plota{compLFI25-S-11} \\
    \end{tabular} 
    \caption{RCA \textbf{25} comparison between measurements(black) and model(blue) results}
    \end{center}
\end{figure*}
\clearpage
\subsection{RCA 26}

\begin{figure*}[h]
    \begin{center}
    \begin{tabular}{c c} 
    \textbf{M-00} & \textbf{M-01} \\
    \plota{compLFI26-M-00}  & \plota{compLFI26-M-01}\\
    \textbf{S-10} & \textbf{S-11}\\
    \plota{compLFI26-S-10}  & \plota{compLFI26-S-11} \\
    \end{tabular} 
    \caption{RCA \textbf{26} comparison between measurements(black) and model(blue) results}
    \end{center}
\end{figure*}
\clearpage
\section{30 GHz}
\subsection{RCA 27}

\begin{figure*}[h]
    \begin{center}
    \begin{tabular}{c c} 
    \textbf{M-00} & \textbf{M-01} \\
    \plota{compLFI27-M-00}  & \plota{compLFI27-M-01}\\
    \textbf{S-10} & \textbf{S-11}\\
    \plota{compLFI27-S-10}  & \plota{compLFI27-S-11} \\
    \end{tabular} 
    \caption{RCA \textbf{27} comparison between measurements(black) and model(blue) results}
    \end{center}
\end{figure*}
\clearpage
\subsection{RCA 28}

\begin{figure*}[h]
    \begin{center}
    \begin{tabular}{c c} 
    \textbf{M-00} & \textbf{M-01} \\
    \plota{compLFI28-M-00}  & \plota{compLFI28-M-01}\\
    \textbf{S-10} & \textbf{S-11}\\
    \plota{compLFI28-S-10}  & \plota{compLFI28-S-11} \\
    \end{tabular} 
    \caption{RCA \textbf{28} comparison between measurements(black) and model(blue) results}
    \end{center}
\end{figure*}
\clearpage

\end{document}